\documentclass[10pt,twocolumn,showpacs,amsmath,amssymb,aps,prx,longbibliography,superscriptaddress]{revtex4-1}
\usepackage{times}
\usepackage[utf8]{inputenc}
\usepackage{color}
\usepackage{array}
\usepackage{verbatim}
\usepackage{multirow}
\usepackage{amsmath}
\usepackage{amssymb}
\usepackage{graphicx}
\usepackage{esint}
\usepackage[unicode=true,
 bookmarks=true,bookmarksnumbered=true,bookmarksopen=false,
 breaklinks=true,pdfborder={0 0 0},backref=false,colorlinks=true]
 {hyperref}
\usepackage{enumitem}
\usepackage{cleveref}
\usepackage[normalem]{ulem}
\usepackage{xcolor}
\usepackage[bb=boondox]{mathalfa}
\usepackage[title]{appendix}
\usepackage{makecell}

\hypersetup{
    linkcolor=red,          
    citecolor=blue,        
    filecolor=magenta,      
    urlcolor=magenta           
}

\usepackage{cleveref}
\crefname{appendix}{Appendix}{Appendices}
\crefname{equation}{Eq.}{Eqs.}
\crefname{figure}{Fig.}{Figs.}
\crefname{table}{Table}{Tables}
\crefname{section}{Section}{Sections}
\crefname{mythe}{Theorem}{Theorems}
\crefname{mydef}{Definition}{Definitions}

\renewcommand{\paragraph}[1]{\vspace{0.2cm}{\bf \textit{#1}}}

\newcommand{\bbb}[1]{\textcolor{blue}{#1}}
\newcommand{\rrr}[1]{\textcolor{red}{#1}}

\definecolor{Gray}{gray}{0.85}
\newcolumntype{a}{>{\columncolor{Gray}}c}

\usepackage{simpler-wick}
\allowdisplaybreaks[1] 
\newcommand{\mbf}{\mathbf}
\newcommand{\mbb}{\mathbb}
\newcommand{\mcl}{\mathcal}
\newcommand{\mbs}{\boldsymbol}
\newcommand{\mrm}{\mathrm}

\newcommand{\td}{\widetilde}
\newcommand{\ovl}{\overline}

\def\qq{\mathbf{q}}

\def\pp{\mathbf{p}}
\def\kk{\mathbf{k}}
\def\rr{\mathbf{r}}
\def\RR{\mathbf{R}}

\def\ttau{\boldsymbol{\tau}}

\def\AA{\mathbf{A}}
\def\uu{\mathbf{u}}
\def\dd{\mathbf{d}}
\def\GG{\mathbf{G}}
\def\KK{\mathbf{K}}
\def\GGamma{\boldsymbol{\Gamma}}

\def\MBZ{\mathrm{MBZ}}
\def\gBZ{\mathrm{gBZ}}

\def\XA{\mathrm{XA}}
\def\YA{\mathrm{YA}}
\def\XO{\mathrm{XO}}
\def\YO{\mathrm{YO}}
\def\ZA{\mathrm{ZA}}
\def\ZO{\mathrm{ZO}}

\def\Aone{\mrm{A1}}
\def\Bone{\mrm{B1}}

\def\up{\uparrow}
\def\down{\downarrow}

\begin{document}
\title{Electron phonon coupling in the topological heavy fermion model of twisted bilayer graphene}
\author{Yi-Jie Wang}
\affiliation{International Center for Quantum Materials, School of Physics, Peking University, Beijing 100871, China}

\author{Geng-Dong Zhou}
\affiliation{International Center for Quantum Materials, School of Physics, Peking University, Beijing 100871, China}

\author{Biao Lian}
\affiliation{Department of Physics, Princeton University, Princeton, New Jersey 08544, USA}

\author{Zhi-Da Song}
\email{songzd@pku.edu.cn}
\affiliation{International Center for Quantum Materials, School of Physics, Peking University, Beijing 100871, China}
\affiliation{Hefei National Laboratory, Hefei 230088, China}
\affiliation{Collaborative Innovation Center of Quantum Matter, Beijing 100871, China
}

\begin{abstract}
    On flat bands of the magic-angle twisted bilayer graphene, exotic correlation physics unfolds. 
    Phonons, through mediating an effective electron-electron interaction, can play a crucial role in selecting various electronic phases. 
    In this study, we derive the full electron-phonon coupling (EPC) vertex from the microscopic tight-binding lattice, and identify the significance of each phonon mode.
    We then project the EPC vertices onto the topological heavy fermion (THF) basis [Song and Bernevig, Phys. Rev. Lett. 129, 047601 (2022)], and show that an anti-Hund's interaction $\hat{H}_{\rm A}$ is induced on each moir\'e-scale local $f$-orbital, with strengths 1 to 4 meV. 
    We analyze the phonon-induced multiplet splittings, which can significantly affect the local correlation. 
    As an example, we elaborate on the phonon-favored symmetry-breaking orders at even-integer fillings. 
    Through systematic self-consistent Hartree-Fock calculations, we uncover a tight competition between $\Gamma$-phonon-favored orbital orders, $K$-phonon-favored inter-valley coherent orders, and the kinetic and Coulomb-favored orders. 
    Contrary to EPC, the carbon atom Hubbard repulsion induces an on-$f$-site Hund's interaction $\hat{H}_{\rm H}$ with strengths 1 to 3 meV that partly counteracts the effect of $\hat{H}_{\rm A}$. 
    The combined influence of $\hat{H}_{\rm A,H}$ on the multiplet splitting and symmetry-breaking states is discussed. 
    In the end, we explore the possibility of finding an exotic Dirac semi-metal formed solely by $c$-electrons at the charge-neutrality point, while $f$-impurities exhibit a symmetric Mott gap by forming non-degenerate singlets under $\hat{H}_{\rm A,H}$. Experimental features that distinguish such a state are discussed. 
\end{abstract}

\maketitle

\section{Introduction}

Flat band electrons in the magic-angle twisted bilayer graphene (MATBG) \cite{BM_2011} exhibit rich interaction-driven physics. 
At integer fillings, correlated symmetry-breaking orders are observed \cite{Cao_2018_CI, Lu_2019_superconductors, Yankowitz_2019_tuning, Xie_2019_spectroscopic, Choi_2019_electronic}, which can spontaneously break the $C_{3z}$ rotation symmetry \cite{Jiang_2019_charge, Kerelsky_2019_maximized}, or exhibit a Kekul\'e pattern indicating inter-valley coherence \cite{Nuckolls_2023_quantum}. 
In the presence of external magnetic fields, correlated Chern insulators can be stabilized \cite{Nuckolls_2020_strongly, Stepanov_untying_2020, Das_symmetry_2021, saito_hofstadter_2021, wu_chern_2021, Park_2021_flavour, choi_correlation_2021}. 
With aligned hexagonal boron nitride (hBN) substrate which breaks $C_{2z}T$ symmetry, intrinsic Chern insulators are observed \cite{serlin_intrinsic_2020}. 
At non-integer fillings, the zero-energy peaks and Coulomb blockade observed in the tunneling spectrum \cite{Wong_2020_cascade, Zondiner_2020_cascade, Nuckolls_2020_strongly, Choi_2021_interaction, Oh_2021_evidence}, as well as the Pomeranchuk effect \cite{Rozen_2021_entropic, Saito_2021_isospin} and the sharp compressibility jumps with increased filling \cite{Wong_2020_cascade, Zondiner_2020_cascade, Rozen_2021_entropic, Saito_2021_isospin, Park_2021_flavour}, together hint at the formation of heavy Fermi liquid. 
A superconducting (SC) phase can also develop \cite{Cao_2018_SC, Lu_2019_superconductors, Yankowitz_2019_tuning, Arora_2020_superconductivity, Stepanov_untying_2020, Saito_independent_2020, Liu_tuning_2021}, enjoying unconventional features including the proximity to the correlated insulating states in the phase diagram, short coherence length \cite{Cao_2018_SC, Lu_2019_superconductors}, a V-shaped tunneling gap \cite{Oh_2021_evidence}, and nematicity \cite{Cao_2021_nematicity}. 
Suppressing the correlated insulators and superconductors, either by heating up or by external tuning such as enhancing screening or applying magnetic fields, a strange metal with near Planckian dissipation appears \cite{Polshyn_2019_large, Cao_2020_strange, Jaoui_2022_quantum}, featuring in the large linear-in-$T$ resistivity with a wide temperature span. 

Due to the non-trivial band topology protected by $C_{2z}T$ and an emergent particle-hole symmetry $P$ \cite{Zou_2018_band, Po_2019_faithful, Song_2019_allMATBG, Ahn_2019_NNfail, TBG2_2021}, finding a lattice representation with a finite orbital number is prohibited, and a theoretical diving into the strongly correlated physics has hence been hindered. 
Recently, a topological heavy fermion (THF) model \cite{Song_2022, Shi_2022_THF, herzogarbeitman_2024_heavy, singh_topological_2024} reveals that, the topological flat bands can be disentangled as an emergent set of moir\'e-scale local orbitals ($f$-orbitals), which are themselves trivially flat with vanishing inter-site overlaps, hybridized with itinerant Dirac fermions ($c$-electrons), which brings about the non-trivial topology. 
The $f$-orbitals dominate the flat band wave-functions in the vast majority of the moir\'e Brillouin zone (MBZ), while the hybridization with $c$-electrons only occurs in a small vicinity of the $\GGamma_M$-point. 
Within this picture, a large Hubbard-type on-$f$-site repulsion $U\sim 58\mrm{meV}$, which originates from the long-range Coulomb repulsion between 2D continuum electrons, is identified as the governing energy scale for the flat band electrons. As $U$ acts to suppress the $f$-charge fluctuation, the occupation number at each site will tend to some integer $N_f$, so that the multiplet of $N_f$-electron states behaves as a local moment, coupled to the bath of $c$-electrons. 
The strong correlation physics are thus able to be understood in a unified framework \cite{Chou_2023_Kondo, Zhou_2024_Kondo, Hu_2023_Kondo, Hu_2023_Symmetric, Datta_2023_heavy, Chou_2023_scaling, Lau_2023_topological, Rai_2023_DMFT, calugaru_thermoelectric_2024}: 
at non-integer fillings, the $f$-moments are Kondo screened at low temperatures, giving rise to the heavy Fermi liquid behaviors; while at integer fillings, the Ruderman-Kittel-Kasuya-Yosida (RKKY) interaction between $f$-moments, mediated by the $c$-bath, can win over the Kondo screening and lead to symmetry-breaking correlated states. 

Interestingly, with only the Coulomb repulsion of 2D electron gas taken into consideration, the flat bands enjoy a rather high emergent $\rm U(4)$ symmetry \cite{Kang_2019_strong, Vafek_2020_RG, Bultinck_2020_2U4, Seo_2019_ferro, TBG3_2021, Song_2022}. Numerous correlated phases with tiny energy differences are thus found closely competing; and consequently, any other sources of interaction that splits the orbital, valley or spin configurations, despite weak when compared to $U$, can play a significant role in finally determining the actual ground state. 
Especially, electron-phonon couplings (EPC) can be of great importance. 
Recent experiments have revealed a prominent EPC between the flat band electrons to the $K$-phonons \cite{Chen_ARPES_23}, which are then theoretically found to favor inter-valley coherent orders that are symmetric under the spinless-time-reversal $T$ at even-integer fillings \cite{Kwan_2023_Kphonon}. 
Such orders include one that preserves the moir\'e translation (dubbed as $\rm TIVC$) and appears at ultra-low strain, and one that breaks it (dubbed as the incommensurate Kekul\'e spiral order, IKS) and appears in strained samples \cite{Kwan_2021_IKS}. 
These orders match the Kekul\'e pattern \cite{Dumitru_2022_STM} observed experimentally for the correlated insulators at $\nu=\pm 2$ \cite{Nuckolls_2023_quantum}, where $\nu$ counts the total electron filling measured from the charge neutrality point (CNP). 
In particular, $\rm TIVC$ will not win as the ground state in the absence of EPC due to penalization from Coulomb interactions \cite{Kang_2019_strong, Bultinck_2020_2U4, Seo_2019_ferro, Vafek_2020_RG, TBG4_2021, TBG6_2021}. 
Moreover, as experimental evidence accumulates that screening Coulomb repulsion could stabilize the SC phase \cite{Stepanov_untying_2020, Saito_independent_2020, Liu_tuning_2021}, phonons appear as a plausible candidate for the pairing glue \cite{Wu_2018_SCop, Wu_2019_phonon, Lian_2019_SCac, Cea_2021_coulomb, Liu_2023_electronkphonon, Angeli_2019_valleyJT, Andrea_2022_local}. 
However, since any phonon-mediated interaction will be far weaker than the large repulsive $U$, $f$-electrons cannot be superconducting in a na\"ive mean-field picture. 
Nevertheless, in the heavy Fermi liquid regime, by splitting the $N_f$-electron multiplet, microscopic interactions such as EPC can significantly reshape the Kondo physics \cite{nozieres_kondo_1980}, which then strongly renormalizes the effective interaction among heavy quasi-particles \cite{yamada_perturbation_1975,yamada_perturbation_1975-1,yoshimori_perturbation_1976,hewson_fermi_1993,hewson_renormalized_2001}. 
In a recent work \cite{Wang_2024_TBGSC}, under the assumption that the Kondo screening is dominated by a $d$-wave two-electron doublet at $\nu\approx-2.5$, we obtained the asymptotic behavior of the renormalized interaction, and showed that the pairing strength to an inter-valley $d$-wave spin-singlet channel can be attractive. 
Such an SC order lies in the strongly correlated regime, which is set by the large $U$ and characterized by the small quasi-particle weight, and belongs to the strong coupling (BEC) limit because the pairing strength is of the same order as the heavy quasi-particle kinetic energy. It is also nematic and Euler-obstructed \cite{Yu_2022_euler,Yu_2023_euler}, leading to pairing nodes in the momentum space. These results provide a potential explanation to the unconventional SC features observed in experiments that are mentioned above. 
Following this line, whether other novel correlated states owing to phonons exist beneath the unexplained experimental findings can be an exciting open question. 

For reference of any future study that are interested in these topics, in this article, we present a detailed analysis on the EPC in the THF model. 
In \cref{sec:ele}, for self-consistency of the article, we shortly review both the continuum model (\cref{sec:ele-cont}) and the THF model (\cref{sec:ele-THF}) for electrons. 
In \cref{sec:epc}, we discuss the phonon motion (\cref{sec:epc-phn}), and the EPC vertex for moir\'e electrons in MATBG, which is derived from the tight-binding carbon atom lattice (\cref{sec:epc-epc}). 
The MATBG $\Gamma$-phonon modes are labeled with $(c/r, \XA/\YA/\ZA/\XO/\YO/\ZO)$, where $\XA/\YA/\ZA/\XO/\YO/\ZO$ stand for single-layer graphene (SLG) acoustic ($\rm A$) and optical ($\rm O$) motion along $\rm X/Y/Z$ directions, and $c/r$ stand for layer co-moving and relative-moving. The MATBG $K$-phonons are labeled with $(l, \Aone/\Bone)$, where $l=1,2$ denotes layer, and $\Aone/\Bone$ denote two orthogonal SLG phonon modes that have $\sqrt{3} \times \sqrt{3}$ pattern, and transform as the $A_1$ and $B_1$ irreducible representations (irreps) under the point group $D_6$, respectively. 
We find that, the out-of-plane $\Gamma$-phonon modes $(c,\ZA)$ and $(c/r,\ZO)$, and all the other $K$-phonons that are not mentioned above, are not significantly coupled to moir\'e electrons captured in the continuum model. 
In the end of \cref{sec:epc-epc}, we also discuss the renormalization (RG) effect from high-energy electron states (with energies above all phonon-frequencies, \textit{e.g.} $\gtrsim$200meV). 
In Ref. \cite{Basko_2008}, it is found that such RG processes are able to amplify the $K$-phonon-mediated interaction strengths by a factor of $\kappa_K = 3.2$, from the bare values derived from the tight-binding approach. 

In \cref{sec:vertex}, we project the EPC vertex onto the $f$-orbitals, which form the major constitute of flat bands, and carry the strong correlation. 
We show that all the co-moving in-plane $\Gamma$-phonon modes, $(c, \XA/\YA/\XO/\YO)$, are not coupled to the $f$-orbitals, due to an approximate chiral symmetry $\td{S}$ that holds only in the vicinity of AA-stacking regions. 
Besides, the layer-breathing $\Gamma$-phonon mode, $(r, \ZA)$, also has negligible EPC on the $f$-orbitals, due to crystalline symmetries and the emergent $P$. 
The only contributing phonon modes are the relative in-plane $\Gamma$-phonon modes, $(r, \XA/\YA/\XO/\YO)$, and $K$-phonons $(l, \Aone/\Bone)$, whose projected EPC vertices are obtained. 
Viewing each $f$-orbital as a giant molecule, the EPC vertices induce the Jahn-Teller effect, which dynamically lifts the orbital and valley degeneracies \cite{Angeli_2019_valleyJT, Andrea_2022_local}. 

In \cref{sec:HA}, by integrating out the phonon fields, we obtain the phonon-mediated effective interaction for electrons on the $f$-orbitals. 
The inter-site interactions are negligible due to the locality of electron orbitals and the locality of phonon fields. 
The on-site interaction, which we denote by $\hat{H}_{\rm A}$, however, has strengths $1\sim4$meV, with the RG factor $\kappa_K$ taken into consideration. 
The multiplet splittings induced by $\hat{H}_{\rm A}$ is explicitly solved in \cref{sec:multi}. 
It is found that, two-electron states with symmetric spatial wave-functions (encoded by the orbital and valley degrees of freedom) are generically more favored in energy, which, by the fermion anti-symmetry, form spin-singlets. 
Especially, at $\kappa_\Gamma=1$ and $\kappa_K=3.2$, an $A_1$ irrep of $D_6$ group ($s$-wave) is the unique two-electron ground state, while the $E_2$ irrep ($d$-wave) afore mentioned comes to the second-lowest states. 
We remark that the energy ordering of the $A_1$ and $E_2$ irreps can eventually be inversed with the carbon atom Hubbard taken into consideration, but we postpone the discussion to \cref{sec:HH}. 
Three and four-electron multiplet splittings are also solved, where the states with the highest total spin are also found to have the highest energy under $\hat{H}_{\rm A}$. 
The three-electron ground states are eight-fold degenerate, guaranteed by symmetries, while the four-electron ground state is a non-degenerate spin-singlet forming $A_1$ irrep. 
Above results imply an anti-Hund's nature of $\hat{H}_{\rm A}$ \cite{dodaro_phases_2018, Angeli_2019_valleyJT, Andrea_2022_local, Wang_2024_TBGSC}, which unveils a striking analogue to the trivalent fullerides $\rm A_3C_{60}$ \cite{Capone_2009_colloquium, Chakravarty_1991_electronic, Auerbach_1994_electronvibron, Capone_2002_strongly}. There, the Jahn-Teller effect on each $\rm C_{60}$ molecule also generates an anti-Hund's on-site interaction, which eventually leads to spin-singlet SC in the strongly correlated regime. 

As an immediate implication of the phonon-mediated $\hat{H}_{\rm A}$, a self-consistent Hartree-Fock study is performed in \cref{sec:mf} to reveal the phonon-favored symmetry-breaking states at even-integer fillings. 
In general, $\Gamma$-phonons favor intra-valley orbital orders, including an orbital polarized ($\rm OP$) order, which is a valley Hall insulator, and a nematic inter-orbital coherent Dirac semi-metal ($\rm NIOC$). 
$K$-phonons favor inter-valley coherent orders, including $\rm TIVC$ and another nematic inter-valley coherent state ($\rm NIVC$). 
To dissect the roles of distinct phonon modes, we also introduce an overall amplification factor for the $\Gamma$-phonon-mediated interactions, $\kappa_\Gamma$. Viewing both $\kappa_\Gamma$ and $\kappa_K$ as tunable parameters, we map out the phase diagram. 

At $\nu=0$, the Kramer's inter-valley coherent state ($\rm KIVC$) and the valley-polarized state ($\rm VP$) appear at small $\kappa_\Gamma \lesssim 1.0$ and small $\kappa_K \lesssim 1.0$, as they are favored by kinetic and Coulomb energies \cite{Bultinck_2020_2U4, TBG4_2021, Song_2022}. Since $\rm KIVC$ is penalized by $K$-phonons, and $\rm VP$ is penalized by $\Gamma$-phonons, the two phases are separated by a boundary with slope $\frac{\Delta\kappa_\Gamma}{\Delta \kappa_K} \approx 0.8$ that goes slightly below $\kappa_\Gamma = \kappa_K = 0$. 
$\rm TIVC$ appears in a large area satisfying $\kappa_K \gtrsim 0.9$ and $0.8\kappa_K \gtrsim \kappa_\Gamma$, which encompasses $\kappa_\Gamma=1$ and $\kappa_K=3.2$. Nevertheless, Kekul\'e pattern is not experimentally observed at this filling \cite{Nuckolls_2023_quantum}. The $\rm OP$ and $\rm NIOC$ orders also appear in the phase diagram but only at weak $\kappa_K$ and large $\kappa_\Gamma$. We briefly discuss how they can further benefit from the effect of hBN substrate and strain, respectively \cite{Parker_2021_strain, Kwan_2021_IKS, Wagner_2022_global}. 
Our results are in qualitative agreement with Refs. \cite{Angeli_2019_valleyJT, Andrea_2022_local, Kwan_2023_Kphonon, shi_2024_moire}, and the quantitative difference can be attributed to different model settings. 

At $\nu=-2$, symmetry-breaking orders polarize two degrees of freedom. At small $\kappa_\Gamma \lesssim 1.6$ and small $\kappa_K \lesssim 1.0$, the lowest symmetry-breaking order is dubbed as $\rm (KIVC\times SP)_{tw}$, which can be understood as $\rm KIVC\times SP$ (where $\rm SP$ denotes spin-polarization) but with a $\pi$ relative spin twist in the two valleys. 
Such a twisted state is equally favored by kinetic and Coulomb energies as $\rm KIVC\times SP$ and $\rm VP\times SP$, but avoids energy penalty from both $\Gamma$- and $K$-phonons. 
Increasing $\kappa_K$, we find two orders that are degenerate at the Hartree-Fock level, $\rm TIVC \times QAH$ and $\rm TIVC \times QSH$, in a large area that includes $\kappa_\Gamma = 1$ and $\kappa_K = 3.2$, which can account for the Kekul\'e pattern observed in ultra-low strained samples at this filling \cite{Nuckolls_2023_quantum, Dumitru_2022_STM}. 
With the $\rm TIVC$ component polarized to benefit from $K$-phonons, two Chern bands with opposite Chern numbers $\pm1$ are populated with both spins. Then, a pair of up and down spins is depleted from either the same Chern sector or two opposite Chern sectors, which generates the quantum anomalous Hall ($\rm QAH$) or quantum spin Hall ($\rm QSH$) response, respectively. 
The result is also in agreement with Refs. \cite{Andrea_2022_local, Kwan_2023_Kphonon, shi_2024_moire}. 
Interestingly, when $\kappa_\Gamma$ and $\kappa_K$ are both strong ($\kappa_\Gamma+\frac{1}{3}\kappa_K \gtrsim \frac{8}{3}$), we identify a new nematic phase, which can be dubbed as $\rm NIOC\times NIVC$. 
Such a state preserves $C_{2z}T$ but spontaneously violates valley U(1), hence can either be a metal with $C_{2z}T$-protected crossings, or annihilate the crossings to form an insulator, depending on parameters. 
By simultaneously forming orbital orders and inter-valley coherence, this state benefits from both $\Gamma$-phonons and $K$-phonons, but do not occur at weak EPC strengths due to a large penalty by kinetic and Coulomb energies. 
Since the $\rm NIOC$ component benefits from an external strain, $\rm NIOC\times NIVC$ can potentially explain the nematic symmetry-breaking orders observed in large-strained samples at this filling \cite{Kerelsky_2019_maximized}. 

In \cref{sec:HH}, we examine a Hund's interaction on the $f$-orbitals $\hat{H}_{\rm H}$ that microscopically originates from the carbon atom Hubbard $U_0$ \cite{Wu_2018_SCop, gonzalez-arraga_electrically_2017, zhang_spin-polarized_2022, Wang_2024_TBGSC}. 
The strength is also of the order $1\sim 3$meV. 
Most crucially, among all the two-electron states, the $A_1$ singlet is penalized by $\hat{H}_{\rm H}$ while the $E_2$ doublet is almost unaffected. Consequently, with both $U_0$ and $\kappa_K$ being large, the $E_2$ doublet will come to the two-electron ground state, furnishing the necessary condition for the $E_2$-doublet-controlled unconventional SC at $\nu \approx -2.5$ suggested by Ref. \cite{Wang_2024_TBGSC}. 
The favored symmetry-breaking orders in the presence of both $\hat{H}_{\rm A}$ and $\hat{H}_{\rm H}$ are also investigated. While the phase diagram of varying $\kappa_\Gamma$ and $\kappa_K$ is enriched, especially with SP phases appearing at weak $\kappa_\Gamma$ and $\kappa_K$, the outcome at $\kappa_\Gamma=1$ and $\kappa_K=3.2$ is seldom affected. At CNP and $\nu=-2$, $\rm TIVC$ and $\rm TIVC\times QAH/QSH$ are found, respectively. 

On the other hand, the four-electron ground state still remains the non-degenerate $A_1$ irrep even with the onset of $\hat{H}_{\rm H}$. In \cref{sec:discuss}, we thus discuss a new exotic phase formed by the light Dirac $c$-electrons only, while $f$-electrons, by forming the non-degenerate singlet, are decoupled from the $c$-bath and become invisible in the low-energy regime. No symmetry-breaking occurs in this strongly correlated state, while $f$-electrons exhibit a full Mott gap. We thus term this state as a Mott semi-metal (Mott-SM). 
Its stability against Kondo screening and the symmetry-breaking effects mediated by RKKY interactions are discussed. Experimental features that distinguish it from other semi-metallic states at CNP are outlined. We also comment that the doped Mott-SM is unlikely to support the experimentally observed SC due to the light Dirac nature.

\section{The electron Hamiltonian and the topological heavy fermion model} \label{sec:ele}

\begin{figure}[tb]
    \centering
    \includegraphics[width=1.0\linewidth]{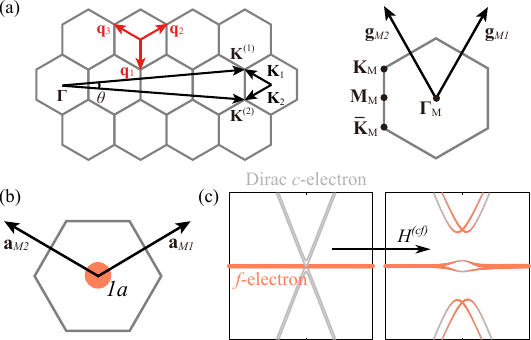}
    \caption{\label{fig:MBZ} (a) Folding MBZ at commensurate twist angle $\theta$. (b) Illustration of an MUC and the location of an $f$-site. 
    (c) $\hat{H}_0$ bands of the THF model. 
    (Left) without $cf$ hybridization, $f$-electrons will be isolated orbitals with vanishing kinetic energy, and $c$-electrons form Dirac cones with quick dispersion. 
    (Right) with $cf$ hybridization turned on, the topological flat bands and the remote band bottoms in the continuum model of MATBG are faithfully reproduced. }
\end{figure}

For the article to be self-contained, we first review the continuum model for moir\'e electrons, and summarize the essential of the THF model. 

\subsection{The continuum model for electron fields}  \label{sec:ele-cont}

We label the two graphene layers as $l = 1,2$, and the two sublattices in each layer as $\alpha = A,B$, so that $\RR^{(l,\alpha)}$ runs over all carbon atoms that belong to layer $l$ and sublattice $\alpha$. $\psi^\dagger_{s}(\RR^{(l,\alpha)})$ creates an electron with spin $s=\up, \down$ in the corresponding tight-binding $p_z$-orbital. 
The low-energy physics involve electrons with momenta close to the graphene valleys, two per layer, which we dub as $\eta \KK^{(l)}$ with $\eta=\pm$,  
\begin{equation} \label{eq:psi_k}
    \psi^\dagger_{\kk, l, \alpha, \eta, s} = \frac{1}{\sqrt{N_g}} \sum_{\RR^{(l,\alpha)}} e^{i(\kk+\eta\KK^{(l)}) \cdot \RR^{(l,\alpha)}} \psi^\dagger_{s}(\RR^{(l,\alpha)}) 
\end{equation}
Here, $N_g$ denotes the number of graphene unit cells per layer, and a high-energy cutoff will be imposed as $|\kk| \lesssim \Lambda$, where $\Lambda \ll |\KK^{(l)}|$ defines a long-wave limit compared to the microscopic lattice. 
Due to the small twist angle $\theta \approx 1.05^\circ$, the same valley of two layers will be displaced by a small momentum $k_\theta = |\KK^{(2)} - \KK^{(1)}| \approx \theta |\KK^{(l)}|$, thus one should choose $\Lambda \gg k_\theta$, so as to capture the moir\'e physics. 

The long-wave limit is represented by the continuum fields,  
\begin{equation}  \label{eq:psi_r}
    \psi^\dagger_{l, \alpha, \eta, s}(\rr) = \frac{1}{\sqrt{N_g \Omega_g}} \sum_{|\kk| \lesssim \Lambda} e^{-i\kk\cdot\rr} \psi^\dagger_{\kk, l, \alpha, \eta, s} \ , 
\end{equation}
where $\Omega_g$ denotes the area of a graphene unit cell. 
Compared to $\psi^\dagger_s(\RR^{(l, \alpha)})$, $\psi^\dagger_{l, \alpha, \eta, s}(\rr)$ has absorbed the fast oscillating phase $e^{i \eta \KK^{(l)} \cdot \rr}$ within each $\sqrt{3}\times\sqrt{3}$ tripled graphene cell as an internal degree of freedom $\eta$.
The continuum model is built by projecting the tight-binding Hamiltonian to the continuum basis \cref{eq:psi_r} (\cref{app:cont-SLG-H0,app:cont-Moire-H0}). The hopping amplitude between two $p_z$-orbitals spatially separated by a 3D vector $\dd$ is parametrized in the following Slater-Koster form \cite{Slater_1954, Neto_2009_RMP, Koshino_epc_2020}, 
{\small
\begin{align} \label{eq:SKhop}
    t(\dd) = V_{\pi} e^{-\frac{|\dd|-a_0/\sqrt{3}}{r_0}} \left[ 1 - \left( \frac{\dd\cdot\hat{\mbf{z}}}{|\dd|} \right)^2 \right] + V_{\sigma} e^{-\frac{|\dd|-c_0}{r_0}} \left( \frac{\dd\cdot\hat{\mbf{z}}}{|\dd|} \right)^2
\end{align}}
where we take $V_\pi \approx -3.1\mrm{eV}$ as the transfer integral of intra-layer nearest-neighbor $\pi$-bonds, $V_\sigma \approx 0.48\mrm{eV}$ as the transfer integral for two orbitals vertically aligned in two layers, and $a_0 = 0.246\mrm{nm}$ as the honeycomb lattice constant, $c_0=0.334\mrm{nm}$ as the typical vertical separation of two graphene layers, $r_0 = 0.184 a_0$ as the decay length of hopping amplitudes. The projected continuum Hamiltonian is the Bistritzer-MacDonald model \cite{BM_2011}, 
\begin{align}   \label{eq:H0}
    \hat{H}_0 =& \sum_{\eta, s} \int\mrm{d}^2\rr \sum_{l,l',\alpha, \alpha'} \psi^\dagger_{l,\alpha,\eta,s}(\rr) \\\nonumber
    & \begin{pmatrix}
        -iv_F \partial_\rr \cdot \mbs{\sigma}^{(\eta)} & w_1 T^{(\eta)}(\rr) \\
        w_1 T^{(\eta)\dagger}(\rr) & -iv_F \partial_\rr \cdot \mbs{\sigma}^{(\eta)} \\
    \end{pmatrix}_{l \alpha; l' \alpha'} \psi_{l',\alpha',\eta,s}(\rr) 
\end{align}
Here, the large $2\times2$ block structure stands for layer indices $l,l'$, and each block is a $2\times 2$ matrix representing sublattices $\alpha, \alpha'$, decomposed with Pauli matrices $\sigma$. We define $\mbs{\sigma}^{(\eta)} = (\eta \sigma_x, \sigma_y)$, and set $\hbar=1$ for convenience. The Dirac velocity of each single layer graphene is determined by $v_F = - \frac{\sqrt{3}}{2} a_0 V_\pi \approx650 \mrm{meV\cdot nm}$. 
The overall inter-layer tunneling strength $w_1$ is determined by Fouriering \cref{eq:SKhop} in a 2D horizontal plane parametrized by $\dd_\parallel$, with fixed height $\dd\cdot\hat{\mbf{z}} = c_0$, at wave vector $\eta\KK^{(l)}$, 
{\small \begin{align}   \label{eq:w1}
    w_1 =  \int \frac{\mrm{d}^2\dd_\parallel}{\Omega_g} ~ t(\dd_\parallel + c_0\hat{\mbf{z}}) e^{-i\eta\KK^{(l)}\cdot\dd_\parallel}  \approx 0.228 V_\sigma \approx 110\mrm{meV} \ . 
\end{align}}%
Note that, since \cref{eq:SKhop} is isotropic in each horizontal plane, $w_1$ is real-valued and only depends on $|\KK^{(l)}|$. 
The inter-layer tunneling matrix reads, 
\begin{align}  \label{eq:Tr}
    T^{(\eta)}(\rr) =& \sum_{j=1}^3 e^{i \eta \qq_j \cdot \rr} \begin{pmatrix}
        u_0 & e^{-i\eta \frac{2\pi}{3}(j-1)} \\
        e^{i\eta \frac{2\pi}{3}(j-1)} & u_0 \\
    \end{pmatrix}
\end{align}
where $\qq_j$ is defined as in \cref{fig:MBZ}, and $u_0 \approx 0.8$ phenomenologically describes the suppression of tunneling amplitudes at AA-stacked regions (centered at $\rr=0$) due to lattice corrugation \cite{Koshino_maximally_2018, Koshino_epc_2020}. 
We also remark that, in obtaining \cref{eq:H0}, $O(\theta)$ corrections in the intra-layer blocks \cite{TBG2_2021} and non-local tunnelings \cite{Kwan_2021_IKS, herzogarbeitman_2024_heavy} in the inter-layer blocks are neglected. 

In the flat bands of $\hat{H}_0$ (\cref{eq:H0}), the dominating energy scale is set by the long-range Coulomb repulsion of moir\'e electrons, modeled by 
\begin{align} \label{eq:HC}
    \hat{H}_{\rm C} = \frac{1}{2} \int \mrm{d}^2\rr \mrm{d}^2\rr' :\hat{\rho}(\rr): V(\rr-\rr') :\hat{\rho}(\rr'): \ ,
\end{align}
where $:\hat{\rho}(\rr): = \sum_{l,\alpha,\eta,s} \left[ \psi^\dagger_{l,\alpha,\eta,s}(\rr) \psi_{l,\alpha,\eta,s}(\rr) - \frac{1}{2} \right] $ measures the electron density with respect to a uniform charge-neutral background, which corresponds to the average charge density at the continuum scale.
The Coulomb potential $V(\rr-\rr')$ is screened by the top and bottom gates that are separated by distance $\xi_0$, 
\begin{align}
    V(\rr) = \frac{e^2}{4\pi \epsilon \epsilon_0 \xi_0} \sum_{n=-\infty}^{\infty} \frac{(-1)^n}{\sqrt{(\rr/\xi_0)^2 + n^2}} \ ,
\end{align}
where $e$ denotes the electron charge, and $\epsilon_0$ and $\epsilon$ denote the vacuum and relative dielectric constants. Typically, there are $\epsilon\approx6$ and $\xi_0 \approx 10$nm. 

Both $\hat{H}_0$ and $\hat{H}_{\rm C}$ (\cref{eq:H0,eq:HC}) conserve the U(1) charge and SU(2) spin in each valley $\eta$ separately, hence we denote the continuous symmetry group as $\rm U(1)^2 \times SU(2)^2$. 
The space group of MATBG is $P622$, which is generated by moir\'e translations $T_{\Delta\RR}$ and spatial rotations $C_{6z}$ and $C_{2x}$. In particular, for an arbitrary moir\'e lattice vector $\Delta\RR$ (spanned by $\mbf{a}_{M1,M2}$, see \cref{fig:MBZ}(b)),
\begin{equation} \label{eq:psi_TR}
    T_{\Delta\RR} \psi^\dagger_{l,\alpha,\eta,s}(\rr) T^{-1}_{\Delta\RR} = e^{- i\eta\KK_l \cdot \Delta\RR} \psi^\dagger_{l,\alpha,\eta,s}(\rr + \Delta\RR) \ , 
\end{equation}
where $\KK_l$ is the moir\'e Brillouin zone (MBZ) corner defined in Fig. \ref{fig:MBZ}(a). 
The phase $e^{-i\eta\KK_l\cdot\Delta\RR}$ appears in \cref{eq:psi_TR} as a consequence of absorbing the fast oscillation $e^{i\eta\KK^{(l)} \cdot \rr}$ to $\psi^\dagger_{l,\alpha,\eta,s}(\rr)$. For a commensurate stacking angle $\theta$, the moir\'e translation is represented on the microscopic lattice as $T_{\Delta\RR} \psi^\dagger_s(\RR^{(l,\alpha)}) T^{-1}_{\Delta\RR} = \psi^\dagger_s(\RR^{(l,\alpha)} + \Delta\RR)$, without any extra phase. By identifying $e^{i\eta\KK_l \cdot \Delta\RR} = e^{i\eta\KK^{(l)} \cdot \Delta\RR}$ for arbitrary $\Delta\RR$ (\cref{fig:MBZ}(a)), \cref{eq:psi_TR} can be directly derived from definitions \cref{eq:psi_k,eq:psi_r}. 
The two valleys are related by a $C_{2z}$ rotation or a spinless time-reversal $T$. 
Each valley thus enjoys a magnetic space group $P6'2'2$ \cite{Song_2019_allMATBG}, generated by $C_{2z}T$, $C_{3z}$, $C_{2x}$, and moir\'e translations $T_{\Delta\RR}$. 
We introduce the Pauli matrix on the layer, sublattice, and valley indices as $\rho_{\mu}$, $\sigma_{\mu}$ and $\tau_{\mu}$ with $\mu=0,x,y,z$, respectively, then the symmetry actions are represented as
\begin{align}   \label{eq:Dg}
    g \psi^\dagger_{l, \alpha, \eta, s}(\rr) g^{-1} = \sum_{l'\alpha'\eta'} \psi^\dagger_{l', \alpha', \eta', s}(g \rr) \left[ D(g) \right]_{l'\alpha'\eta', l\alpha\eta} 
\end{align}
with
\begin{align}   \label{eq:DC3z}
    D(C_{3z}) &= \rho_0 e^{i\frac{2\pi}{3} \sigma_z \tau_z} \ , \\ \label{eq:DC2x}
    D(C_{2x}) &= \rho_x \sigma_x \tau_0 \ , \\ \label{eq:DC2z}
    D(C_{2z}) &= \rho_0 \sigma_x \tau_x \ , \\ \label{eq:DT}
    D(T) &= \rho_0 \sigma_0 \tau_x \ .
\end{align}

Besides crystalline symmetries, an emergent unitary particle-hole symmetry $P$, is also identified \cite{Song_2019_allMATBG, TBG2_2021}. $P$ inverses the spatial coordinates, $P\rr=-\rr$, and 
\begin{align}   \label{eq:DP}
    D(P) = -i\rho_y \sigma_0 \tau_0 \ . 
\end{align}
It can be verified that the bilinear $\hat{H}_0$ (\cref{eq:H0}) obeys $P \hat{H}_0 P = - \hat{H}_0$, while the quartic $\hat{H}_{\rm C}$ (\cref{eq:HC}) obeys $P \hat{H}_{\rm C} P^{-1} = \hat{H}_{\rm C}$. 
One can thus construct a many-body charge conjugation $\mcl{P}_c$, with $P$ followed by a $C_{2z}T$ action and an exchange between $\psi^\dagger_{l,\alpha,\eta,s}(\rr)$ and $\psi_{l,\alpha,\eta,s}(\rr)$. 
$\mcl{P}_c$ commutes with both $\hat{H}_0$ and $\hat{H}_{\rm C}$, hence the many-body physics of MATBG at $\pm \nu$ fillings will be identical. 
In the limit of $u_0 \to 0$, $\hat{H}_0$ further obeys a chiral symmetry $C$ with $C\rr=\rr$ and $D(C) = \sigma_z$ (hence the name chiral limit), so that $C \hat{H}_0 C^{-1} = -\hat{H}_0$ and $C \hat{H}_{\rm C} C^{-1} = \hat{H}_{\rm C}$ \cite{Tarnopolsky_2019_chiral}. 
$P$ and $C$ thus combine to an emergent inversion symmetry in this limit \cite{wang_chiral_2021}. 
In realistic systems, $P$ is broken by the omitted $O(\theta)$ corrections \cite{TBG2_2021} and non-local tunnelings \cite{Kwan_2021_IKS, herzogarbeitman_2024_heavy}. 

Finally, we remark that, if truncating the Hilbert space to the flat bands, and ignoring the small flat band width, $C_{2z}P$ becomes an exact commuting symmetry for the projected Hamiltonian. Since $C_{2z}P$ acts locally in the real space, one can combine $C_{2z}P$ with the generators of $\rm U(1)^2 \times SU(2)^2$ rotations, so as to generate a $\rm U(4)$ group. We term it as the flat-$\rm U(4)$ \cite{Kang_2019_strong, Vafek_2020_RG, Bultinck_2020_2U4, Seo_2019_ferro, TBG3_2021, Song_2022}.

\subsection{The topological heavy fermion model} \label{sec:ele-THF}

To further zoom in to the flat bands, we exploit the THF model \cite{Song_2022, Shi_2022_THF, herzogarbeitman_2024_heavy, singh_topological_2024}, which serves as a faithful representation of all single-electron states within an energy window of $\sim 70$meV, which includes the two flat bands and the bottom of four lowest remote bands (see \cref{fig:MBZ}(c)). 
Two highly localized $f$-orbitals (labeled by $\beta=1,2$) per moir\'e unit cell (MUC) are constructed per valley $\eta$ and per spin $s$, using the Bloch states within the energy window. They significantly overlap with the flat bands in the majority of the MBZ ($96\%$), except in the vicinity of $\GGamma_M$, where they overlap with the remote band bottoms instead, as a consequence of the non-trivial topology. The details of Wannierization can be found in Ref. \cite{Song_2022}. 
They form an $E$ irrep of the point group $6'2'2$ at Wyckoff position $1a$ (namely, the AA-stacked region). 
We denote the corresponding creation operator as $f^\dagger_{\RR\beta\eta s}$, where $\RR$ labels the Wannier center. 
All the crystalline symmetries (\cref{eq:DC3z,eq:DC2x,eq:DC2z}), time-reversal $T$ (\cref{eq:DT}), and the emergent particle-hole $P$ (\cref{eq:DP}) act in a closed form on the $f$-orbitals \cite{Song_2022}, 
\begin{align}   \label{eq:Dfg}
    g f^\dagger_{\RR \beta \eta s} g^{-1} = \sum_{\beta'\eta'} f^\dagger_{(g\RR) \beta' \eta' s} \left[ D^f(g) \right]_{\beta'\eta', \beta\eta} 
\end{align}
with
\begin{align} \label{eq:DfC3z}
    D^f(C_{3z}) &= e^{i\frac{2\pi}{3} \sigma^z \tau^z} \ , \\ \label{eq:DfC2x}
    D^f(C_{2x}) &= \sigma^x \tau^0 \ , \\ \label{eq:DfC2z}
    D^f(C_{2z}) &= \sigma^x \tau^x \ , \\ \label{eq:DfT}
    D^f(T) &= \sigma^0 \tau^x \ , \\  \label{eq:DfP}
    D^f(P) &= i\sigma^z \tau^z \ ,
\end{align}
where we have introduced the Pauli matrix on the orbital ($\beta$) and valley ($\eta$) indices under the THF basis as $\sigma^{\mu}$ and $\tau^\mu$, with $\mu=0,x,y,z$. 

The Wannier functions $w^{(\eta)}_{l\alpha,\beta}(\rr)$ define the unitary transformation from the continuum basis to the $f$-orbitals, 
\begin{align}   \label{eq:f}
    f^\dagger_{\RR\beta\eta s} = \sum_{l,\alpha} e^{-i\eta\KK_l\cdot \RR} \int\mrm{d}^2\rr ~ w^{(\eta)}_{l\alpha,\beta}(\rr-\RR) \psi^\dagger_{l,\alpha,\eta,s}(\rr)
\end{align}
The layer and valley dependent phase $e^{-i\eta\KK_l\cdot\RR}$ in \cref{eq:f} is chosen to compensate the similar phase in \cref{eq:psi_TR}, so that $f^\dagger_{\RR\beta\eta s}$ transforms under a moir\'e translation as $T_{\Delta\RR} f^\dagger_{\RR\beta\eta s} T_{\Delta\RR}^{-1} = f^\dagger_{(\RR+\Delta\RR)\beta\eta s}$. The Wannier functions can be well approximated by the following Gaussian form \cite{Song_2022, calugaru_twisted_2023}. For the $\beta=1$ orbital, 
\begin{align}   \label{eq:w_Gauss_A}
    w^{(\eta)}_{lA,1}(\rr) & = \frac{\alpha_1}{\sqrt{2\pi}} \frac{1}{\lambda_1} e^{i\frac{\pi}{4} \zeta_l \eta} e^{- \frac{\rr^2}{2\lambda_1^2}} \\
    \label{eq:w_Gauss_B}
    w^{(\eta)}_{lB,1}(\rr) & = -\zeta_l \frac{\alpha_2}{\sqrt{2\pi}} \frac{x+i\eta y}{\lambda_2^2} e^{i\frac{\pi}{4} \zeta_l \eta} e^{- \frac{\rr^2}{2\lambda_2^2}}
\end{align}
Here, $\zeta_l = [\rho_z]_{l,l}$. 
$\alpha_{1,2} = 0.8193, 0.5734$ control the relative portion of $A,B$ sublattices, and in particular, the $\beta=1$ orbital comprises of mostly the $A$ sublattice. 
$\lambda_{1,2} = 0.1791a_M, 0.1910a_M$ control the Gaussian width, with $a_M \sim \frac{a_0}{\theta}$ denoting the moir\'e lattice constant. 
The $\beta=2$ Wannier function is related to the $\beta=1$ Wannier function by conjugating \cref{eq:f} with $C_{2z}T$, 
\begin{align} \label{eq:w_Gauss_2}
    w^{(\eta)}_{lA,1}(\rr) = w^{(\eta)*}_{lB,2}(-\rr), \quad w^{(\eta)}_{lB,1}(\rr) = w^{(\eta)*}_{lA,2}(-\rr) \ .
\end{align}
We also note that, the Wannier functions with the same $\beta$ but opposite valley $\eta$ are related by the time-reversal $T$, hence sharing the same density profile in real space. 

With the $f$-components projected out, the remaining Bloch states form four dispersive bands per valley per spin at $\GGamma_M$, with a quadratic Dirac point in the middle (\cref{fig:MBZ}(c)). 
They are termed as $c$-electrons. 
Due to the fast Dirac velocity, to faithfully represent the low-energy electron states, it suffices to $\kk\cdot\pp$-expand the $c$-electron Hamiltonian and the $cf$-hybridization in the vicinity of $\GGamma_M$. 
We denote the corresponding creation operators as $c^\dagger_{\kk b \eta s}$, with $|\kk|<\Lambda_c$ and $b=1,2,3,4$. The cutoff $\Lambda_c$ can be formally extended to infinity without altering the low-energy physics, but in practice, one can take $\Lambda_c$ as a few times of $k_\theta$. 
Projecting the kinetic energy $\hat{H}_0$ \cref{eq:H0} to these THF basis, one obtains 
\begin{widetext}
\begin{align} \label{eq:Hf0}
    \hat{H}_0 &= \frac{1}{\sqrt{N_M}} \sum_{|\kk|<\Lambda_c} \sum_{\RR} \sum_{b\beta \eta s} e^{-\frac{\lambda^2|\kk|^2}{2}} \left[  e^{-i\kk\cdot\RR} c^\dagger_{\kk b \eta s} H_{b\beta}^{(cf,\eta)}(\kk) f_{\RR \beta \eta s} + \mrm{H.c.} \right] + \sum_{|\kk|<\Lambda_c} \sum_{bb'\eta s} c^\dagger_{\kk b \eta s} H_{bb'} ^{(c,\eta)}(\kk) c_{\kk b' \eta s} \\ \nonumber
    & H^{(cf,\eta)}(\kk) = \begin{pmatrix}
        \gamma \sigma^0 + v_\star'(\eta\kk_x \sigma^x + \kk_y\sigma^y)  \\ 
        0_{2\times2} \\
    \end{pmatrix}, \quad H^{(c,\eta)}(\kk) = \begin{pmatrix}
        0_{2\times2} & v_\star(\eta \kk_x \sigma^0 + i \kk_y \sigma^z)  \\
        v_\star(\eta \kk_x \sigma^0 - i \kk_y \sigma^z)  & M \sigma^x \\
    \end{pmatrix}
\end{align}
\end{widetext}
Here, $N_M \sim \theta^2 N_g$ denotes the number of MUC's, $\lambda = 0.3375a_M$ is a damping factor proportional to the size of $f$-orbitals, and decouples high-energy $c$-electrons from the flat bands. The Dirac velocity of $c$-electrons is found as $v_\star = -430.3\mrm{meV\cdot nm}$. $M=3.697\mrm{meV}$ couples the two Dirac cones to produce a quadratic Dirac point, and $\gamma = -24.75\mrm{meV}$, $v_\star' = 162.2\mrm{meV\cdot nm}$ describe $cf$-hybridization. 

Projecting the long-range Coulomb interaction \cref{eq:HC} to the THF basis, we obtain
\begin{widetext}
\begin{align}  \label{eq:HfC}
    \hat{H}_{\rm C} &= \hat{H}_U + \hat{H}_W + \hat{H}_J + \hat{H}_V \ , \\ \label{eq:HU}
    \hat{H}_U &= \frac{U}{2} \sum_{\RR} \left[ \hat{N}_f(\RR) - 4 \right]^2 + \frac{U_2}{2} \sum_{\langle\RR,\RR'\rangle} \left[ \hat{N}_f(\RR) - 4 \right] \left[ \hat{N}_f(\RR') - 4 \right] \ , \\  \label{eq:HW}
    \hat{H}_W &= \frac{1}{N_M} \sum_{\RR} \sum_{|\kk|,|\kk'|<\Lambda_c} e^{i(\kk-\kk')\cdot\RR} \left[ \hat{N}_f(\RR) - 4 \right]  \left[ \sum_{b\eta s} W_b : c^\dagger_{\kk' b \eta s} c_{\kk b \eta s} : \right] \ , \\ \label{eq:HJ}
    \hat{H}_J &= - \frac{J}{N_M} \sum_{\RR} \sum_{|\kk|,|\kk'|<\Lambda_c} e^{i(\kk-\kk') \cdot \RR} \sum_{\beta\eta s} \sum_{\beta'\eta's'} \frac{\eta\eta'+(-1)^{\beta+\beta'}}{2}  : f^\dagger_{\RR\beta\eta s} f_{\RR\beta' \eta's'} :  : c^\dagger_{\kk' (\beta'+2)\eta's'} c_{\kk (\beta+2) \eta s} : \ , \\ \label{eq:HV}
    \hat{H}_V &= \frac{V}{2 N_M}  \sum_{\kk,\kk',\qq} \sum_{b\eta s} \sum_{b'\eta's'} :c^\dagger_{(\kk-\qq) b \eta s} c_{\kk b \eta s}: :c^\dagger_{(\kk'+\qq) b'\eta' s'} c_{\kk' b'\eta' s'}: \ .
\end{align}
\end{widetext}
Here, $\hat{H}_{U,W,V}$ are density-density interactions, and $\hat{H}_J$ is an exchange interaction between $f$-orbital and $c$-electrons. $\hat{N}_f(\RR) = \sum_{\beta\eta s} f^\dagger_{\RR\beta\eta s} f_{\RR\beta\eta s}$ counts the occupation number of an $f$-orbital, and $:f^\dagger_{\RR\beta\eta s} f_{\RR\beta'\eta' s'}: = f^\dagger_{\RR\beta\eta s} f_{\RR\beta'\eta's'} - \frac{1}{2} \delta_{\beta\eta s, \beta'\eta's'}$, $:c^\dagger_{\kk b\eta s} c_{\kk'b'\eta' s'}: = c^\dagger_{\kk b\eta s} c_{\kk'b'\eta' s'} - \frac{1}{2} \delta_{\beta\eta s, \beta'\eta's'}$. 
For $\epsilon=6$ and $\xi=10\mrm{nm}$, it is found that $U = 57.95\mrm{meV}$, $U_2 = 2.329\mrm{meV}$, $W_1=W_2 = 44.03\mrm{meV}$, $W_3=W_4 = 50.20\mrm{meV}$, $J = 16.38\mrm{meV}$, $V = 48.33\mrm{meV}$. 
Other interaction terms that does not conserve the $f$-charge number are omitted, as they are related to high-energy excitations that at least cost an energy $\sim U$. 

Finally, it can be noted that in the limit of $M\to0$, $\hat{H}_0$ takes a bipartite form, namely, regarding $c^\dagger_{\kk b \eta s}$ with $b=1,2$ and the rest of electronic fields as two subsystems, only mutual couplings exist. This corresponds to a new unitary chiral symmetry $S$. One thus obtains $C_{2z}PS$ as a commuting symmetry for both $\hat{H}_0$ and $\hat{H}_{\rm C}$, which is local in real space and flips valley. Combining $C_{2z}PS$ with the $\rm U(1)^2 \times SU(2)^2$ generators, a $\rm U(4)$ group is generated. It has been shown that this $\rm U(4)$ group is consistent with the flat-U(4) group defined for the continuum model, when projected onto the flat bands \cite{Song_2022}. Surprisingly, the U(4) group in the THF model holds for the entire model beyond the flat bands, without any necessity of projection. We will not distinguish the two groups in below. The flat-U(4) generators projected onto the $f$-orbitals read 
\begin{align}
    \sigma^0 \tau^0 s^\mu, \quad \sigma^y \tau^x s^\mu, \quad \sigma^y \tau^y s^\mu, \quad \sigma^z \tau^z s^\mu
\end{align}
with $s^\mu$ ($\mu=0,x,y,z$) denoting the Pauli matrices for spins. Specially, $\hat{H}_J$ can be understood as a \textit{ferromagnetic} interaction between the flat-U(4) moment of $f$-electrons and that of $c$-electrons for $b=3,4$ \cite{Song_2022}. 
In the present article, we keep $M$ as its original finite value, hence the flat-U(4) symmetry holds only approximately.

\section{Electron-phonon couplings for generic moir\'e electrons} \label{sec:epc}

In this section and in \cref{app:cont}, we derive the EPC vertices from the microscopic tight-binding lattice, where the phonon vibration alters interatomic distances, hence changes the electron hopping amplitudes. The results are then projected to the continuum model for electrons, which only involves electrons near the valley bottoms. 
The derivation follows Ref.~\cite{Koshino_epc_2020} and generalizes to include arbitrary phonon modes, including optical phonons and out-of-plane modes. 
For clarity, most of the details are presented in \cref{app:cont}. 
Here, we only summarize the main results, with brief explanation to the form and strengths of the EPC vertices that we obtain. 

Apart from affecting the interatomic hoppings, longitudinal acoustic phonons also modulate the local area, creating a deformation potential that couples to the local electron density \cite{Suzuura_2002, Hwang_acoustic_2008, Mariani_2010_temperature-dependent, Efetov_2010_controlling}. We analyze the effect of this EPC mechanism in \cref{app:deform}, and find that, when projected to the local $f$-orbitals, it only contributes a weak attractive Hubbard interaction that can be absorbed to the large repulsive $\hat{H}_U$ (\cref{eq:HU}). It does not induce any multiplet splitting, hence will not be the focus of the present article.

\subsection{The continuum model for phonon fields} \label{sec:epc-phn}

On an equal footing as electron fields, phonon fields are obtained by quantizing lattice vibration. We denote the displacement vector of the carbon atom located at $\RR^{(l,\alpha)}$ as $\uu_i(\RR^{(l,\alpha)})$, with $i=x,y,z$. The Fourier component of graphene lattice momentum $\qq$ reads
\begin{align}   \label{eq:def_uq}
    \uu^{(l,\alpha)}_{\qq, i} = \frac{1}{\sqrt{N_g}} \sum_{\RR^{(l,\alpha)}} e^{-i\qq \cdot \RR^{(l,\alpha)}} \uu_i(\RR^{(l,\alpha)}) \ . 
\end{align}
Note that $\qq$ is defined up to the graphene Brillouin zone of each layer ($\mrm{gBZ}^{(l)}$). 
If the two graphene layers are decoupled, $l$ and $\qq\in\mrm{gBZ}^{(l)}$ will both remain good quantum numbers. For a phonon to affect the low-energy electronic physics, it should scatter an electron in the valley $\eta$ either to the same valley, or to the opposite valley $\ovl{\eta}$. 
Momentum conservation thus implies that the phonon lattice momentum should be either $|\qq| \lesssim \Lambda$, or $|\qq - \eta\KK^{(l)}| \lesssim \Lambda$ ($\eta = \pm$). We will term them as $\Gamma$-phonons and $K$-phonons, respectively. Viewed from the low-energy electrons, how the $\Gamma$-phonon bands and $K$-phonon bands are connected in the graphene Brillouin zone is irrelevant, hence they can be effectively deemed as two independent new species. 

Turning on the inter-layer couplings, different $l$ and $\qq$ will hybrid. 
Due to the long-wave nature of the moir\'e couplings, the momentum hybridization should only span several times of $k_\theta$, much smaller than $|\KK^{(l)}|$, as in the case of moir\'e electrons. Therefore, the $\Gamma$-phonons and $K$-phonons are still well isolated. We now analyze the two species separately. 

For $\Gamma$-phonons, $\uu_{\qq,i}^{(l, \alpha)}$ can hybrid with $\uu_{\qq,i'}^{(\ovl{l}, \alpha')}$, as well as $\uu_{\qq+\mbf{g}, i'}^{(l', \alpha')}$, where $\mbf{g}$ denotes a non-zero moir\'e reciprocal lattice vector (\cref{fig:MBZ}(a)). 
The former type, referred to as non-moir\'e couplings, is also present in non-moir\'e multilayers, {\it e.g.}, Bernal-stacked bilayer graphene, while the latter type accounts for moir\'e foldings. 
In the present article, we will ignore the moir\'e foldings of phonons, which can be justified by the weak van der Waals nature of the inter-layer phonon dynamic matrix elements \cite{Nam_lattice_2017}. 
Recovering the moir\'e foldings can reconstruct the phonon density of states and the phonon wave-functions for an energy scale $\lesssim$1meV \cite{Cocemasov_tbg_phonon_2013, Koshino_moire_phonon_2019, xie_lattice_2023,Miao_2023_truncated}, which is small compared to the characteristic phonon frequency that couples to the moir\'e electrons. 
As we are only interested in the phonon-mediated effective interaction among electrons, which integrates over the full phonon spectrum, such reconstruction should not result in qualitative error. Therefore, we can still label $\Gamma$-phonons with $\qq$ ($|\qq| \lesssim \Lambda$). 
The non-moir\'e couplings, on the other hand, can be expected to be of the same order as in the non-moir\'e multilayers \cite{Yan_phonon_spec_2008, Cocemasov_tbg_phonon_2013}.

We therefore consider the following single-layer modes $\chi_g@\Gamma \in \{ \XA,\YA,\ZA,\ZO,\XO,\YO \}$ as the complete basis for $\Gamma$-phonons, 
\begin{align}  \label{eq:uG_q}
    \begin{pmatrix}
        u_{\qq,l,\mrm{IA}} \\
        u_{\qq,l,\mrm{IO}} \\
    \end{pmatrix} = \begin{pmatrix}
        \frac{1}{\sqrt{2}} & \frac{1}{\sqrt{2}} \\ 
        \frac{1}{\sqrt{2}} & -\frac{1}{\sqrt{2}} \\ 
    \end{pmatrix}  \begin{pmatrix}
        \uu_{\qq, i}^{(l,A)} \\
        \uu_{\qq, i}^{(l,B)} \\
    \end{pmatrix} , 
\end{align} 
where $|\qq| \lesssim \Lambda$, and $\rm IA,IO$ with $\rm I\!=\!X,Y,Z$ denote the acoustic (A) or optical (O) vibration along $i \!=\! x,y,z$ directions. 
In this work, the splitting between the XA and YA motion to the longitudinal and transverse modes will be neglected, and we will take the corresponding sound velocity as $v_s = 1.7\times 10^4\mrm{m/s} = \frac{3.6\mrm{meV}}{k_\theta}$, which is an average of the original longitudinal ($2.1\times10^4$m/s) and transverse ($1.4\times10^4$m/s) sound velocities in SLG \cite{Suzuura_2002, Kaasbjerg_2012}. Such an approximation for acoustic modes is expected to be valid for the purpose of estimating the effective interaction, as the splitting will not change the order of magnitude of the induced interaction, which we find weak when compared to the interaction strength from optical modes ($\sim 20\%$, see \cref{sec:epc-epc,sec:HA})).

To account for the non-moir\'e inter-layer couplings, it is also convenient to introduce the layer co-moving ($c$) and relative-moving ($r$) basis 
\begin{align} \label{eq:uG_q_cr}
    \begin{pmatrix}
    u_{\qq, c, \chi_g@\Gamma} \\
    u_{\qq, r, \chi_g@\Gamma} \\
    \end{pmatrix} = \begin{pmatrix}
        \frac{1}{\sqrt{2}} & \frac{1}{\sqrt{2}} \\ 
        \frac{1}{\sqrt{2}} & -\frac{1}{\sqrt{2}} \\ 
    \end{pmatrix} \begin{pmatrix}
    u_{\qq, 1, \chi_g@\Gamma} \\
    u_{\qq, 2, \chi_g@\Gamma} \\
    \end{pmatrix}\ ,
\end{align}
For the acoustic $\Gamma$-phonons, co-moving modes are not affected by stacking, thus as in SLG, in-plane ($c, \XA/\YA$) phonons disperse linearly with $\qq$ as $\omega_{\qq, c, \XA/\YA} = v_s |\qq|$, and out-of-plane ($c, \ZA$) phonons disperse quadratically. As will be shown in \cref{sec:epc-epc}, $(c, \ZA)$ does not contribute to EPC, hence we will not make further discussion on its dispersion. 
For the in-plane relative modes ($r, \XA/\YA$), in the limit of $\qq\to0$, they can be deemed as a relative glide between two continuum sheets, which simply shifts the moir\'e pattern hence does not cost energy \cite{xie_lattice_2023}. For the microscopic lattice at incommensurate small twist-angle, the same argument also applies. 
Therefore, $\omega_{\qq, r, \XA/\YA}$ should vanish linearly at $\qq \to 0$ as well. This behavior in MATBG is to be contrasted with that in non-moir\'e bilayers, where a gap typically $\sim$2meV is opened at $\qq=0$ \cite{Yan_phonon_spec_2008, Cocemasov_tbg_phonon_2013}. 
Since the inter-layer coupling is weak, the velocity of $(r, \XA/\YA)$ should also approach that of $(c, \XA/\YA)$ for larger $|\qq| \lesssim \Lambda$ \cite{Yan_phonon_spec_2008, Cocemasov_tbg_phonon_2013, Koshino_moire_phonon_2019}, hence we take $\omega_{\qq, r, \XA/\YA} = v_s |\qq|$ as well. 
For the $(r, \ZA)$ mode, since it alters the layer spacing, which does affect the total energy, its frequency will have a finite gap at $\qq=0$. According to calculations on various bilayer graphenes such as Refs. \cite{Yan_phonon_spec_2008, Cocemasov_tbg_phonon_2013}, we set $\omega_{\qq, r, \ZA} = 11\mrm{meV}$, where the weak dispersion at small finite $\qq$ is neglected. 
For optical $\Gamma$-phonons, as their frequencies are high (all $\gtrsim 100\mrm{meV}$), we will neglect the small splittings due to inter-layer couplings, as well as the weak dispersions with $\qq$, and set $\omega_{\qq, c/r, \XO/\YO} = 180\mrm{meV}$ and $\omega_{\qq, c/r, \ZO} = 100\mrm{meV}$ \cite{Yan_phonon_spec_2008, Cocemasov_tbg_phonon_2013, Chen_ARPES_23}. 

For $K$-phonons in MATBG, in contrast to the $\Gamma$-phonons, the simple strategy to separate out non-moir\'e couplings would be invalid. For example, if we naively restrict $\uu^{(1,\alpha)}_{\KK^{(1)}, i}$ to be coupled to $\uu^{(2,\alpha)}_{\KK^{(1)}, i'}$ only, then since $\KK^{(1)}$ is not a $C_{3z}$-invariant point in $\gBZ^{(2)}$, such an assignment necessarily violates $C_{3z}$ symmetry; while if all the $C_{3z}$-related momenta are included, momenta will only be preserved up to MBZ. Nevertheless, since all the $K$-phonon frequencies are high (in-plane $\gtrsim 110\mrm{meV}$, and out-of-plane $\gtrsim 60\mrm{meV}$), any small inter-layer couplings and weak dispersions can be safely ignored, too. Consequently, eigen modes can still be labeled by single-layer modes.
Importantly, as will be shown in \cref{sec:epc-epc}, in the vicinity of each zone corner $\eta\KK^{(l)}$, only one mode per layer will contribute significantly to EPC, with wave-function
\begin{align} \label{eq:uK_eta}
    u_{\qq, l, \eta K} &= (-i \eta) \frac{\uu_{\qq+\eta\KK^{(l)}, x}^{(l,A)} + \uu_{\qq+\eta\KK^{(l)}, x}^{(l,B)} }{2} \\\nonumber
    & - \frac{\uu_{\qq+\eta\KK^{(l)}, y}^{(l,A)} - \uu_{\qq+\eta\KK^{(l)}, y}^{(l,B)}}{2} \ .
\end{align}
In particular, $u_{0,l,\eta K}$ forms an $A_1$ irrep of the little group $D_3$ at $\eta\KK^{(l)}$; and superposing the two $A_1$ irreps at opposite $\eta$ induces an $A_1$ irrep and a $B_1$ irrep for the SLG point group $D_6$. 
The corresponding vibration pattern induce EPC by changing the nearest-neighbor $\pi$-bond lengths in the single-layer graphene. $u_{\qq,l,\eta K}$ with small $|\qq|\lesssim \Lambda$ describes the slow modulation of such vibration. 
However, for the same reason as electron fields, the moir\'e translation will act as $T_{\Delta\RR} u_{\qq,l,\eta K} T_{\Delta\RR}^{-1} = e^{i\eta\KK_l\cdot \Delta\RR} e^{i\qq \cdot \Delta\RR} u_{\qq,l,\eta K}$ with an extra $e^{i\eta\KK_l \cdot\Delta\RR}$ phase, which would lead to inconvenience of a continuum description. 
To remove this phase, we therefore consider the following $K$-phonon basis (\cref{app:cont-SLG-K}), 
\begin{align}  \label{eq:uK_q}
    ~& \begin{pmatrix}
        u_{\qq,l,\Aone} \\
        u_{\qq,l,\Bone} \\
    \end{pmatrix} = 
    \begin{pmatrix}
        \frac{1}{\sqrt{2}} & \frac{1}{\sqrt{2}} \\
        \frac{i}{\sqrt{2}} & \frac{-i}{\sqrt{2}} \\
    \end{pmatrix} \begin{pmatrix}
        u_{\qq-\KK_l, l, + K} \\
        u_{\qq+\KK_l, l, - K} \\
    \end{pmatrix} \ ,
\end{align}
where by a momentum shift on the right-hand side, $\qq$ (with $|\qq| \lesssim \Lambda$) on the left-hand side is measured from an origin that is equivalent to $\GGamma$ under MBZ folding (\cref{fig:MBZ}(a)). Accordingly, we define the single-layer $K$-phonon modes as $\chi_g@K \in \{ \Aone, \Bone \}$, and we set the frequencies as $\omega_{\qq, l, \Aone/\Bone} = 150\mrm{meV}$ \cite{Yan_phonon_spec_2008, Cocemasov_tbg_phonon_2013, Chen_ARPES_23}. 

For notation brevity, we will use $\chi@\Gamma = (c/r, \chi_g@\Gamma)$ and $\chi@K = (l, \chi_g@K)$ with $l=1,2$ to collectively denote the $\Gamma$-phonon and $K$-phonon modes in MATBG, respectively. If $\chi$ or $\chi_g$ is used alone without specifying $@\Gamma$ or $@K$, it includes both $\Gamma$-phonon and $K$-phonon modes.
By definitions in \cref{eq:uG_q_cr,eq:uK_q}, both $\Gamma$-phonons and $K$-phonons obey $u_{\qq, \chi} = u^\dagger_{-\qq, \chi}$, and transform under moir\'e translation $T_{\Delta\RR}$ as 
\begin{align}  \label{eq:uq_TR}
    T_{\Delta\RR} u_{\qq, \chi} T^{-1}_{\Delta\RR} = e^{i\qq\cdot\Delta\RR} u_{\qq, \chi} \ .
\end{align}
This allows defining the continuum phonon fields for both species, 
\begin{align}   \label{eq:u_r}
    u_{\chi}(\rr) = \frac{1}{\sqrt{N_g}} \sum_{|\qq| \lesssim \Lambda} e^{i\qq\cdot\rr} u_{\qq, \chi} \ , 
\end{align}
so that $u_{\chi}(\rr)$ is real-valued, and obeys $T_{\Delta\RR} u_{\chi}(\rr) T^{-1}_{\Delta\RR} = u_{\chi}(\rr+\Delta\RR)$. 
Phonon fields also obey other crystalline symmetries and time-reversal $T$, which is analyzed in Appendix \ref{app:sym}. 

To obtain the phonon propagator, we first convert $u_{\qq,\chi}$ to the Matsubara frequency domain $u_{\qq,\chi}(iq_0)$, where $iq_0$ stands for bosonic Matsubara frequencies. With decomposition
{\small \begin{align}  \label{eq:uq_b}
    u_{\qq,\chi}(iq_0) = \sqrt{\frac{1}{2m_0 \omega_{\qq,\chi}}}\left( b_{\qq,\chi}(iq_0) + b^\dagger_{-\qq,\chi}(-iq_0) \right) \ ,
\end{align}}%
where $m_0$ denotes the carbon atom mass, the phonon action reads, 
\begin{align}
    S_{\rm ph} = \sum_{\qq,\chi} \sum_{q_0} b^\dagger_{\qq,\chi}(iq_0) (-iq_0 + \omega_{\qq,\chi}) b_{\qq,\chi}(iq_0) \ ,
\end{align} 
The phonon propagator is thus given by 
\begin{align}   \label{eq:D}
    D^{(\chi)}(\qq, iq_0) &= - \langle u_{\qq,\chi}(iq_0) u^\dagger_{\qq,\chi}(iq_0) \rangle_{\rm ph} \\\nonumber
    &= \frac{1}{m_0} \frac{-1}{(\omega_{\qq,\chi})^2 + (q_0)^2} \ ,
\end{align}
where $\langle \cdots \rangle_{\rm ph}$ stands for the functional average weighted by $e^{-S_{\rm ph}}$. 
We will work in the convention that $u_{\qq,\chi}$ instead of $b_{\qq,\chi}$ is chosen as the basic phonon fields, hence our definition of the propagator \cref{eq:D} and the EPC vertex \cref{eq:Hepc_cont} can be different from some literatures. Nevertheless, the convention is self-consistent, and there should be no ambiguity in the effective interaction (\textit{e.g.} \cref{eq:V_measure,eq:Sint}), which is constructed by two EPC vertices and a propagator line.

\subsection{Electron-phonon couplings} \label{sec:epc-epc}

\begin{table*}[t]
    \centering
    \begin{tabular}{c|c|c|c|c|c|c|c|c}
    \hline 
        $\chi$ & $(c,\XA/\YA)$ & $(c,\XO/\YO)$ & $(r,\XA/\YA)$ & $(r,\XO/\YO)$ & $(c,\ZA)$ & $(r,\ZA)$ & $(c/r,\ZO)$ & $(l, \Aone/\Bone)$ \\ 
    \hline 
        $\mcl{V}^{(\chi)}$ (meV) & 0.96 & 2.96 & $\infty$ & 2.96 & - & 0.38 & $<0.01$ & 4.26 \\ 
    \hline 
        $J_{\rm a}$ (meV) & $<0.01$ & $<0.01$ & 0.07 (0.08) & 0.41 (0.63) & - & $<0.01$ & $<0.01$ & - \\
        $J_{\rm b}$ (meV) & $<0.01$ & $<0.01$ & 0.29 (0.20) & 1.31 (1.55) & - & $<0.01$ & $<0.01$ & - \\
        $J_{\rm c}$ (meV) & $<0.01$ & $<0.01$ & 0 & 0 & - & $<0.01$ & $<0.01$ & - \\
        $J_{\rm d}$ (meV) & - & - & - & - & - & - & - & 1.30 (1.39) \\
        $J_{\rm e}$ (meV) & - & - & - & - & - & - & - & 1.19 (1.76) \\
    \hline 
    \end{tabular}
    \caption{\label{tab:int} Phonon-mediated interaction strengths obtained at the bare tight-binding level, namely $\kappa_\Gamma = \kappa_K = 1$. $\mcl{V}^{(\chi)}$ is a rough measure of the interaction felt by typical moir\'e electrons at the AA-stacked region. ``-'' indicates no contribution. ``$<0.01$'' indicates extremely weak contribution whose absolute strength does not exceed $0.01\mrm{meV}$. 
    ``$\infty$'' indicates the artificial result that $\mcl{V}^{(\chi)}$ cannot distinguish the inter-energy-band scatterings from intra-energy-band scatterings, which are finite as discussed in the main text. 
    $J_{\rm a,b,c,d,e}$ are the phonon-mediated anti-Hund's on-$f$-site interaction parameters. Values estimated from the numeric Wannier functions (no brackets) and from the Gaussian fit (in brackets) are both presented. In particular, $J_{\rm a,b,d,e}$ are multiplet splitting parameters, while $J_{\rm c}$ simply indicates a Hubbard-type attraction that can be absorbed to $U$. 
    Due to the particle-hole symmetry $P$ of the $f$-orbital Wannier functions, $J_{\rm c} = 0$ for $(r,\XA/\YA)$ and $(r,\XO/\YO)$. }
\end{table*}

According to \cref{eq:SKhop}, when the interatomic distance $\dd$ is changed under lattice vibration, the hopping amplitude $t(\dd)$ will be altered, and the (bare) EPC vertex can be derived, with the EPC strengths determined by the tight-binding parameters. 
Leaving the derivation details to \cref{app:cont}, we summarize the results here with sketchy explanations. 

In general, the EPC vertex in the continuum model takes the following form
\begin{align}   \label{eq:Hepc_cont}
    \hat{H}_\mrm{epc} =& \frac{1}{\sqrt{N_g}} \sum_{s} \sum_{\chi} \sum_{|\qq| \lesssim \Lambda} \int\mrm{d}^2\rr ~ u_{\qq, \chi} \sum_{\alpha, \alpha', l, l', \eta, \eta'} \\\nonumber
    ~& \psi^\dagger_{l,\alpha,\eta,s}(\rr) C^{(\chi)}_{l\alpha\eta, l'\alpha'\eta'}(\qq;\rr) \psi_{l',\alpha',\eta',s}(\rr)  e^{i\qq\cdot\rr}
\end{align} 
$C^{(\chi)}_{l\alpha\eta, l'\alpha'\eta'}(\qq,\rr)$ can be understood as a $\qq$- and $\rr$-dependent tensor with one phonon index ($\chi$), and two electron indices ($l\alpha\eta$ and $l'\alpha'\eta'$). $C^{(\chi)}$ with given $\chi$, $\qq$, and $\rr$ is hence an $8\times 8$ matrix that can be decomposed with layer ($\rho$), sublattice ($\sigma$) and valley ($\tau$) Pauli matrices. 
The Hermiticity of $\hat{H}_{\rm epc}$ dictates that $C^{(\chi)}(\qq,\rr) = C^{(\chi)\dagger}(-\qq,\rr)$. 

In order to estimate the relevance of each phonon mode $\chi$ in the continuum model, we introduce 
\begin{align}   \label{eq:V_measure}
    \mcl{V}^{(\chi)} = \frac{\theta^2}{m_0} \lim_{\qq \to 0} \frac{||C^{(\chi)}(\qq;0) \otimes C^{(\chi)}(-\qq;0)||}{\omega_{\qq, \chi}^2}
\end{align}
as a rough measure of the zero-frequency interaction strength felt by a typical moir\'e electron at $\rr=0$. Here, $\frac{1}{m_0} \frac{1}{\omega^2_{\qq,\chi}} = -D^{(\chi)}(\qq,0)$ originates from the zero-frequency phonon propagator, and $||\cdots||$ denotes the Frobenius norm. 
The $\theta^2 \sim \frac{\Omega_g}{\Omega_M} $ factor is introduced for the expanded MUC size $\Omega_M$ with respect to the microscopic graphene unit cells $\Omega_g$, and hence the lowered characteristic energy scale. 

We start with $\Gamma$-phonons. $(c, \XA/\YA)$ modulate the graphene $\pi$-bond lengths in a long-wave manner, resulting in intra-layer electron scatterings with amplitudes $\sim \frac{v_F |\qq|}{r_0}$, while leaving the moir\'e pattern and hence the inter-layer tunnelings unchanged. Correspondingly, it can be derived that
\begin{align} 
    \label{eq:C_c_XA}
    C^{(c,\XA)}(\qq;\rr) =& \frac{v_F}{4r_0} \cdot (-iq_x \rho_0 \sigma_x \tau_0 +iq_y \rho_0 \sigma_y \tau_z) , \\
    \label{eq:C_c_YA}
    C^{(c,\YA)}(\qq;\rr) =& \frac{v_F}{4r_0} \cdot (iq_x \rho_0 \sigma_y \tau_z + iq_y \rho_0 \sigma_x \tau_0)
\end{align}
It can thus be estimated that $\mcl{V}^{(c,\XA/\YA)} = \frac{\theta^2}{m_0 r_0^2} \left(\frac{1}{\sqrt{2}} \frac{v_F}{v_s} \right)^2 \sim 0.96 \mrm{meV}$. 
As for $(c, \XO/\YO)$, they vibrate the $\pi$-bonds within each microscopic unit cell, leading to intra-layer scatterings
\begin{align} 
    \label{eq:C_c_XO}
    C^{(c,\XO)}(\qq;\rr) =& \frac{\sqrt{3} v_F}{a_0 r_0} \cdot \rho_0 \sigma_y \tau_z \ , \\ 
    \label{eq:C_c_YO}
    C^{(c,\YO)}(\qq;\rr) =& \frac{\sqrt{3} v_F}{a_0 r_0} \cdot \rho_0 (-\sigma_x) \tau_0 \ .
\end{align}
In principle, $(c,\XO/\YO)$ can also wobble the opposite sublattice of opposite layers, inducing inter-layer scatterings. 
The corresponding amplitude (derived in \cref{app:cont-Moire-in}) can be quickly estimated as follows. For the inter-layer tunneling \cref{eq:w1}, if sublattice $\alpha$ of layer 1 is shifted rigidly by a horizontal vector $\uu$ relative to the sublattice $\alpha'$ of layer 2, then substituting $t(\dd_{\parallel} + c_0\hat{\mbf{z}}) \to t(\dd_{\parallel} + \uu + c_0\hat{\mbf{z}})$ generates an additional phase $e^{i\eta\KK^{(l)} \cdot \uu}$ to $w_1$, which when expanded to the linear order of $\uu$ results in EPC strength proportional to $w_1 |\KK^{(l)}|$. 
Since $w_1|\KK^{(l)}| \sim \frac{1}{30}\frac{v_F}{r_0 a_0}$, the inter-layer processes will be negligible compared to the intra-layer processes. Similar analysis also applies to the relative in-plane optical modes, $(r, \XO/\YO)$, and 
\begin{align}
    \label{eq:C_r_XO}
    C^{(r,\XO)}(\qq;\rr) =& \frac{\sqrt{3} v_F}{a_0 r_0} \cdot \rho_z \sigma_y \tau_z \ , \\
    \label{eq:C_r_YO}
    C^{(r,\YO)}(\qq;\rr) =& \frac{\sqrt{3} v_F}{a_0 r_0} \cdot \rho_z (-\sigma_x) \tau_0 \ . 
\end{align}
$(c, \XO/\YO)$ and $(r,\XO/\YO)$ are thus of the same significance for generic moir\'e electrons captured by the continuum model, with $\mcl{V}^{(c,\XO/\YO)} = \mcl{V}^{(r,\XO/\YO)} = \frac{\theta^2}{m_0r_0^2} \left( \frac{2\sqrt{6} v_F / a_0}{\omega_{0,c,\XO/\YO}} \right)^2 \sim 2.96\mrm{meV}$. 

For the relative glide $(r, \XA/\YA)$, however, aside from inducing intra-layer scatterings as $(c, \XA/\YA)$, they also shift the moir\'e pattern horizontally, with the amplitude magnified by a factor of $\frac{1}{\theta}$ \cite{Lian_2019_SCac}, leading to 
{\small
\begin{align} 
    \label{eq:C_r_XA}
    C^{(r,\XA)}(\qq;\rr) =& \frac{v_F}{4r_0} (-iq_x \rho_z \sigma_x \tau_0 +iq_y \rho_z \sigma_y \tau_z) - \frac{w_1}{\theta} \partial_y T(\rr) \\ 
    \label{eq:C_r_YA}
    C^{(r,\YA)}(\qq;\rr) =& \frac{v_F}{4r_0} (iq_x \rho_z \sigma_y \tau_z +iq_y \rho_z \sigma_x \tau_0) + \frac{w_1}{\theta} \partial_x T(\rr) 
\end{align}}%
where $T(\rr)$ denotes the layer-off-diagonal part in the first-quantized matrix of $\hat{H}_0$, defined by $[T(\rr)]_{1\alpha\eta, 2\alpha'\eta'} = \delta_{\eta,\eta'} [T^{(\eta)}(\rr)]_{\alpha,\alpha'}$, and $[T(\rr)]_{2\alpha\eta, 1\alpha'\eta'} = [T(\rr)]^*_{1\alpha'\eta', 2\alpha\eta}$. 
In contrast with intra-layer scatterings, the inter-layer term is not linear in $\qq$, but remains finite as $\qq \to 0$, leading to a diverging $\mcl{V}^{(r,\XA/\YA)}$. 
Nevertheless, as we show in Appendix \ref{app:divergence}, if $C^{(r, \XA/\YA)}(\qq;\rr)$ is projected to the moir\'e energy bands, only inter-energy-band elements can remain finite at $\qq=0$, while intra-energy-band elements must still vanish linearly with $\qq$. 
The finite inter-energy-band mixing originates from the re-construction of Bloch wave-functions when the moir\'e pattern is shifted as a whole. 
Here, we will only focus on the intra-energy-band EPC vertices on the flat bands. As will be shown in Sec. \ref{sec:vertex-rXAYA}, when projected to the THF $f$-orbitals, the EPC vertex will be forced by crystalline symmetries ($C_{2z}T$ and $C_{3z}$) to vanish linearly as $\qq\to0$. 
Since the $f$-orbitals overlap significantly with the flat bands, we will regard the surviving terms as a qualitatively valid estimation for the desired EPC vertex. 
However, with crystalline symmetries broken, the above argument can be invalid, and a more careful treatment will be in demand. 

For the out-of-plane $\Gamma$-phonons, to the first order of $\uu$, none of them changes the $\pi$-bond lengths to induce intra-layer scatterings. 
For the $(c, \ZA)$ mode, neither will it alter the local layer spacing. As a result, only $(c,\ZO)$ and $(r,\ZA/\ZO)$ are involved in the EPC vertex by inducing inter-layer scatterings, with amplitudes $\sim \frac{w_1}{r_0}$. For instance, it is found for $(r, \ZA)$ modes that (\cref{app:cont-Moire-z}), 
\begin{align}  \label{eq:C_r_ZA}
    C^{(r,\ZA)}(\qq;\rr) = \zeta_1 ~ T(\rr) \ .
\end{align}
with
{\small \begin{align}
    \zeta_1 = \left. \partial_c \int \frac{\mrm{d}^2\dd_\parallel }{\Omega_g} ~ t(\dd_\parallel + c\hat{\mbf{z}}) e^{-i\eta\KK^{(l)} \cdot \dd_\parallel} \right|_{c=c_0} \approx -1.11 \frac{w_1}{r_0}
\end{align}}%
Despite the weak scattering amplitude, $\mcl{V}^{(r, \ZA)} = \frac{\theta^2}{m_0} \left( \frac{2\sqrt{2} u_0 \zeta_1 }{\omega_{0,r,\ZA}} \right)^2 \sim 0.38 \mrm{meV}$ is considerable because the phonon frequency $\omega_{\qq,r,\ZA} = 11\mrm{meV}$ is also ultra-low. 
In contrast, for $(c/r, \ZO)$, due to the much higher $\omega_{\qq,c/r,\ZO} = 100$meV, one can evaluate $\mcl{V}^{(c/r,\ZO)} = \frac{\theta^2}{m_0} \left( \frac{2\sqrt{2}  u_0 \zeta_1 }{\omega_{0,c/r,\ZO}} \right)^2 \sim 0.005 \mrm{meV}$, which will be negligible. Therefore, for general moir\'e electrons, it suffices to keep $(r,\ZA)$ among all out-of-plane $\Gamma$-phonons.

For $K$-phonons, it can be shown that, only the $(l, \Aone)$ and $(l, \Bone)$ modes defined in \cref{eq:uK_q} contribute to the intra-layer EPC vertex, by altering nearest-neighbor $\pi$-bond lengths (\cref{app:cont-SLG-K}). The corresponding EPC vertex reads 
\begin{align}   \label{eq:C_l_A1}
    C^{(l,\Aone)}(\qq;\rr) =& \frac{\sqrt{6} v_F}{a_0 r_0} \cdot \rho_l \sigma_x \tau_x e^{-i\tau_z \KK_l\cdot\rr } \ , \\
    \label{eq:C_l_B1}
    C^{(l,\Bone)}(\qq;\rr) =& \frac{\sqrt{6} v_F}{a_0 r_0} \cdot \rho_l \sigma_x (-\tau_y) e^{-i\tau_z \KK_l\cdot\rr } \ .
\end{align}
Here, $\rho_l=\frac{\rho_0+(-1)^{l-1}\rho_z}{2}$ is a projector to the layer $l$, and the layer- and valley-dependent phases compensate the discrepancy between the gauge choices of \cref{eq:psi_TR,eq:uq_TR}. 
$\mcl{V}^{(l, \Aone/\Bone)} = \frac{\theta^2}{m_0r_0^2} \left( \frac{2\sqrt{6} v_F / a_0}{\omega_{0,c,\XO/\YO}} \right)^2 \sim 4.26\mrm{meV}$, contributing more significantly than the $(c/r, \XO/\YO)$ modes. 
For the inter-layer scatterings, the amplitudes induced by in-plane and out-of-plane $K$-phonons will be $\sim w_1|\KK^{(l)}|$ and $\sim \frac{w_1}{r_0}$, respectively (\cref{app:cont-Moire-in,app:cont-Moire-z}), which are both much smaller than $\frac{v_F}{a_0 r_0}$. Due to the high frequency of $K$-phonons, the resultant effective interaction will be negligible. \par 

In obtaining $\hat{H}_{\rm epc}$ (\cref{eq:Hepc_cont}), all the $O(\theta)$ corrections and non-local inter-layer scatterings have been omitted, just as in deriving $\hat{H}_0$ (\cref{eq:H0}) (see \cref{app:cont}). $\hat{H}_{\rm epc}$ enjoys the same space group $P622$ and time-reversal symmetry $T$, but $K$-phonons break the independent $\rm U(1)^2 \times SU(2)^2$ rotations in two valleys to a global $\rm U(1) \times SU(2)$ charge-spin conservation. 
We sum up that, for a generic moir\'e electron captured by the continuum model, its coupling to all the in-plane $\Gamma$-phonons, $(c/r, \XA/\XO/\YA/\YO)$, and out-of-plane $(r, \ZA)$, and in-plane $K$-phonon $(l, \Aone/\Bone)$, can be non-negligible, with the effective interaction strength estimated as several milli electron volts. 

We also remark that, the EPC strengths presented in this section are \textit{bare} values derived from the tight-binding approach \cref{eq:SKhop}. It has been shown that an interplay between EPC and the Coulomb interactions among electrons of energies higher than the phonon frequencies ($\gtrsim \omega_{0,c/r,\XO/\YO}$) can renormalize the phonon-mediated interaction strengths among low-energy electrons away from the bare values \cite{Basko_2008}. In particular, Ref.~\cite{Basko_2008} finds that, to the order of leading logarithmic corrections, the $(l,\Aone/\Bone)$ phonon-mediated interaction strengths can be enhanced by a factor around $3.2$, whereas the optical $\Gamma$-phonon-mediated interactions are unchanged. 
We will thus introduce two dimensionless RG factors, $\kappa_{\Gamma}$ and $\kappa_K$, for the overall $\Gamma$-phonon mediated interactions and $K$-phonon mediated interactions, respectively.

\section{Electron-phonon couplings on the \texorpdfstring{$f$}{f}-orbitals} \label{sec:vertex}

\begin{figure}[bt]
    \centering
    \includegraphics[width=0.75\linewidth]{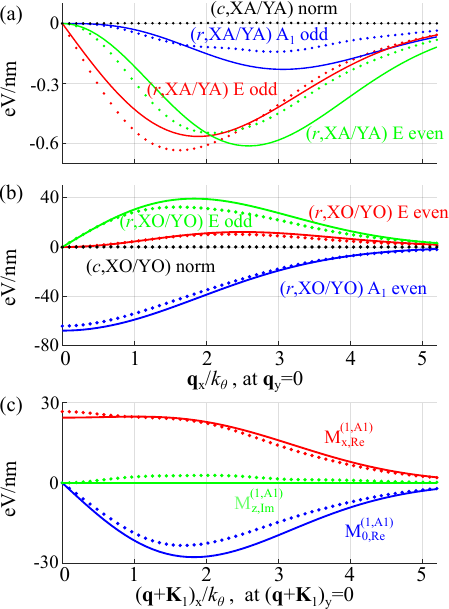}
    \caption{\label{fig:M_element} The projected EPC vertices, decomposed with parametrization functions tabulated in \cref{tab:rXAYA_irrep,tab:rXOYO_irrep,tab:lA1_irrep}. Dots are obtained from numerical Wannier functions, and solid lines are the Gaussian formulae. 
    (a) Parametrization functions for $(r,\XA/\YA)$ along $\qq_y=0$. There are two $E$ irreps with different $\qq$-parity. For each irrep, only one of the two basis functions, which is the one non-vanishing along $\qq_y=0$, is plotted. The $A_2$ component is also not shown, because it vanishes at $\qq_y=0$ by the odd $C_{2x}$-dependence. As analyzed in the main-text, its leading $\qq$-dependence along other directions starts from $O(|\qq|^3)$, which will be weak. 
    We also plot the Frobenius norm of $\sqrt{||M^{(c,\XA)}_{+,+}(\qq)||^2 + ||M^{(c,\YA)}_{+,+}(\qq)||^2}$ as black dots, to show that $(c,\XA/\YA)$ are indeed negligible on $f$-orbitals. 
    (b) Parametrization functions for $(r,\XO/\YO)$ along $\qq_y=0$, with notations following the same as in (a). 
    (c) Parametrization functions for $(1, \Aone)$. }
\end{figure}

As the $f$-orbitals are the major constituent to the flat bands, and provide the main arena where the correlated physics takes place, \textit{e.g.}, heavy Fermi liquid \cite{Chou_2023_Kondo, Zhou_2024_Kondo, Hu_2023_Kondo, Datta_2023_heavy, Rai_2023_DMFT, Chou_2023_scaling, Lau_2023_topological} and symmetry-breaking insulators \cite{Song_2022}, we project the EPC vertices obtained in the last section to the Wannier functions of the $f$-orbitals. Since analyzing the asymptotic $\qq$-dependence of the projected vertices will be our main goal, we fix $\kappa_\Gamma = \kappa_K = 1$ in this section. 

Owing to the small Gaussian width, the overlap between adjacent sites is exponentially small (see \cref{app:inter-vertex} for discussion on inter-$f$-site EPC vertex). It thus suffices to focus on the on-site transitions induced by phonons, 
\begin{align}   \label{eq:Hepc}
    \hat{H}_{\mrm{epc}} =& \frac{1}{\sqrt{N_g}} \sum_{\RR,s} \sum_{\beta\beta'\eta\eta'} \sum_{\chi} \sum_{|\qq| \lesssim \Lambda} e^{i\qq\cdot\RR} \\\nonumber
    & ~~~~ \times  u_{\qq,\chi} \times f^\dagger_{\RR\beta'\eta' s} M^{(\chi)}_{\beta'\eta',\beta\eta}(\qq) f_{\RR\beta\eta s} 
\end{align}
where the projected EPC vertex is 
\begin{align}  \label{eq:M}
    M^{(\chi)}_{\beta\eta, \beta'\eta'}(\qq) &= \int\mrm{d}^2\rr  ~ e^{i\qq\cdot\rr} \times \\\nonumber 
    & \left[\sum_{l\alpha,l'\alpha'} w^{(\eta)*}_{l\alpha,\beta}(\rr) ~ C^{(\chi)}_{l\alpha\eta,l'\alpha'\eta'}(\qq;\rr) ~ w^{(\eta')}_{l'\alpha',\beta'}(\rr) \right] 
\end{align}

Since $\Gamma$-phonons ($K$-phonons) induce intra-valley (inter-valley) scatterings exclusively, only valley diagonal elements $M^{(\chi@\Gamma)}_{\beta\eta, \beta'\eta}(\qq)$ (valley off-diagonal elements $M^{(\chi@K)}_{\beta\eta, \beta'\ovl{\eta}}(\qq)$) can be non-vanishing. 
In below, we will denote such $2\times2$ blocks as $M^{(\chi@\Gamma)}_{\eta,\eta}(\qq)$ (and $M^{(\chi@K)}_{\eta,\ovl{\eta}}(\qq)$), and analyze their long-wave behavior. 

\begin{table*}[tb]
    \centering
    \begin{tabular}{c|c|c|c}
    \hline\hline
        $D_3$ irrep & $\qq$-parity & Basis function & Gaussian formula \\
    \hline
        $A_1$ & odd & $\left[ M^{(r,\XA)}_{y,Im}(\qq) - M^{(r,\YA)}_{x,Im}(\qq) \right] / 2$ & $- \left( \frac{v_F}{16r_0} - \frac{3w_1 k_\theta^2 \lambda_2^2}{32\theta}  \right) \alpha_2^2\lambda_2^2 e^{-\frac{1}{4}\qq^2\lambda_2^2} \times \mrm{Re}(\qq_x + i\qq_y)^3$ \\
    \hline
        $A_2$ & odd & $\left[ M^{(r,\XA)}_{x,Im}(\qq) + M^{(r,\YA)}_{y,Im}(\qq) \right] / 2$ & $\left( \frac{v_F}{16r_0} - \frac{3w_1 k_\theta^2 \lambda_2^2}{32\theta}  \right) \alpha_2^2\lambda_2^2 e^{-\frac{1}{4}\qq^2\lambda_2^2} \times \mrm{Im}(\qq_x + i\qq_y)^3$ \\
    \hline
        $E$ & odd & $ \begin{pmatrix}
            \left[ M^{(r,\XA)}_{x,Im}(\qq) - M^{(r,\YA)}_{y,Im}(\qq) \right] / 2 \\
            \left[ M^{(r,\XA)}_{y,Im}(\qq) + M^{(r,\YA)}_{x,Im}(\qq)\right] / 2 \\
        \end{pmatrix} $ & $\left(- \frac{v_F}{4r_0} + \frac{3w_1k_\theta^2\lambda_1^2}{8\theta} \right)\alpha_1^2 e^{-\frac{1}{4}\qq^2\lambda_1^2} \times \begin{pmatrix}
            \qq_y \\ \qq_x \\
        \end{pmatrix}$ \\
    \hline
        $E$ & even & $ \begin{pmatrix}
            M^{(r,\XA)}_{z,Re}(\qq) \\
            M^{(r,\YA)}_{z,Re}(\qq) \\
        \end{pmatrix}$ & $\left(- \frac{v_F}{4r_0} + \frac{3w_1k_\theta^2\lambda'^2}{8\theta}\right)\frac{\alpha_1 \alpha_2 \lambda'^4}{\lambda_1\lambda_2^2} e^{-\frac{1}{4}\qq^2\lambda'^2} \times \begin{pmatrix}
            \qq_x^2 - \qq_y^2 \\ -2\qq_x\qq_y \\
        \end{pmatrix}$ \\
    \hline\hline
    \end{tabular}
    \caption{\label{tab:rXAYA_irrep} Symmetry-allowed parametrization functions for the $(r,\XA/\YA)$ vertex in the $\eta=+$ valley, $M^{(r,\XA/\YA)}_{+,+}(\qq)$. 
    We have defined $\frac{2}{{\lambda'}^2} = \frac{1}{{\lambda_1}^2} + \frac{1}{{\lambda_2}^2}$, hence $\lambda' = 0.1848 a_M$.
    $\qq$ denotes the phonon momentum, and the $\qq$-dependence is classified according to irreps of the $D_3$ group. 
    The parity of $\qq$-dependence is however constrained by Hermiticity. The Gaussian formulae agree well with the numeric results, as shown by \cref{fig:M_element}. The projected EPC vertices decay with $\qq$ in a Gaussian form, with the typical decaying length set by the Wannier spread, $\lambda_{1,2}$.}

    \centering
    \begin{tabular}{c|c|c|c}
    \hline\hline
        $D_3$ irrep & $\qq$-parity & Basis function & Gaussian formula \\
    \hline
        $A_1$ & even & $\left[ M^{(r,\XO)}_{x,Re}(\qq) + M^{(r,\YO)}_{y,Re}(\qq) \right] / 2$ & $- \frac{\sqrt{3} v_F}{r_0 a_0}  \alpha_1^2 e^{-\frac{1}{4} \qq^2 \lambda_1^2 }$ \\
    \hline
        $A_2$ & even & $\left[ M^{(r,\XO)}_{y,Re}(\qq) - M^{(r,\YO)}_{x,Re}(\qq) \right] / 2$ & 0 \\
    \hline
        $E$ & even & $\begin{pmatrix}
            \left[ M^{(r,\XO)}_{x,Re}(\qq) - M^{(r,\YO)}_{y,Re}(\qq) \right] / 2 \\
            \left[ M^{(r,\XO)}_{y,Re}(\qq) + M^{(r,\YO)}_{x,Re}(\qq) \right] / 2 \\
        \end{pmatrix}$ & $ \frac{\sqrt{3} v_F}{4 r_0 a_0} \alpha_2^2 \lambda_2^2 e^{-\frac{1}{4} \qq^2 \lambda_2^2}  \times \begin{pmatrix}
            \qq_x^2 - \qq_y^2 \\ 2\qq_x \qq_y \\
        \end{pmatrix}$ \\
    \hline
        $E$ & odd & $\begin{pmatrix}
            M^{(r,\XO)}_{z,Im}(\qq) \\ M^{(r,\YO)}_{z,Im}(\qq) \\
        \end{pmatrix}$ & $\frac{\sqrt{3} v_F}{r_0 a_0} \frac{\alpha_1\alpha_2\lambda'^4} {\lambda_1\lambda_2^2} e^{-\frac{1}{4} \qq^2 {\lambda'}^2} \times \begin{pmatrix}
            -\qq_y \\ \qq_x \\ 
        \end{pmatrix}$ \\
    \hline\hline
    \end{tabular}
    \caption{\label{tab:rXOYO_irrep} Symmetry-allowed parametrization functions for the $(r,\XO/\YO)$ vertex in the $\eta=+$ valley, $M^{(r,\XO/\YO)}_{+,+}(\qq)$. 
    Notations follow \cref{tab:rXAYA_irrep}.} 
    
    \centering
    \begin{tabular}{c|c}
    \hline\hline
        Parametrization function & Gaussian formula \\
    \hline
        $M^{(l,\Aone)}_{0,Re}(\qq)$ & $-\frac{\sqrt{6} v_F}{a_0r_0} \frac{\alpha_1\alpha_2 {\lambda'}^4}{2\lambda_1 \lambda^2_2} e^{-\frac{1}{4} {\lambda'}^2 (\qq+\KK_l)^2} \left( \qq+\KK_l \right)_x$ \\
        $M^{(l,\Aone)}_{z,Im}(\qq)$ & $\frac{\sqrt{6} v_F}{a_0r_0} \frac{\alpha_1\alpha_2 {\lambda'}^4}{2\lambda_1 \lambda^2_2} e^{-\frac{1}{4} {\lambda'}^2 (\qq+\KK_l)^2} \left( \qq+\KK_l \right)_y$ \\
        $M^{(l, \Aone)}_{x,Re}(\qq)$ & $\frac{\sqrt{6} v_F}{a_0r_0} \left( \frac{\alpha_1^2}{2} e^{-\frac{1}{4} \lambda_1^2 (\qq+\KK_l)^2} - \frac{\alpha_2^2}{2} \left( 1 - \frac{1}{4} \lambda_2^2 (\qq+\KK_l)^2 \right) e^{-\frac{1}{4} \lambda_2^2 (\qq+\KK_l)^2}) \right)$ \\
    \hline\hline
    \end{tabular}
    \caption{\label{tab:lA1_irrep} Symmetry-allowed parametrizing functions for the $(l,\Aone)$ vertex that flips $\eta=-$ to $\eta=+$, $M^{(l,\Aone)}_{+,-}(\qq)$. $C_{3z}$ symmetry further relates $M^{(l,\Aone)}_{0,Re}(\qq)$ and $M^{(l,\Aone)}_{z,Im}(\qq)$ through \cref{eq:MK_C3z}. The projected vertex for $(l,\Bone)$ modes can be obtain by $M^{(l,\Bone)}_{+,-}(\qq) = i M^{(l,\Aone)}_{+,-}(\qq)$. }
\end{table*}

\subsection{\texorpdfstring{$\Gamma$}{Gamma}-phonons} \label{sec:vertex-G}

Due to the time-reversal symmetry, $M^{(\chi@\Gamma)}_{+,+}(\qq)$ and $M^{(\chi@\Gamma)}_{-,-}(\qq)$ must be related. As detailed in Appendix \ref{app:sym}, $T$ acts on phonon fields as $T u_{\qq,\chi} T^{-1} = u_{-\qq, \chi}$ for arbitrary $\chi$. Therefore, for \cref{eq:Hepc} to preserve $T$, with electron fields transforming according to \cref{eq:DfT}, there must be 
\begin{align}    \label{eq:M_T_Gamma}
    M^{(\chi@\Gamma)}_{+, +}(\qq) = M^{(\chi@\Gamma)*}_{-, -}(-\qq)
\end{align}
Moreover, \cref{eq:Hepc} must obey Hermiticity, hence 
\begin{align}   \label{eq:M_Herm_Gamma} 
    M^{(\chi@\Gamma)}_{\eta, \eta}(\qq) = M^{(\chi@\Gamma)\dagger}_{\eta, \eta}(-\qq)
\end{align}
Combining \cref{eq:M_T_Gamma} with \cref{eq:M_Herm_Gamma}, there must be
\begin{align}   \label{eq:M_THerm_Gamma}
      M^{(\chi@\Gamma)}_{+, +}(\qq) = M^{(\chi@\Gamma)T}_{-, -}(\qq) \ . 
\end{align} 
It thus suffices to obtain $M^{(\chi@\Gamma)}_{+,+}(\qq)$, which can be parametrized in terms of Pauli matrices as, 
\begin{align}   \label{eq:M_para_G}
    M^{(\chi@\Gamma)}_{+, +}(\qq) &= \sum_{\mu=0,x,y,z} \left( M^{(\chi@\Gamma)}_{\mu, Re}(\qq) + i M^{(\chi@\Gamma)}_{\mu, Im}(\qq) \right) \sigma^\mu 
\end{align} 
with real smooth functions $M^{(\chi@\Gamma)}_{\mu, Re}(\qq)$ and $M^{(\chi@\Gamma)}_{\mu, Im}(\qq)$. Due to the Hermiticity requirement (\cref{eq:M_Herm_Gamma}), $M^{(\chi@\Gamma)}_{\mu, Re}(\qq)$ and $M^{(\chi@\Gamma)}_{\mu, Im}(\qq)$ must be even and odd in $\qq$, respectively.

\subsubsection{\texorpdfstring{$(c, \XA/\YA/\XO/\YO)$}{(c, XA/YA/XO/YO)}} \label{sec:vertex-G-1}

We first show that, for the co-moving in-plane $\Gamma$-phonons $(c, \XA/\YA/\XO/\YO)$, \cref{eq:M} asymptotically vanishes for all $\qq$. 
This can be understood from the Gaussian fit \cref{eq:w_Gauss_A,eq:w_Gauss_B,eq:w_Gauss_2}, where both the $\beta=1,2$ Wannier functions hold as the exact eigenstates of the following operator $\td{S} = \rho_y \sigma_z $ with the same eigenvalue $-1$, 
\begin{align} \label{eq:tdS_eig}
    \sum_{l'\alpha'} [\td{S}]_{l\alpha, l'\alpha'} ~  w^{(+)}_{l'\alpha',\beta}(\rr)  = (-1)  w^{(+)}_{l\alpha,\beta}(\rr) \ .
\end{align} 
Here, the summation over $l'\alpha'$ indices is interpreted as a matrix product, and we can denote \cref{eq:tdS_eig} as $\td{S} |\beta\rangle = - |\beta\rangle$ for shorthand. 
Now, for any operator $O$ that anti-commutes with $\td{S}$, $O|\beta\rangle$ must have eigenvalue $+1$, hence being orthogonal to any $|\beta'\rangle$ with eigenvalue $-1$. Thereby, $\langle\beta'|O|\beta\rangle$ must vanish. 
It can be immediately checked that, $C^{(c, \XA/\YA/\XO/\YO)}(\qq;\rr)$ (\cref{eq:C_c_XA,eq:C_c_YA,eq:C_c_XO,eq:C_c_YO}) anti-commutes with $\td{S}$ for all $\qq$, hence this criterion applies, forcing the matrix elements \cref{eq:M} to be exactly vanishing, as long as the Gaussian fit is adopted. 
For the numeric Wannier functions, due to the large overlap with the Gaussian fit, we verify that the matrix elements \cref{eq:M} are indeed negligibly small, as illustrated by the black dotted lines in \cref{fig:M_element}(a,b). 

It is also worth noting that, if the moir\'e potential in $\hat{H}_0$ (\cref{eq:H0}) is expanded to the first order of $k_\theta \rr$ in the vicinity of AA-center, the resulting local Hamiltonian will also anti-commute with $\td{S}$. In Ref. \cite{Shi_2022_THF}, it has been shown that the local Hamiltonian can be interpreted as two coupled Dirac fermions moving under opposite pseudomagnetic fields. The $f$-orbitals, as eigenstates of $\td{S}$, are thus interpreted as the chiral zero modes under the pseudomagnetic fields.

\subsubsection{\texorpdfstring{$(r, \ZA)$}{(r, ZA)}}

Now we show that $M^{(r, \ZA)}_{+,+}(\qq)$ vanishes at $\qq=0$, forced by $C_{3z}$ and the emergent particle-hole $P$, despite the original $C^{(r,\ZA)}(\qq;\rr)$ vertex (\cref{eq:C_r_ZA}) does not. 
To begin with, due to \cref{eq:DfC3z}, the two $f$-orbitals in the same valley $\eta$ with opposite $\beta$ have opposite $C_{3z}$ angular momentum. 
As detailed in \cref{app:sym}, $u_{0,r,\ZA}$ has zero $C_{3z}$ angular momentum, hence cannot induce any orbital-off-diagonal transitions, forcing $M^{(r,\ZA)}_{x/y,Re/Im}(0)$ to vanish. 
On the other hand, by conjugating \cref{eq:f} with $P$ (\cref{eq:DfP}), it can be shown that the Wannier functions must obey
\begin{align} \label{eq:DP_eig}
    \sum_{l'\alpha'} [D(P)]_{l\alpha,l'\alpha'} ~ w^{(+)}_{l'\alpha',\beta}(\rr) = \left(i (-1)^{\beta-1} \right) w^{(+)}_{l\alpha,\beta}(-\rr)
\end{align}
Recall that $P$ anti-commutes with $\hat{H}_0$, which implies that the moir\'e potential $T(\rr)$, and thus $C^{(r,\ZA)}(\qq;\rr)$ by \cref{eq:C_r_ZA}, must anti-commute with $D(P)$, with an extra coordinate inversion that flips $\rr$ to $-\rr$, 
\begin{align} \label{eq:DP_anticom}
    D(P) T(\rr) D(P)^{-1} = - T(-\rr) \ .
\end{align}
Therefore, at $\qq=0$, after substituting \cref{eq:DP_eig,eq:DP_anticom} into \cref{eq:M}, and integrating over $\rr$, it will be immediate that the orbital diagonal elements spanned by $M^{(r,\ZA)}_{0/z,Re/Im}(0)$ must vanish as well. 

Note that for finite $\qq$, due to the $e^{i\qq\cdot\rr}$ factor, $M^{(r,\ZA)}_{+,+}(\qq;\rr)$ can grow to acquire finite values. However, as smooth functions, $M^{(r, \ZA)}_{+,+}(\qq)$ can be at most of the $O(|\qq|)$ order at $\qq\to0$. 
With the finite gap $\omega_{0,r,\ZA}$ in the phonon dispersion, the effective interaction mediated by the $(r,\ZA)$ phonons among $f$-orbitals will be extremely weak. Following the procedures that are to be introduced in \cref{sec:HA}, the corresponding interaction matrix elements are calculated as $\lesssim 0.01\mrm{meV}$. 
We will thus treat $M^{(r, \ZA)}_{+,+}(\qq)$ as negligible, too.

\subsubsection{\texorpdfstring{$(r, \XA/\YA)$}{(r, XA/YA)}}   \label{sec:vertex-rXAYA}

We first prove that $C_{2z}T$ and $C_{3z}$ enforce $M^{(r, \XA/\YA)}_{+,+}(\qq)$ to vanish at $\qq \to 0$, which will free us from the artificial divergence in obtaining the effective interaction. 

As detailed in \cref{app:sym}, $(C_{2z}T) u_{\qq, \chi} (C_{2z}T)^{-1} = - u_{\qq, \chi}$ for both $\chi=(r,\XA/\YA)$, where the $-1$ sign arises because $C_{2z}$ rotation also rotates the displacement vectors, and $C_{2z}T$ leaves the momentum $\qq$ unchanged. 
Moreover, combining \cref{eq:DfC2z,eq:DfT}, there is $D^f(C_{2z}T) = \sigma^x \tau^0$.
For \cref{eq:Hepc} to stay invariant under $C_{2z}T$, $M^{(\chi)}_{+, +}(\qq)$ thus must anti-commute with $\sigma^x K$, hence for arbitrary $\qq$, only $M^{(r, \XA/\YA)}_{z, Re}(\qq)$ and $M^{(r, \XA/\YA)}_{0/x/y, Im}(\qq)$ can survive. 
As is clarified below Eq. (\ref{eq:M_para_G}), $M^{(r, \XA/\YA)}_{0/x/y, Im}(\qq)$ are odd in $\qq$, thus must vanish at $\qq\to0$. Consequently, we only need to examine the behavior of $M^{(r, \XA/\YA)}_{z, Re}(\qq)$ at $\qq\to0$, which describes intra-orbital transition that does not change the electron $C_{3z}$ angular momentum. 
Nevertheless, $u_{0,r,\XA}$ and $u_{0,r,\YA}$ form a vector operator, which carries $C_{3z}$ angular momentum $\pm 1$ mod 3 (see \cref{app:sym}). 
The mismatch thus dictates the orbital-diagonal $M^{(r, \XA/\YA)}_{z, Re}(\qq)$ component to vanish at $\qq=0$ as well. 

We then prove in \cref{app:vertex-rXAYA} that, due to the emergent particle-hole symmetry $P$ obeyed by the Wannier functions (\cref{eq:DP_eig}), $M^{(r,\XA/\YA)}_{0,Im}(\qq)$ is further forbidden at arbitrary $\qq$. The remaining six functions, namely $M^{(r, \XA/\YA)}_{z, Re}(\qq)$ and $M^{(r, \XA/\YA)}_{x/y, Im}(\qq)$, must form representations of the $D_3$ group, which is the point group obeyed by $f$-electrons in a single valley. 
As shown in \cref{tab:rXAYA_irrep}, the off-diagonal elements ($M^{(r, \XA/\YA)}_{x/y, Im}(\qq)$, $\qq$-odd) are characterized into the $A_1 + A_2 + E$ irreps, while the diagonal elements ($M^{(r, \XA/\YA)}_{z, Re}(\qq)$, $\qq$-even) span an $E$ irrep. 

Using the numeric Wannier functions, the parametrization functions can be evaluated explicitly. 
Alternatively, we also attempt to analytically solve the asymptotic behavior at $\qq\to0$ through the Guassian fit \cref{eq:w_Gauss_A,eq:w_Gauss_B,eq:w_Gauss_2}, which is elaborated in \cref{app:vertex-rXAYA}. 
However, due to the complicated $\rr$-dependence of the inter-layer term in $C^{(r, \XA/\YA)}(\qq;\rr)$ (\cref{eq:C_r_XA,eq:C_r_YA}), the results can be greatly complicated. 
We thus exploit the narrow spread of $f$-orbitals, and expand the $\rr$-dependent terms to the linear order of $(\qq_j\cdot\rr)$ near the AA-stacking center ($\rr=0$). The Gaussian formulae obtained by this approach are summarized in \cref{tab:rXAYA_irrep}. 
The error of the approximation relies on the relative width of Wannier functions, $\frac{\lambda_{1,2}}{a_M}$, and we compare the approximate analytic results with the numeric results in \cref{fig:M_element}(a). 
It can be seen that, for the $\qq$-odd and $\qq$-even $E$ irrep functions, which correspond to adding dynamic $\sigma^{x/y}$ and $\sigma^z$ orbital Zeeman fields, they have $O(|\qq|)$ and $O(|\qq|^2)$ dependence to the leading order, respectively. The linear and quadratic coefficients agree qualitatively well between the two approaches. 
For the $A_1$ and $A_2$ irrep functions, the leading $\qq$-dependence starts at $O(|\qq|^3)$. 
Inducing dynamic $\sigma^{x/y}$ fields as the odd $E$ irrep, they have a much weaker contribution.

\subsubsection{\texorpdfstring{$(r, \XO/\YO)$}{(r, XO/YO)}}  \label{sec:vertex-rXOYO}

Following a symmetry analysis parallel to that in the last subsection (also see \cref{app:vertex-rXOYO}), the only non-vanishing parametrizing functions are $M^{(r, \XO/\YO)}_{z,Im}(\qq)$ ($\qq$-odd) and $M^{(r, \XO/\YO)}_{x/y, Re}(\qq)$ ($\qq$-even). 
Concretely, $M^{(r, \XO/\YO)}_{0,Re}(\qq)$ is allowed by $C_{2z}T$ but forbidden by the emergent $P$, while all the other components are forbidden by $C_{2z}T$. 
The non-vanishing functions are categorized into $D_3$ irreps in \cref{tab:rXOYO_irrep}, and the Gaussian formulae and the numeric results are compared in \cref{fig:M_element}(b). 
For the two $E$ irreps and the $A_1$ irrep, the two methods produce consistent results. For the $\qq$-even $A_2$ irrep function, which vanishes in the Gaussian results, it can be verified that the leading order $\qq$-dependence starts at $O(|\qq|^6)$, which should be weak. They do not appear in the Gaussian results.

\subsection{\texorpdfstring{$K$}{K}-phonons}

For $K$-phonons, the Hermiticity of \cref{eq:Hepc} requires that 
\begin{align}   \label{eq:M_Herm_K}
    M^{(\chi@K)}_{+, -}(\qq) = M^{(\chi@K)\dagger}_{-,+}(-\qq)
\end{align}
while the time-reversal $T$ requires that
\begin{align}   \label{eq:M_T_K}
    M^{(\chi@K)}_{+, -}(\qq) = M^{(\chi@K)*}_{-,+}(-\qq)
\end{align}
Together, they constrain $M^{(\chi@K)}_{+, -}(\qq)$ to be a symmetric matrix, 
\begin{align} \label{eq:M_THerm_K}
    M^{(\chi@K)}_{+, -}(\qq) = M^{(\chi@K) T}_{+, -}(\qq)
\end{align}
hence it suffices to include symmetric Pauli matrices in the parametrization, 
\begin{align}   \label{eq:M_para_K}
    M^{(\chi@K)}_{+,-}(\qq) = \sum_{\mu=0,x,z}  \left[ M^{(\chi@K)}_{\mu, Re}(\qq) + i  M^{(\chi@K)}_{\mu, Im}(\qq) \right] \sigma^\mu  
\end{align}
Unlike the $\Gamma$-phonons, here, the $\qq$-parity of parametrization functions is not restricted. 

For $\chi@K = (l, \Aone/\Bone)$ of our interest, deriving from \cref{eq:C_l_A1,eq:C_l_B1}, there is $M^{(l, \Bone)}_{+, -}(\qq) = i M^{(l, \Aone)}_{+, -}(\qq)$. 
This can be better understood from the $u_{\qq,l,\eta K}$ basis \cref{eq:uK_eta}, where the same $u_{\qq,l,-K}$ component is responsible for the electron scattering from the $-$ valley to $+$ valley in both modes, except for a different phase (\cref{app:vertex-K}). We thus focus on $M^{(l, \Aone)}_{+, -}(\qq)$. 

As an $A_1$ irrep, $(C_{2z}T) u_{\qq,l,\Aone} (C_{2z}T)^{-1} = u_{\qq,l,\Aone}$ (\cref{app:sym}). Therefore, the $C_{2z}T$ symmetry of \cref{eq:Hepc} dictates $M^{(l,\Aone)}_{+,-}(\qq)$ to commute with $\sigma^x K$, namely only $M^{(l,\Aone)}_{0/x,Re}(\qq)$ and $M^{(l,\Aone)}_{z,Im}(\qq)$ can appear. Moreover, as $C_{3z} u_{\qq,l,\eta K} C_{3z}^{-1} = u_{C_{3z}\qq,l,\eta K}$ (\cref{app:sym}), the following constraints are imposed, 
\begin{align} \label{eq:MK_C3z}
    e^{i\frac{2\pi}{3} \sigma^z} M_{+,-}^{(l, \Aone)}(\qq) e^{i\frac{2\pi}{3} \sigma^z} = M_{+,-}^{(l, \Aone)}(C_{3z}\qq + C_{3z}\KK_l - \KK_l) 
\end{align}
Note that the $C_{3z}$ rotation center for the parametrization functions has been displaced to $\qq = -\KK_l$, as explained around \cref{eq:uK_eta}. \cref{eq:MK_C3z} thus implies that $M^{(l,\Aone)}_{0,Re}(-\KK_l)$ and $M^{(l,\Aone)}_{z,Im}(-\KK_l)$ must vanish. 
The Gaussian results are tabulated in \cref{tab:lA1_irrep}, and the numeric results are compared to the Gaussian formulae in \cref{fig:M_element}(c), which again agree well.

\section{The phonon-mediated anti-Hund's interaction}  \label{sec:HA}

Given the on-site EPC vertex \cref{eq:Hepc}, if a phonon is emitted at site $\RR$ and then absorbed at another site $\RR'$, it will effectively mediate an interaction among electrons at the two sites. The interaction is in general frequency dependent, and can be written in terms of the following action,  
{\small \begin{align}   \nonumber
    & S_{\rm A} = \frac{\mrm{T}}{2 N_g } \sum_{\chi} \sum_{|\qq| \lesssim \Lambda} e^{i\qq \cdot (\RR - \RR')} \times \sum_{\RR,\RR'}  \sum_{\beta_{1,2}\beta'_{1,2}} \sum_{\eta_{1,2}\eta'_{1,2}} \sum_{s,s'} \sum_{k_0,p_0,q_0} \\\nonumber
    & f^{\dagger}_{\RR\beta_1\eta_1 s} (ik_0+iq_0) f^\dagger_{\RR'\beta_1'\eta_1's'}(ip_0 - iq_0) f_{\RR'\beta_2'\eta_2's'}(ip_0) f_{\RR\beta_2\eta_2s} (ik_0)  \\  \label{eq:Sint}
    & \times M^{(\chi)}_{\beta_1\eta_1, \beta_2\eta_2}(\qq) D^{(\chi)}(\qq, iq_0) M^{(\chi)}_{\beta'_1\eta'_1, \beta'_2\eta'_2}(-\qq) 
\end{align}}%
where the Boltzmann constant is set to $k_B=1$ for convenience, $\rm T$ is the temperature, $ik_0$ and $ip_0$ are fermionic Matsubara frequencies, and $f^\dagger_{\RR\beta\eta s}(ik_0)$ denotes a Grassman number. 
Nevertheless, for the $f$-electrons in MATBG, the kinetic energy is almost quenched - for the uncorrelated flat bands, the band width is $\lesssim$3meV, and in the Kondo regime, the renormalized band width for heavy quasi-particles can be estimated as $T_{\rm K} \lesssim 1\mrm{meV}$. In contrast, the characteristic phonon frequencies are much higher. The acoustic in-plane $\Gamma$-phonons that are significantly coupled to the $f$-orbitals span a momentum range over several times of $k_\theta$ (\cref{fig:M_element}(a)), which corresponds to a frequency range of several times of $v_s k_\theta$ ($v_s k_\theta \sim 3.6\mrm{meV}$), and the optical in-plane $\Gamma$-phonons and $K$-phonons possess even higher frequencies ($\gtrsim$150meV). 
Therefore, the phonon retardation should be un-important, and we will neglect the frequency dependence. 
Moreover, as shown in the last section, the $\qq$-dependent quantity $M^{(\chi)}_{\beta_1\eta_1, \beta_2\eta_2}(\qq) D^{(\chi)}(\qq, 0) M^{(\chi)}_{\beta_1'\eta_1',\beta_2'\eta_2'}(-\qq) $ forms a smooth Gaussian function with large width $\frac{1}{\lambda_{1,2}} > \frac{1}{a_M}$. Thereby, summing over $\qq$ will cause \cref{eq:Sint} to decay exponentially with $|\RR - \RR'|$. We estimate the strengths of inter-site interactions in \cref{app:inter-int}, and find them to be of the order of $10^{-2}$meV or less. To sum up, it suffices to focus on the instantaneous on-site interaction, which can now be written as, 
{\small \begin{align} \label{eq:HA}
    \hat{H}_{\rm A} =& \frac{1}{2} \sum_{\RR,\eta,\eta',s,s'} \sum_{\alpha\alpha'\beta\beta'} V^{\eta'\eta}_{\beta'\alpha',\beta\alpha} f^\dagger_{\RR\alpha'\eta s} f^\dagger_{\RR\beta'\eta' s'} f_{\RR\beta\eta' s'} f_{\RR\alpha\eta s} \\ \nonumber
    +& \frac{1}{2} \sum_{\RR,\eta,s,s'} \sum_{\alpha\alpha'\beta\beta'} \td{V}^{\eta}_{\beta'\alpha',\beta\alpha} f^\dagger_{\RR\alpha'\ovl{\eta} s} f^\dagger_{\RR\beta'\eta s'} f_{\RR\beta\ovl{\eta} s'} f_{\RR\alpha\eta s} 
\end{align}}%
with
{\small \begin{align}   \label{eq:V}
    V^{\eta'\eta}_{{\beta'}{\alpha'}, {\beta}{\alpha}} = - \frac{1}{N_g m_0} \sum_{\chi@\Gamma} \sum_{|\qq|<\Lambda}  \frac{M^{(\chi@\Gamma)}_{\beta'\eta',\beta\eta'}(-\qq) M^{(\chi@\Gamma)}_{\alpha'\eta,\alpha\eta}(\qq)} {\omega^2_{\qq,\chi}}
\end{align}
\begin{align}   \label{eq:Vtd}
    \td{V}^{\eta}_{{\beta'}{\alpha'},{\beta}{\alpha}} = - \frac{1}{N_g m_0} \sum_{\chi@K} \sum_{|\qq|<\Lambda} \frac{M^{(\chi@K)}_{\beta'\eta,\beta\ovl{\eta}}(-\qq) M^{(\chi@K)}_{\alpha'\ovl{\eta},\alpha\eta}(\qq)} {\omega^2_{\qq,\chi}}
\end{align}}%

$V^{\eta'\eta}$ and $\td{V}^{\eta}$ originate from $\Gamma$-phonons and $K$-phonons, respectively. In particular, $\td{V}^{\eta}$ breaks the $\rm SU(2) \times SU(2)$ spin rotation of two valleys down to the global spin SU(2) symmetry; but the separate $\rm U(1) \times U(1)$ charges of two valleys are recovered after phonon fields are integrated out. We will term the U(1) charge generated by $\tau^z$ as the valley charge, and denote the corresponding quantum number as $N_v$. 

Dictated by the Hermiticity, as well as the crystalline $D_6$ symmetry and time-reversal $T$, the free parameters in the $V^{\eta'\eta}$ and $\td{V}^{\eta}$ matrices are further constrained. 
To begin with, $C_{2y}T = C_{2x} \cdot C_{2z}T$, which acts on the $f$-basis as a complex conjugation (combining \cref{eq:DfC2z,eq:DfT,eq:DfC2x}), will require $V^{\eta'\eta}$ and $\td{V}^{\eta}$ to be real-valued. 
Next, the spinless time-reversal $T$ (\cref{eq:DfT}) requires that $V^{\eta\eta} = V^{\ovl{\eta}\ovl{\eta}}$, $V^{\eta\ovl{\eta}} = V^{\ovl{\eta}\eta}$, and $\td{V}^{\ovl{\eta}} = \td{V}^{\eta}$. By replacing one of the two EPC vertices in Eq. (\ref{eq:V}) with its time-reversal counterpart given by Eq. (\ref{eq:M_THerm_Gamma}), it can be further shown that $V^{\eta\eta}_{\beta'\alpha', \beta\alpha} = V^{\eta\ovl{\eta}}_{\beta'\alpha,\beta\alpha'}$. Therefore, now it suffices to determine $V^{\eta\eta}$ and $\td{V}^{\eta}$ for one specific $\eta$. 

As an $f^\dagger_{\RR\beta\eta s}$ electron carries angular momentum $\eta \beta$ mod 3 (\cref{eq:DfC3z}), all elements in $V^{\eta\eta}$, except $V^{\eta\eta}_{\beta\beta, \beta\beta}$, $V^{\eta\eta}_{\beta\ovl{\beta}, \beta\ovl{\beta}}$, and $V^{\eta\eta}_{\beta\ovl{\beta}, \ovl{\beta}\beta}$ will vanish due to the $C_{3z}$ symmetry. By $C_{2x}$ symmetry, which only flips $\beta$ (\cref{eq:DfC2x}), these three groups of elements will all be independent of $\beta$, which is also in agreement with the requirement of Hermiticity. $V^{\eta\eta}$ is thus parametrized as
\begin{align}   \label{eq:Vee}
    V^{\eta\eta}_{\beta'\alpha', \beta'\alpha} =& \begin{pmatrix}
        -J_{\rm a} & & & \\
         & J_{\rm a} & -J_{\rm b} & \\
         & -J_{\rm b} & J_{\rm a} & \\
         &  & & -J_{\rm a} \\
    \end{pmatrix}_{\beta'\alpha',\beta\alpha} - J_{\rm c} ~ \delta_{\beta'\alpha', \beta\alpha}
\end{align}
where the columns and rows are sorted as $(\beta\alpha)$ = (11), (12), (21), (22). 
Note that $J_{\rm c}$ can be understood as an attractive Hubbard term.

For $\td{V}^\eta$, similarly, $C_{3z}$ symmetry requires all elements except $\td{V}^\eta_{\beta\beta, \beta\beta}$, $\td{V}^{\eta}_{\beta\beta, \ovl{\beta}\ovl{\beta}}$, and $\td{V}^\eta_{\beta\ovl{\beta}, \ovl{\beta}\beta}$ to vanish, while $C_{2x}$ renders these three groups all independent of $\beta$. 
Recall that $M^{(\chi@K)}_{\eta,\ovl{\eta}}(\qq)$ must be a symmetric matrix constrained by the Hermiticity and $T$ (\cref{eq:M_THerm_K}), swapping $\alpha$ with $\alpha'$ in Eq. (\ref{eq:Vtd}) results in identical values, which further restricts that $\td{V}^{\eta}_{\beta\beta, \ovl{\beta}\ovl{\beta}} = \td{V}^\eta_{\beta\ovl{\beta}, \ovl{\beta}\beta}$. Thereby, $\td{V}^\eta$ is parameterized as
\begin{align}   \label{eq:Ve}
    \td{V}^{\eta}_{\beta'\alpha', \beta'\alpha} =& \begin{pmatrix}
        -J_{\rm e} & & & -J_{\rm d} \\
         & & -J_{\rm d} & \\
         & -J_{\rm d} & & \\
        -J_{\rm d} & & & -J_{\rm e} \\
    \end{pmatrix}_{\beta'\alpha',\beta\alpha} 
\end{align}
The signs are chosen so that $J_{\rm a,b,c,d,e}$ are positive. 

We tabulate the bare tight-binding values (namely, $\kappa_\Gamma = \kappa_K = 1$) of $J_{\rm a,b,c,d,e}$ in Table \ref{tab:int}, obtained from both the numeric and Gaussian Wannier wave-functions. 
In particular, we find $J_{\rm c}=0$, which results from the following two facts already analyzed in \cref{sec:vertex-G} - first, among all $\Gamma$-phonons, $f$-orbitals are only coupled to $(r,\XA/\YA)$ and $(r,\XO/\YO)$ modes; second, the projected vertex of these phonon modes does not contain $\sigma^0$ components due to the emergent $P$ symmetry. 
In realistic systems where $P$ holds only approximately, $J_{\rm c}\not=0$. 
Nevertheless, a finite $J_{\rm c}$ can always be absorbed into the on-site Hubbard $U$, which acts $\rm U(8)$-symmetrically in all orbital, valley, and spin indices. 
We will henceforth set $J_{\rm c}=0$, and focus on the $\rm U(8)$-splittings induced by $J_{\rm a, b, d, e}$. 

In the next section, we directly diagonalize $\hat{H}_{\rm A}$ (\cref{eq:HA}) in each subspace of fixed $f$-electron number $N_f$. 
With $J_{\rm a,b,d,e}>0$, in general, $N_f$-electron states with a higher total spin are also found with a higher energy. Therefore, we term the phonon-mediated $\hat{H}_{\rm A}$ as an anti-Hund's interaction.

\section{Phonon-induced multiplet splittings}  \label{sec:multi}

\begin{figure}
    \centering
    \includegraphics[width=\linewidth]{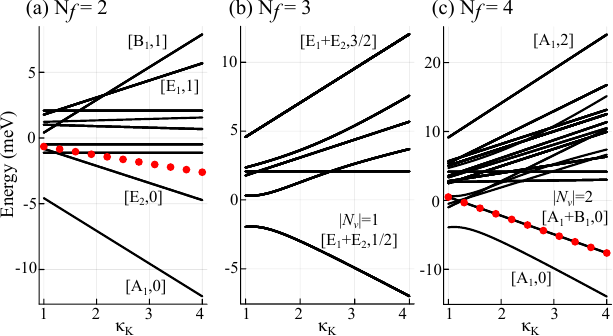}
    \caption{\label{fig:split-0} (a-c) Multiplet splitting induced by $\hat{H}_{\rm A}$ in $N_f=2,3,4$ subspaces, with $\kappa_\Gamma=1$ fixed. 
    Each black line indicates a multiplet level, with symmetry labels tabulated in \cref{tab:2e,tab:3e,tab:4e}, respectively. 
    The symmetry irreps of several lowest-energy or highest-energy states are explicitly marked. 
    In (a) and (c), the energy expectation value of a $\rm TIVC \times QAH/QSH$ order and a $\rm TIVC$ order are also plotted as red dots for comparison, respectively.
    These Hartree-Fock orders are superpositions of energy eigenstates. }
\end{figure}

\begin{table*}[tb]
    \centering
    \begin{tabular}{c|c|c|c|c|c}
    \hline\hline
        $|N_v|$ & $|L|$ & $[\rho, j]$ & degeneracy & rep. w.f. & energy \\
    \hline
        \multirow{4}{*}{0} & \multirow{4}{*}{0} & $[A_1, 0]$ & 1 & $\frac{1}{2} (f^\dagger_{1+\up} f^\dagger_{1-\down} + f^\dagger_{2+\up} f^\dagger_{2-\down}) - [\up \leftrightarrow \down]$ & $-J_{\rm a}-J_{\rm b}-J_{\rm d}-J_{\rm e}$ \\
        &  & $[B_2, 0]$ & 1 & $\frac{1}{2} (f^\dagger_{1+\up} f^\dagger_{1-\down} - f^\dagger_{2+\up} f^\dagger_{2-\down}) - [\up \leftrightarrow \down]$ & $-J_{\rm a}+J_{\rm b}+J_{\rm d}-J_{\rm e}$ \\
        &  & $[B_1, 1]$ & 3 & $\sqrt{\frac{1}{2}} ( f^\dagger_{1+\up} f^\dagger_{1-\up} + f^\dagger_{2+\up} f^\dagger_{2-\up} )$ & $-J_{\rm a} - J_{\rm b} + J_{\rm d} + J_{\rm e}$ \\
        &  & $[A_2, 1]$  & 3 & $\sqrt{\frac{1}{2}} ( f^\dagger_{1+\up} f^\dagger_{1-\up} - f^\dagger_{2+\up} f^\dagger_{2-\up} ) $ & $- J_{\rm a} + J_{\rm b} - J_{\rm d} + J_{\rm e}$ \\
    \hline
        \multirow{2}{*}{0} & \multirow{2}{*}{2} & $[\mrm{E}_2, 0]$ & 2 & $\sqrt{\frac{1}{2}} f^\dagger_{1+\up} f^\dagger_{2-\down} - [\up \leftrightarrow \down]$ & $+J_{\rm a} - J_{\rm d}$ \\ 
        &  & $[\mrm{E}_1, 1]$ & 6 & $f^\dagger_{1+\up} f^\dagger_{2-\up}$ & $+J_{\rm a} + J_{\rm d}$ \\ 
    \hline
        \multirow{2}{*}{2} & \multirow{2}{*}{0} & $[\mrm{A}_1+\mrm{B}_1, 0]$ & 2 & $\sqrt{\frac{1}{2}} f^\dagger_{1+\up} f^\dagger_{2+\down} - [\up \leftrightarrow \down]$ & $+J_{\rm a} - J_{\rm b}$ \\ 
        &  & $[\mrm{A}_2+\mrm{B}_2, 1]$ & 6 & $f^\dagger_{1+\up} f^\dagger_{2+\up}$  & $+J_{\rm a} + J_{\rm b}$ \\ 
    \hline
        2 & 2 & $[\mrm{E}_1 + \mrm{E}_2, 0]$ & 4 & $f^\dagger_{1+ \up} f^\dagger_{1+ \down}$ & $-J_{\rm a}$ \\ 
    \hline\hline
    \end{tabular}
    \caption{\label{tab:2e} Two-electron eigenstates of $\hat{H}_{\rm A}$, labeled by the $D_6$ representation $\rho$ and the global SU(2) spin-$j$. 
    For each irrep $[\rho, j]$, we choose the state with the highest $N_v$, $L$, and $j_z$ as the representative state, and write down its wave-function (``rep. w.f.''). 
    The wave-function for other states in the same irrep can be generated by acting $C_{2z}$, $C_{2x}$, and spin-lowering operators on the rep. w.f.. 
    $[\up \leftrightarrow \down]$ indicates a similar term to its former term, except with spin indices swapped. }
\end{table*}

Apart from $N_v$, $\hat{H}_{\rm A}$ (\cref{eq:HA,eq:Vee,eq:Ve}) preserves an extra U(1) charge generated by $\sigma^z \tau^z$, which can be regarded as a continuous spatial rotation promoted from $C_{3z}$. We dub the corresponding quantum number as the angular momentum $L$. 
Besides, $\hat{H}_{\rm A}$ also respects the point group $D_6$ and global spin SU(2) symmetry. 
Since $C_{2z}$ ($\sigma^x \tau^x$) inverses $N_v$ by anti-commuting with $\tau^z$, and $C_{2x}$ ($\sigma^x$) inverses $L$ by anti-commuting with $\sigma^z \tau^z$, symmetry-protected multiplets can be found in blocks labeled by $|N_v|$ and $|L|$. 
Within each block, we label each degenerate multiplet by $[\rho, j]$, where $\rho$ denotes a representation of $D_6$, and $j$ denotes the total spin. 
Note that in sectors with $|N_v|\not=0$ or $|L|\not=0$, $\rho$ can be composed of several different $D_6$ irreps, which differ from each other only by a valley U(1) or angular momentum U(1) rotation. 

In the next, we first show that $N_f$-electrons states and $(8-N_f)$-electron states are in particle-hole conjugation, and then obtain the phonon-induced multiplet splittings among two, three, and four-electrons states. All the one-electron states are trivially degenerate, because they are free from interactions. The eight-fold multiplet spanned by $f^\dagger_{\beta\eta s}|0\rangle$ has $|N_v|=|L|=1$, and can be labeled by $[E_1+E_2, \frac{1}{2}]$. As this section focuses on one $f$-site, the subscript $\RR$ will be kept implicit. 

\subsection{Particle-hole conjugation}

It has been shown that the kinetic and Coulomb energies of MATBG own an emergent many-body charge conjugation symmetry, $\mcl{P}_c$, which is a manifestation of the local symmetry $C_{2z}T P$ in the second quantized form \cite{TBG3_2021}. 
Correspondingly, the many-body physics at $\pm \nu$ fillings are expected to be in particle-hole conjugation. 
It is thus desired to investigate whether $\hat{H}_{\rm A}$ respects $\mcl{P}_c$ or not. 
On the $f$-basis, $\mcl{P}_c$ is represented as 
\begin{align} \label{eq:Pc}
    \mcl{P}_c f^\dagger_{\beta\eta s} \mcl{P}_c^{-1} &= (- i \zeta_\beta \eta) f_{\ovl{\beta}\eta s}  \\\nonumber
    \mcl{P}_c f_{\beta\eta s} \mcl{P}_c^{-1} &= (i\zeta_\beta\eta) f^\dagger_{\ovl{\beta}\eta s}
\end{align}
so that $\mcl{P}_c \hat{n}_{\beta\eta s} \mcl{P}_c^{-1} = 1 - \hat{n}_{\ovl{\beta}\eta s}$, where $\hat{n}_{\beta\eta s} = f^\dagger_{\beta\eta s} f_{\beta\eta s}$. 
It can be thus verified that, $\hat{H}_{\rm A}$ transforms under $\mcl{P}_c$ as, 
\begin{align} \label{eq:HA_Pc}
    \mcl{P}_c \hat{H}_{\rm A} \mcl{P}_c^{-1} & = \hat{H}_{\rm A} - \left( J_{\rm a} + J_{\rm b} + J_{\rm d} + J_{\rm e} \right) \left(\hat{N}_f - 4 \right) 
\end{align}
The constant energy can always be absorbed by re-defining the energy zero-point, and the $\hat{N}_f$ term amounts to adding an additional chemical potential to $f$-electrons only. 
On an $f$-site with fixed $f$-electron number, $\hat{N}_f$ acts as an identity, hence guaranteeing the multiplet energy spacings are in exact particle-hole conjugation for $N_f$ and $(8-N_f)$-electron states. 
Nevertheless, we remark that on the lattice, such a term can lead to a weak particle-hole asymmetry between $\pm\nu$ fillings.

\subsection{Two-electron states}  \label{sec:multi-2e}

Two-electron states of $\hat{H}_{\rm A}$ are solved in \cref{tab:2e}. It is found that each $[\rho, j]$ happens to appear only once for all blocks of $|N_v|$ and $|L|$, hence we will directly use $[\rho, j]$ to label all the two-electron multiplets. 

To see the anti-Hund's nature of $\hat{H}_{\rm A}$, we also interpret the two-electron states as two-electron scattering channels, and in the $U\to0$ limit, those channels with negative energies lead to mean-field SC orders of bare $f$-electrons on the lattice. 
We observe that, $J_{\rm a}$ favors electron pairs occupying the same $\beta$-index ($f^\dagger_{\beta\eta s} f^\dagger_{\beta\eta' s'}$) by an energy $2J_{\rm a}$, which share the same real-space density profile as discussed in \cref{sec:ele-THF}, against electron pairs with opposite $\beta$-index ($f^\dagger_{\beta\eta s} f^\dagger_{\ovl{\beta} \eta' s'}$). 
Moreover, $J_{\rm b,d,e}$ induce pair scatterings - 
$J_{\rm b}$ acts on electron pairs with $|L|=0$, and favors $C_{2x}$-even wave-functions ($A_1,B_1$) against odd ones ($A_2, B_2$); 
$J_{\rm d}$ acts on inter-valley pairs with $|N_v|=0$, and favors $C_{2z}$-even wave-functions ($A_1, A_2, E_2$) against odd ones ($B_1, B_2, E_1$); 
$J_{\rm e}$ acts on electron pairs with both $|N_v|=|L|=0$, and favors wave-functions with the same $C_{2x}$ and $C_{2z}$ parity ($A_1, B_2$) against opposite ones ($B_1, A_2$). 
Summed up, scattering channels that are symmetric in the spatial part (encoded in the $\beta$ and $\eta$ indices) in general have lower energies under the phonon-mediated interactions. By the fermion anti-symmetry, they form spin-singlet pairs. 

As the most spatially symmetric state, $[A_1, 0]$ wins as the unique phonon-favored ground state. 
As shown in \cref{fig:split-0}(a), if the RG factor $\kappa_K=3.2$ is taken into consideration as in Ref. \cite{Basko_2008}, the $[E_2, 0]$ doublet comes as the second lowest states. 
$[A_1, 0]$ and $[E_2, 0]$ correspond to the mean-field $s$-wave and $d$-wave inter-valley spin-singlet SC orders when put on the lattice, respectively. 
These results are consistent with previous researches on the phonon-mediated SC orders in the absence of $U$ \cite{Wu_2018_SCop, Liu_2023_electronkphonon, Angeli_2019_valleyJT, Andrea_2022_local}. 

We also emphasize the above results have only included the effect of EPC. 
In \cref{sec:HH-multi}, with the carbon atom Hubbard repulsion taken into consideration, we will show that the $[E_2, 0]$ doublet can become the two-electron ground states in MATBG for a realistic parameter range.

\subsection{Three-electron and Four-electron states}  \label{sec:multi-3e4e}

The three and four-electron spectra with fixed $\kappa_\Gamma=1$ and increasing $\kappa_K$ are shown in \cref{fig:split-0}(b,c), respectively, and the symmetry labels and the analytic form of the wave-function and energies are summarized in \cref{app:multi}. 

Since any three-electron state at least carries a non-vanishing valley charge $|N_v|$, a non-vanishing angular-momentum $|L|$, and a half-integer spin, by $C_{2z}$, $C_{2x}$, and spin SU(2) rotations, each multiplet level must be at least 8-fold degenerate. 
The ground states are found in the $|N_v|=1$, $|L|=1$ sector, forming an $[E_1 + E_2, \frac{1}{2}]$ irrep. There are three such irreps in total, which are allowed by symmetry to mix one another, hence the detailed form of the ground state wave-function relies on parameters. 
Other irreps appear only for once. 

For the $N_f=4$ subspace, crucially, the ground state is a non-degenerate singlet found in the $|N_v| = |L| = 0$ block, forming an $[A_1, 0]$ irrep. Similar to the three-electron ground states, there are three $[A_1, 0]$ irreps that are allowed to mix with one another, hence the specific wave-function of the four-electron ground state relies on specific parameters as well. We will discuss in \cref{sec:discuss} about the possibility that this non-degenerate ground state leads to a Mott-SM at CNP. 

For both $N_f=3,4$, we find the highest-spin states ($j=\frac{3}{2}$ and $2$, respectively) to have the highest energy, which confirms the anti-Hund's nature of $\hat{H}_{\rm A}$.

\section{Phonon-favored symmetry-breaking orders at even integer-fillings} \label{sec:mf}

\begin{table}[tb]
    \centering
    \begin{tabular}{c|c|c|c|c|c}
    \hline 
        $\xi$ & Phonon-favored $\mcl{P}_\xi$ & Order & valley U(1) & $C_{2z} T$ & $C_{3z}$ \\ 
    \hline
        $\rm a$ & $\{ \sigma^z \}$ & $\rm OP$ & \checkmark & $\times$ & \checkmark \\
        $\rm b$ & $\{ \sigma^x , \sigma^y \tau^z \}$ & $\rm NIOC$ & \checkmark & \checkmark & $\times$ \\
        $\rm d$ & $\{ \sigma^x \tau^x , \sigma^x \tau^y \}$ & $\rm TIVC$ & $\times$ & \checkmark & \checkmark \\
        $\rm e$ & $\{ \tau^x, \tau^y, \sigma^z\tau^x, \sigma^z\tau^y \}$ & $\rm NIVC$ & $\times$ & \checkmark & $\times$ \\
    \hline
    \end{tabular}
    \caption{\label{tab:Pxi} The phonon-favored electron orders. Condensation of an arbitrary order in $\mcl{P}_\xi$ respects $T$, but can break different crystalline symmetries. }
\end{table}

\begin{figure*}[bt]
    \centering
    \includegraphics[width=0.9\linewidth]{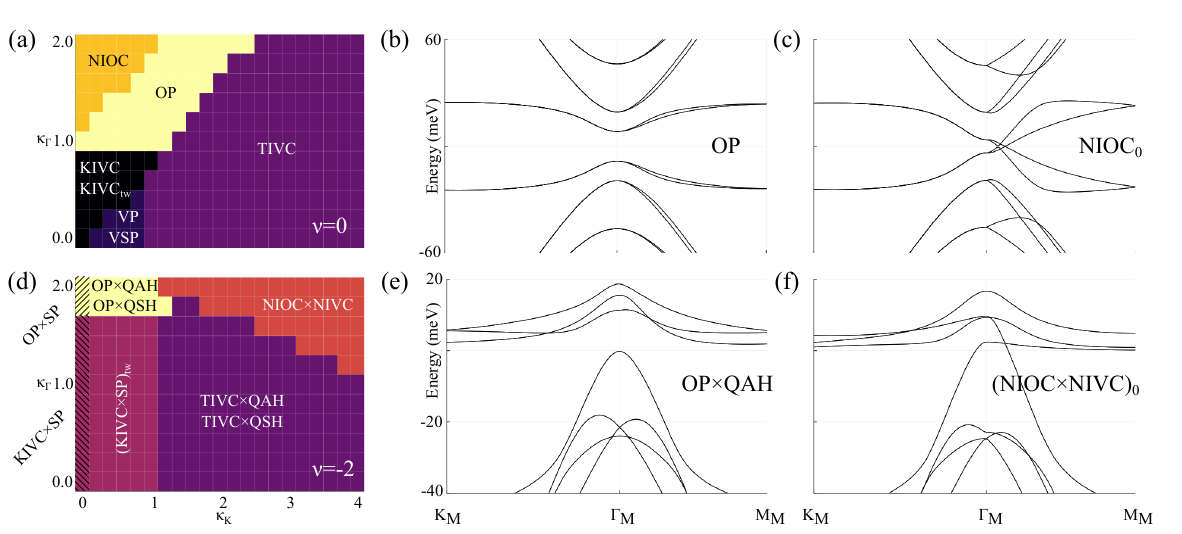}
    \caption{\label{fig:epc-phd} Self-consistent Hartree-Fock mean-field phase diagram in the presence of $\hat{H}_{\rm A}$ at even integer fillings, (a) $\nu=0$ and (d) $\nu=-2$.
    $J_{\rm a}=0.48\mrm{meV}$, $J_{\rm b}=1.60\mrm{meV}$, $J_{\rm d}=1.30\mrm{meV}$, and $J_{\rm e}=1.19\mrm{meV}$ are adopted, and $\kappa_{\Gamma,K}$ are scanned. 
    (b,c) Single-particle excitation spectra of the $\rm OP$ and $\rm NIOC_0$ orders at $\nu=0$. (e,f) Single-particle excitation spectra of the $\rm OP \times QAH$ and $\rm (NIOC\times NIVC)_0$ orders at $\nu=-2$. }
\end{figure*}

In this section, we investigate the phonon-favored symmetry-breaking orders at even-integer fillings. Due to the large Hubbard $U$ and the formation of local SU(8) $f$-moments (namely with $\beta,\eta,s$ indices), symmetry-breaking can be understood as the spatial aligning of the  $f$-moments through the RKKY mechanism \cite{Song_2022, Rai_2023_DMFT, Hu_2023_Kondo}. 
At the mean-field level, it thus manifests as an almost full polarization of the eight flavors at each $f$-site, while the symmetry-breaking order in the $c$-electrons and $cf$-hybridization only arises subsidiarily, and remains weak (for quantitative discussions see \cref{app:HFcalc}). 
Therefore, it suffices to consider the following on-site order parameter, 
\begin{align} \label{eq:O}
    \langle f^\dagger_{\RR\beta\eta s} f_{\RR\beta' \eta's'} \rangle = O_{\beta\eta s,\beta'\eta's'} \ , 
\end{align}
where the $\RR$ dependence is absent by assuming translation symmetry. 
Since $\mrm{Tr}O = \nu_f + 4$ simply counts the total $f$-electron filling, it is more instructive to separate out the traceless part $Q$, defined through 
\begin{align}   \label{eq:Q}
    O = \frac{\nu_f+4}{8}\left(1 + Q \right)
\end{align} 
Two orders with $\pm Q$ are related by the spontaneously broken symmetry, hence have the same energy. We will not discriminate the sign of $Q$. 

$\hat{H}_{\rm A}$, by splitting the otherwise degenerate many-body levels, dictates that certain flavor polarization is energetically more favored.
To see this, we carry out Hartree-Fock decomposition to $\hat{H}_{\rm A}$, using \cref{eq:O}. 
Leaving the derivation details to Appendix \ref{app:HF}, we find that for $\xi ={\rm a,b,d,e}$, the Hartree-Fock energy of the $J_\xi$ term takes the following generic form
\begin{align}   \label{eq:Hxi}
    E_{\xi} &= \frac{J_{\xi}}{2} \left(\frac{\nu_f + 4}{8}\right)^2 \frac{1}{|\mcl{P}_{\xi}|} \\\nonumber
    & \qquad \times \sum_{P_\xi \in \mcl{P}_{\xi}} \left[ - \left(\mrm{Tr} [Q \cdot P_\xi]\right)^2 + \mrm{Tr} \left[ (Q \cdot P_\xi)^2 \right] + 8 \right]
\end{align}
Here, $\mcl{P}_{\xi}$ is a set of $8\times 8$ Pauli matrices, listed in Table \ref{tab:Pxi}, and $|\mcl{P}_{\xi}|$ denotes the set cardinality. 
The elements $P_\xi \in \mcl{P}_\xi$ are all traceless, and $P^2_\xi = 1$. The $-(\mrm{Tr}[Q\cdot P_\xi])^2$ term arises through the Hartree channel, which dominates in the large-$N$ limit (here $N=8$ denotes the total flavor number), indicating that any order $Q$ that is proportional to $P_{\xi}$ can lower the Hartree energy; while the $\mrm{Tr}[(Q\cdot P_\xi)^2]+8$ terms arise through the Fock channel, and has a subleading contribution. 
We thus refer to $\mcl{P}_{\xi}$ as the set of phonon-favored symmetry-breaking orders. 

In general, $\Gamma$-phonons favor orbital orders that respect the valley U(1) symmetry. $\mcl{P}_{\rm a}$ corresponds to $\rm OP$ that preserves $T$ but breaks $C_{2z}T$. Since the $\beta=1,2$ orbitals overlap mostly with the $\alpha=A,B$ sublattices, respectively (see discussions below \cref{eq:w_Gauss_A,eq:w_Gauss_B}), $\rm OP$ can also be understood as the sublattice polarization on the microscopic level. 
$\mcl{P}_{\rm b}$ corresponds to $\rm NIOC$ that respects $C_{2z}T$ but violates $C_{3z}$. 
Concretely, $\sigma^x$ anti-commutes with $\sigma^y \tau^z$, and $\sigma^x$ and $\sigma^y \tau^z$ are related by $C_{3z} = e^{i\frac{2\pi}{3} \sigma^z\tau^z}$ (\cref{eq:DfC3z}). Therefore, any order with $Q = \sigma^x \cos\varphi + \sigma^y \tau^z \sin\varphi$ ($\varphi \in \mbb{R}$) can benefit equally from $E_{\rm b}$, and $\varphi$ can be interpreted as a label for the orientation angle of $C_{3z}$-breaking. 

On the other hand, $K$-phonons favor inter-valley coherence. 
$\mcl{P}_{\rm d}$ corresponds to $\rm TIVC$ \cite{TBG4_2021, Song_2022, Kwan_2023_Kphonon}, while $\mcl{P}_{\rm e}$ features a distinct $\rm NIVC$ that violates $C_{3z}$. 
Among $\mcl{P}_{\rm e}$, $\tau^x$ and $\sigma^z \tau^y$ can be continuously deformed to $\tau^y$ and $\sigma^z \tau^x$ by valley U(1) rotation $e^{i\vartheta \tau^z}$, respectively, and $\tau^x$ and $\sigma^z \tau^y$ are linked through $C_{3z}$ (\cref{eq:DfC3z}). 
A general order benefiting from $\hat{H}_{\rm e}$ thus takes the form of $Q = e^{i\vartheta \tau^z} (\tau^x \cos\varphi + \sigma^z \tau^y \sin\varphi) e^{-i\vartheta \tau^z}$, where $\varphi$ again labels the orientation of $C_{3z}$-breaking (\cref{sec:mf-0}). 

Besides the competition among various phonon-favored orders, the actual ground state also significantly relies on the kinetic and Coulomb energies. 
Therefore, we carry out self-consistent Hartree-Fock calculations, with varying $\kappa_\Gamma$ and $\kappa_K$, and map the phase diagram at $\nu=0$ and $\nu=-2$. 
The kinetic $\hat{H}_0$ and Coulomb $\hat{H}_{\rm C}$ energies are given by \cref{eq:Hf0,eq:HfC}, respectively. 
We also discuss the nature of each symmetry-breaking order appearing in the phase diagram, which can be either insulating or (semi-)metallic with $C_{2z}T$ protected crossings. 

We comment that, phonons can also induce interactions between $c$ and $f$ sectors or among $c$-electrons, and their strengths can be estimated as of the same order and comparably weaker than the on-site $J_{\rm a,b,d,e}$, because the Bloch state spread of $c$-electrons is similar to and comparably wider than the $f$ Wannier orbitals \cite{calugaru_twisted_2023}. 
However, due to the much weaker symmetry-breaking strength in these sectors, the energy correction through these channels will be much weaker than through the on-site splitting $\hat{H}_{\rm A}$ that we consider above. 
Moreover, as these neglected channels obey the same symmetry as $\hat{H}_{\rm A}$, they do not induce extra degeneracy splittings as well. 
Therefore, further incorporating them should not qualitatively affect our results, although quantitative corrections such as the shift of phase boundaries in the parameter space can occur.

\subsection{Self-consistent phase-diagram at \texorpdfstring{$\nu=0$}{v=0}} \label{sec:mf-0}

\begin{table*}[tb]
    \centering
    \begin{tabular}{l|c|c||c||c|c|c|c}
    \hline\hline
        Order & $Q$ & $E_0$ (meV) & Gap & $E_{\rm a}$ & $E_{\rm b}$ & $E_{\rm d}$ & $E_{\rm e}$ \\
    \hline
        $\rm KIVC / KIVC_{tw}$ & $\sigma^y \tau^{x,y} / \sigma^y \tau^{x,y} s^{\mu}$ & 0 & Insu & 0 & 0 & \bbb{$J_{\rm d}$} & \bbb{$J_{\rm e}$} \\
        $\rm VP / VSP$ & $\tau^z / \tau^zs^z$ & 0.087 & Insu & \bbb{$2J_{\rm a}$} & \bbb{$2J_{\rm b}$} & 0 & 0 \\
        $\rm SP$ & $s^z$ & 0.087 & Insu & \bbb{$2J_{\rm a}$} & \bbb{$2J_{\rm b}$} & \bbb{$2J_{\rm d}$} & \bbb{$2J_{\rm e}$}\\ 
        $\rm OSP$ & $\sigma^z s^\mu$ & 2.196 & Insu & \bbb{$2J_{\rm a}$} & 0 & 0 & \bbb{$2J_{\rm e}$} \\
    \hline
        $\rm TIVC$ & $\sigma^x \tau^{x,y}$ & 2.266 & Insu & 0 & \bbb{$2J_{\rm b}$} & \rrr{$-3 J_{\rm d}$} & \bbb{$J_{\rm e}$} \\
        $\rm NIVC_{0}$ & $\sigma^0 \tau^{x,y}$ & 4.216 & SM & \bbb{$2J_{\rm a}$} & \bbb{$J_{\rm b}$} & \bbb{$J_{\rm d}$} & \rrr{$-J_{\rm e}$} \\
        $\rm NIVC_{\frac{\pi}{2}}$ & $\sigma^z \tau^{x,y}$ & 4.091 & SM & \bbb{$2J_{\rm a}$} & \bbb{$J_{\rm b}$} & \bbb{$J_{\rm d}$} & \rrr{$-J_{\rm e}$} \\
        $\rm OP$ & $\sigma^z$ & 2.196 & Insu & \rrr{$-6J_{\rm a}$} & 0 & 0 & \bbb{$2J_{\rm e}$} \\
        $\rm NIOC_{0}$ & $\sigma^x$ & 4.130 & SM & 0 & \rrr{$-3J_{\rm b}$} & \bbb{$2J_{\rm d}$} & \bbb{$J_{\rm e}$} \\
        $\rm NIOC_{\frac{\pi}{2}}$ & $\sigma^y \tau^z$ & 4.233 & SM & 0 & \rrr{$-3J_{\rm b}$} & \bbb{$2J_{\rm d}$} & \bbb{$J_{\rm e}$} \\
    \hline\hline
    \end{tabular}
    \caption{\label{tab:0_energy} Perturbative energy correction to various Hartree-Fock orders at $\nu = 0$ due to the phonon-mediated interaction. 
    $E_0$ denotes the self-consistent energy per MUC, measured from the KIVC energy at $\kappa_\Gamma = \kappa_K = 0$. We note that, $\rm OP$ and $\rm TIVC$, $\rm NIOC$ and $\rm NIVC$ are also related by flat-U(4) rotations, respectively, hence have close $E_0$. 
    The subscript of $\rm NIOC$ and $\rm NIVC$ denotes the $C_{3z}$-breaking angle $\varphi=0$ or $\frac{\pi}{2}$. 
    The red or blue color emphasizes the corresponding order is favored by anti-Hund's or Hund's interactions. }
\end{table*}

A fully polarization of $O$ at CNP can be sketched by a projector to four out of the eight flavors - $O = \frac{1+Q}{2}$, and $Q^2=1$. 
We first discuss the self-consistent kinetic and Coulomb energy of various orders at $\kappa_\Gamma = \kappa_K = 0$, which we dub as $E_0$. 
In accordance with existing literatures, \textit{e.g.}, Refs. \cite{Bultinck_2020_2U4, TBG4_2021, Song_2022}, $\rm KIVC$ ($Q = \sigma^y \tau^{x,y}$) has the lowest energy among all symmetry-breaking orders, and we set its energy as the reference point, $E_0 = 0$. 
Due to the independent $\rm SU(2)\times SU(2)$ spin rotations in two valleys, acting $e^{i\frac{\pi}{2}\frac{\tau^0-\tau^z}{2} s^\mu}$ on a $\rm KIVC$ order produces an exact degenerate order, which we dub as $\rm KIVC_{tw}$ with $Q = \sigma^y \tau^{x,y} s^\mu$. 
Due to the approximate flat-U(4) symmetry, $\rm VP$ ($Q = \tau^z$) and $\rm KIVC$ are related by $e^{i\frac{\pi}{4} \sigma^y \tau^{x,y}}$, hence $\rm VP$ comes in close competition with a slightly higher energy $E_0 \sim 0.1 \mrm{meV}$ per MUC. 
At the Hartree-Fock level, $\rm SP$ ($Q = s^z$) and the valley-spin-polarized state ($\rm VSP$, $Q = \tau^z s^z$) are degenerate with $\rm VP$, albeit not related by continuous symmetries. 
In comparison, other orders, including the phonon-favored orders listed in Table \ref{tab:Pxi}, are disfavored by kinetic and Coulomb energies and have higher $E_0$. 
Generically speaking, semi-metallic orders with Dirac points in the charge excitation spectrum have higher energy penalty than insulating orders with a full charge gap. 

Turning on $\kappa_\Gamma$ and $\kappa_K$ shuffles the energy of each order. 
The self-consistent mean-field phase diagram is shown in \cref{fig:epc-phd}(a), where the energy competition between various orders can be qualitatively understood from $E_0 + \kappa_\Gamma (E_{\rm a} + E_{\rm b})+ \kappa_K (E_{\rm d} + E_{\rm e})$, as tabulated in \cref{tab:0_energy}. Here, $E_\xi$ describes the perturbative energy correction from the $J_{\xi}$ term ($\xi = {\rm a,b,d,e}$) in $\hat{H}_{\rm A}$, and is calculated through \cref{eq:Hxi} using the projector ansatz $O = \frac{1+Q}{2}$. 
Specifically, $\rm KIVC/KIVC_{tw}$ and $\rm VP/VSP$ are penalized by $K$-phonons and $\Gamma$-phonons through the Fock channel in \cref{eq:Hxi}, respectively, while SP order receives maximal energy penalization from both $\Gamma$- and $K$-phonons. 
Therefore, for small $\kappa_\Gamma \lesssim 1.0$ and $\kappa_K \lesssim 1.0$, $\rm KIVC/KIVC_{tw}$ and $\rm VP/VSP$ compete to give rise to a phase boundary with slope $\frac{\Delta\kappa_\Gamma}{\Delta\kappa_K} \approx 0.8$ that extends slightly beneath $\kappa_\Gamma = \kappa_K = 0$, while $\rm SP$ is completely ruled out. 
We also remark that, although the independent $\rm SU(2) \times SU(2)$ symmetry is lost with finite $\kappa_K$, $\rm KIVC$ and $\rm KIVC_{tw}$ still remain degenerate at the Hartree-Fock level. 

Beyond the threshold $\kappa_\Gamma \gtrsim 1.0$ and $\kappa_K\gtrsim 1.0$, phonon-favored orders start to appear. Along the direction of increasing $\kappa_\Gamma$, $\rm OP$ appears first owing to the lower $E_0$, while $\rm NIOC$ appears next by benefiting more energies from $\kappa_\Gamma$. Along the direction of increasing $\kappa_K$, only $\rm TIVC$ is observed, which spans the major region of $0.8 \kappa_K \gtrsim \kappa_\Gamma$. $\rm NIVC$ is not observed throughout the phase diagram, by having a much higher $E_0$ and benefiting much fewer energy from increasing $\kappa_K$. Our results along the axis of $\kappa_\Gamma=0$ and $\kappa_\Gamma=\kappa_K$ are also qualitatively consistent with previous literatures, \textit{e.g.}, Refs. \cite{Kwan_2023_Kphonon, shi_2024_moire}. 

Different symmetry-breaking orders exhibit rather distinct experimental features. Specifically, by breaking $C_{2z}T$, $\rm OP$ gaps out the Dirac points and forms an insulator with full charge gap (\cref{fig:epc-phd}(b)). 
By populating each valley with a Chern band with opposite Chern numbers, $\rm OP$ also exhibits valley Hall response. 
In experimental setups with aligned hBN substrate, $C_{2z}T$ is externally broken, which necessarily manifests as a 
\begin{align}
    \hat{H}^{(f)}_{0, \rm hBN} = m_z \sum_{\RR} \sum_{\beta \eta s} f^\dagger_{\RR \beta \eta s} [\sigma^z]_{\beta,\beta'} f_{\RR \beta' \eta s}
\end{align}
term when projected on the $f$-orbitals. Without loss of generality, we can take $m_z > 0$. Thereby, an $\rm OP$ order with $Q=-\sigma^z$ will acquire a negative energy correction at the first order of $m$, hence become more competitive. 
On the other hand, $\rm NIOC$ preserves $C_{2z}T$ and forms a Dirac semi-metal (\cref{fig:epc-phd}(c)) \cite{Bultinck_2020_2U4, shi_2024_moire}. The orientation angle of the Dirac points can be continuously tuned by the $C_{3z}$-breaking parameter $\varphi$, while the self-consistent energy only varies slightly ($\sim$0.1meV, see the last two rows in \cref{tab:0_energy}). We also comment that, an external strain necessarily induces the following term
\begin{align}
    \hat{H}^{(f)}_{0,\mrm{str}} &= m_x \sum_{\RR} \sum_{\beta\eta s} f^\dagger_{\RR\beta\eta s} \times \\\nonumber
    & \qquad \left[\sigma^x \cos\varphi_0 + \sigma^y \tau^z \sin\varphi_0\right]_{\beta\eta,\beta'\eta'} f_{\RR\beta'\eta's}
\end{align}
that explicitly breaks $C_{3z}$ \cite{herzogarbeitman_2024_heavy}. Assuming $m_{x}>0$ without loss of generality, an $\rm NIOC$ order with $\varphi=\varphi_0+\pi$ can further lower the energy. 
The above arguments on the effect of hBN and strain are also consistent with previous literatures \textit{e.g.} Refs. \cite{Parker_2021_strain, Wagner_2022_global, Kwan_2021_IKS}, which investigated the continuum model in the absence of EPC. 

For $\rm TIVC$, by breaking valley U(1), Dirac points are hybridized and gapped out, leading to an insulator. More importantly, $\rm TIVC$ exhibits Kekul\'e pattern in the charge density distribution \cite{Dumitru_2022_STM}. 
Although our Hartree-Fock mean-field results finds $\rm TIVC$ at $\kappa_\Gamma=1$ and $\kappa_K=3.2$ (\cref{fig:epc-phd}(a)), in experiments Kekul\'e pattern is not reported at CNP \cite{Nuckolls_2023_quantum}. 
In ultra-low strained samples (inter-layer strain $0.03\%$), an insulating feature is observed, while in mildly strained samples ($0.1\sim 0.2\%$), semi-metallic behaviors are observed \cite{Nuckolls_2023_quantum}. 
While $\rm KIVC$ and $\rm NIOC$ may potentially explain the difference \cite{Parker_2021_strain}, in \cref{sec:discuss} we discuss another strongly correlated semi-metal phase that may exist at CNP, which possesses a symmetric Mott gap in the $f$-electrons. The description of this exotic phase goes beyond the Hartree-Fock scope. 

\subsection{Self-consistent phase-diagram at \texorpdfstring{$\nu=-2$}{v=-2}}

\begin{table*}[tb]
    \centering
    \begin{tabular}{l|c|c|c||c||c||c|c|c|c}
    \hline\hline
        Order & $Q_1$ & $Q_2$ & $Q_3$ & $E_0$ (meV) & Gap & $E_{\rm a}$ & $E_{\rm b}$ & $E_{\rm d}$ & $E_{\rm e}$ \\
    \hline 
        $\rm VP \times SP$ & $\tau^z$ & $s^\mu$ & $\tau^z s^\mu$ & 0.056 & Insu & \bbb{$J_{\rm a}$} & \bbb{$J_{\rm b}$} & 0 & 0 \\
        $\rm KIVC \times SP$ & $\sigma^y \tau^{x,y}$ & $s^\mu$ & $\sigma^y \tau^{x,y} s^\mu$ &  0 & Insu & 0 & 0 & \bbb{$0.5 J_{\rm d}$} & \bbb{$0.5 J_{\rm e}$} \\ 
        $\rm (KIVC \times SP)_{tw}$ & $\sigma^y \tau^{x,y} s^z$ & $\tau^z s^x$ & $\sigma^y \tau^{y,z} s^y$ & 0 & Insu & 0 & 0 & 0 & 0  \\
    \hline
        $\rm OP\times SP $ & $\sigma^z$ & $s^\mu$ & $\sigma^z s^\mu$ & 0.482 & Insu & \rrr{$-J_{\rm a}$} & 0 & 0 & \bbb{$J_{\rm e}$} \\
        $\rm TIVC \times SP$ & $\sigma^x \tau^{x,y}$ & $s^\mu$ & $\sigma^x \tau^{x,y} s^\mu$ & 0.521 & Insu & 0 & \bbb{$J_{\rm b}$} & \rrr{$-0.5 J_{\rm d}$} & \bbb{$0.5 J_{\rm e}$} \\
    \hline 
        $\rm (NIOC \times NIVC)_0$ & $\sigma^x$ & $\tau^{x,y}$ & $\sigma^x \tau^{x,y}$ & 1.991 & Metal / Insu & 0 & \rrr{$-0.5 J_{\rm b}$} & \rrr{$-0.5 J_{\rm d}$} & \rrr{$-0.25 J_{\rm e}$} \\
        $\rm (NIOC \times NIVC)_{\frac{\pi}{2}}$ & $\sigma^y \tau^z$ & $\sigma^z\tau^{x,y}$ & $\sigma^x \tau^{y,x}$ & 1.977 & Metal / Insu & 0 & \rrr{$-0.5 J_{\rm b}$} & \rrr{$-0.5 J_{\rm d}$} & \rrr{$-0.25 J_{\rm e}$} \\
    \hline
        $\mrm{OP \times QAH / QSH}$ & $\sigma^z$ & $\sigma^z \tau^z / \sigma^z \tau^z s^\mu$ & $\tau^z / \tau^z s^\mu$ & {0.482} & {Insu} & {\rrr{$-J_{\rm a}$}} & 0 & 0 & 0 \\ 
        $\mrm{TIVC \times QAH / QSH}$ & $\sigma^x \tau^y $ & $\sigma^z \tau^z / \sigma^z \tau^z s^\mu$  & $\sigma^y \tau^x / \sigma^y \tau^x s^\mu$  & {0.485} & {Insu} & 0 & 0 & \rrr{$-0.5 J_{\rm d}$} & 0 \\
    \hline\hline
    \end{tabular}
    \caption{\label{tab:-2_energy} Perturbative energy correction to various Hartree-Fock orders at $\nu = -2$ due to the phonon-mediated interaction. $E_0$ denotes the self-consistent energy per MUC, measured from the $\rm KIVC \times SP$ energy at $\kappa_\Gamma = \kappa_K = 0$. 
    The red or blue color emphasizes the corresponding order is favored by anti-Hund's or Hund's interactions. }
\end{table*}

The order parameter at $\nu=-2$ can be modeled as a product of two \textit{commuting} projections, $O = \frac{1+Q_1}{2} \frac{1+Q_2}{2}$, with $Q_1^2 = Q^2 = 1$, $Q_1 \not= Q_2$ and $[Q_1, Q_2] = 0$, so that the number of occupied flavours are halved twice. 
Such an order can thus be dubbed as a product of two orders that correspond to $\frac{1+Q_1}{2}$ and $\frac{1+Q_2}{2}$. 
Nevertheless, such a notation can be non-unique - by terming $Q_3 = Q_1 Q_2$, $Q_1$ and $Q_3$ (as well as $Q_2$ and $Q_3$) define the same order parameter. We simply keep this redundancy. 

We start with the kinetic and Coulomb energies $E_0$ in the limit of $\kappa_\Gamma = \kappa_K = 0$, where the symmetry-breaking ground state is $\rm KIVC\times SP$, in accordance with literatures \cite{Bultinck_2020_2U4, TBG4_2021, Song_2022}. By the $\rm SU(2)\times SU(2)$ symmetry, twisting a $\rm KIVC\times SP$ order with $Q_1 = \sigma^y \tau^x$ and $Q_2 = s^z$ with $e^{i\frac{\pi}{2}\frac{\tau^0-\tau^z}{2} s^y}$ produces an exactly degenerate new order with $Q_1 = \sigma^y \tau^y s^y$ and $Q_2 = \tau^z s^x$, which we dub as $\rm (KIVC\times SP)_{tw}$. 
$\rm VP \times SP$ comes in close competition with $E_0 \sim 0.05\mrm{meV}$, due to the emergent flat-U(4) symmetry. 
Other states typically suffer a penalty of $E_0 \gtrsim 0.5\mrm{meV}$.  

With $\kappa_\Gamma$ and $\kappa_K$ turned on, the self-consistent mean-field phase diagram is shown in \cref{fig:epc-phd}(d). Crucially, at small $\kappa_\Gamma$ and $\kappa_K$, while $\rm KIVC\times SP$ and $\rm VP\times SP$ are penalized by $K$-phonons and $\Gamma$-phonons, respectively, $\rm (KIVC\times SP)_{tw}$ asymptotically escapes both phonons, hence survives as the ground state. 
Beyond the threshold $\kappa_\Gamma \gtrsim 1.6$, $\rm (KIVC\times SP)_{tw}$ gives way to two degenerate orders with $Q_1 = \sigma^z$ and $Q_2 = \sigma^z \tau^z$ or $Q_2 = \sigma^z \tau^z s^\mu$. We dub them as $\rm OP \times QAH/QSH$, respectively, as a $\sigma^z \tau^z$ ($\sigma^z \tau^z s^\mu$) order exhibits quantum anomalous Hall (quantum spin Hall) response. They benefit from $\Gamma$-phonons by possessing the $\rm OP$ component. The corresponding charge excitation spectrum is insulating, as shown in \cref{fig:epc-phd}(e). 

Likewise, beyond the threshold of $\kappa_K \gtrsim 1.0$, $\rm (KIVC\times SP)_{tw}$ gives way to $\rm TIVC \times QAH/QSH$, which benefits from $K$-phonons with the $\rm TIVC$ component. This is also discussed by previous literatures, and the results agree qualitatively well \cite{Kwan_2023_Kphonon, shi_2024_moire}. As a comparison, although $\rm TIVC \times SP$ also benefits from the $J_{\rm d}$ term induced by $K$-phonons, the $\rm SP$ component is disfavored by other terms such as $J_{\rm b,e}$, hence does not appear in the phase diagram. Both orders own a full charge gap and exhibits the charge Kekul\'e pattern \cite{Dumitru_2022_STM}. 

Interestingly, in the regime where both $\kappa_\Gamma$ and $\kappa_K$ are large ($\kappa_\Gamma+\frac{1}{3}\kappa_K \gtrsim \frac{8}{3}$), a new phase shows up, which we dub as $\rm NIOC \times NIVC$. This order has $Q_1 = \sigma^x \cos\varphi + \sigma^y \tau^z \sin \varphi$, $Q_2 = e^{i\vartheta \tau^z}( \tau^x \cos\varphi + \sigma^z \tau^y \sin\varphi ) e^{-i\vartheta \tau^z}$, and $Q_3 = e^{i\vartheta \tau^z}( \sigma^x \tau^x ) e^{-i\vartheta \tau^z}$, hence simultaneously benefits from $\Gamma$-phonons and $K$-phonons. $\varphi$ tunes the $C_{3z}$-breaking angle, and is locked in $Q_1$ and $Q_2$ (mod $\pi$) to guarantee that $[Q_1,Q_2]=0$. 

The single-particle spectrum for $\rm NIOC \times NIVC$ at $\varphi=0$ is illustrated in \cref{fig:epc-phd}(f), which can exhibit $C_{2z}T$-protected band crossings near the Fermi level. This also explains the large kinetic and Coulomb penalty $E_0 \sim 1.9 \mrm{meV}$, which renders $\rm NIOC \times NIVC $ disfavored at intermediate $\kappa_{\Gamma}$ and $\kappa_K$. Tuning the $C_{3z}$-breaking angle $\varphi$ again rotates the orientation of band crossings, while the self-consistent energy is still scarcely affected ($\lesssim$0.05meV, see \cref{tab:-2_energy}). 
However, since the valley U(1) symmetry is broken, with parameters tuned, \textit{e.g.}, at larger $\kappa_\Gamma$ and $\kappa_K$, the two Dirac vortices can also annihilate and produce an insulating spectrum. 

Taking $\kappa_\Gamma=1$ and $\kappa_K=3.2$, we find the symmetry-breaking order as $\rm TIVC \times QAH/QSH$, which can explain the Kekul\'e pattern observed in ultra-low strained samples ($0.03\%$) at this filling \cite{Nuckolls_2023_quantum}, in consistence with previous theoretical proposals \cite{Kwan_2023_Kphonon}. The IKS order observed in mildly strained samples ($0.1\sim0.2\%$) \cite{Nuckolls_2023_quantum} breaks translation symmetry, hence are not captured by the current work. Besides, this point is also close to the phase boundary to $\rm NIOC \times NIVC$, which can also lower its energy through the $\rm NIOC$ component in presence of an external strain. Therefore, $\rm NIOC \times NIVC$ provides a potential explanation to the nematic correlated state with an insulating tunneling gap observed in large-strained samples ($0.6\sim0.7\%$) at this filling \cite{Kerelsky_2019_maximized}. 

\section{The carbon atom Hubbard induced Hund's interaction}  \label{sec:HH}

\begin{figure*}[bt]
    \centering
    \includegraphics[width=0.9\linewidth]{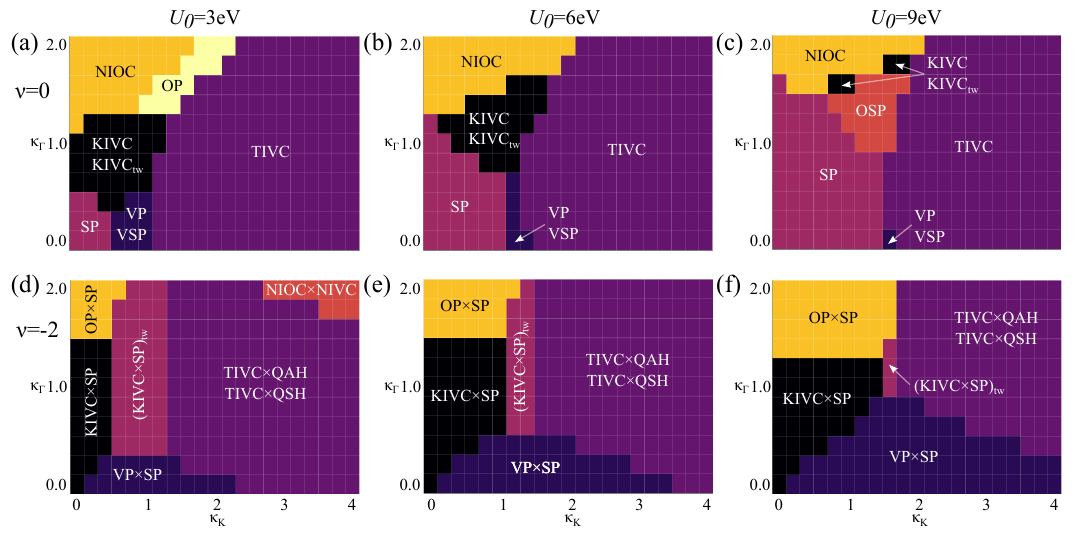}
    \caption{\label{fig:hund-phd} Self-consistent Hartree-Fock mean-field phase diagram in the presence of both $\hat{H}_{\rm A}$ and $\hat{H}_{\rm H}$ at even integer fillings, (a-c) $\nu=0$ and (d-f) $\nu=-2$. $U_0 = 3,6,9\mrm{eV}$ is taken in the left, middle, and right panels, respectively. 
    $J_{\rm a}=0.48\mrm{meV}$, $J_{\rm b}=1.60\mrm{meV}$, $J_{\rm d}=1.30\mrm{meV}$, and $J_{\rm e}=1.19\mrm{meV}$ are adopted in all calculations, and $\kappa_{\Gamma,K}$ are scanned. }
\end{figure*}

Apart from EPC, another non-negligible source of microscopic interaction results from the Hubbard repulsion $U_0$ on the microscopic carbon atoms \cite{Wu_2018_SCop, gonzalez-arraga_electrically_2017,zhang_spin-polarized_2022}, which is \textit{not} captured by the long-range Coulomb interaction $\hat{H}_{\rm C}$ \cref{eq:HC}. 
In terms of the continuum electron fields, $U_0$ can be modeled as a contact interaction within the same sublattice and the same layer 
\begin{align}   \label{eq:HU0}
    \hat{H}_{\rm H} =& \frac{U_0 \Omega_g}{2} \int \mrm{d}^2\rr \sum_{l,\alpha} \sum_{\eta_1\eta_1'\eta_2\eta_2'} \sum_{ss'} \delta_{\eta_1+\eta_1', \eta_2+\eta_2'} \times \\\nonumber
    & ~ \psi^\dagger_{l,\alpha,\eta_1,s}(\rr) \psi^\dagger_{l,\alpha,\eta_1',s'}(\rr) \psi_{l,\alpha,\eta_2's'}(\rr) \psi_{l,\alpha,\eta_2,s}(\rr)
\end{align}
where $U_0 = 3\sim 10\mrm{eV}$ \cite{wehling_strength_2011,schuler_optimal_2013}. The large graphene momenta encoded by valley indices are conserved, and inter-valley scatterings ($\eta_1 = \eta_2'=\ovl{\eta}_1' = \ovl{\eta}_2$) occur with the same amplitude as intra-valley ($\eta_1 = \eta_2=\ovl{\eta}_1' = \ovl{\eta}_2'$) scatterings, due to the contact nature of the microscopic Hubbard interaction. 

Owing to the expanded length scale, the interaction strength felt by a moir\'e electron will be largely suppressed by a factor of $\frac{\Omega_g}{\Omega_M} \sim \theta^2$, hence much weaker than $U$ in size. Nevertheless, when projected to an $f$-orbital, $\hat{H}_{\rm H}$ can manifest as another source of $\rm U(8)$-splitting other than the phonon-mediated $\hat{H}_{\rm A}$. The results have been calculated in Ref.~\cite{Wang_2024_TBGSC}, and we simply incorporate it here. 
By obeying the same symmetry constraints as $\hat{H}_{\rm A}$, projected $\hat{H}_{\rm H}$ necessarily adopts the same form as \cref{eq:HA,eq:Vee,eq:Ve}, and we dub the corresponding parameters as $K_{\rm a,b,c,d,e}$. $K_{\rm c}$ denotes a Hubbard interaction as $J_{\rm c}$, while $K_{\rm a,b,d,e}$ denote the multiple splitting parameters. 
We find that $K_{\rm c} = -0.216\times 10^{-3} U_0$, which is a weak correction to be absorbed to $U$. We thus set $K_{\rm c} = 0$, and focus on the traceless part. Using the same numeric Wannier functions, it is obtained 
\begin{align} 
    K_{\rm a} =& -0.113 \times 10^{-3} U_0 \ , \\
    K_{\rm b} = K_{\rm d} =& - 0.103 \times 10^{-3} U_0 \ , \\
    K_{\rm e} =& - 0.328 \times 10^{-3} U_0 \ , 
\end{align}
$K_{\rm a,b,d,e} < 0$ implies that $\hat{H}_{\rm H}$ is of the Hund's nature, and tends to cancel the effect of $\hat{H}_{\rm A}$. Nevertheless, due to the different microscopic origin, the relative ratio between $K_{\rm a,b,d,e}$ is very different from that between $J_{\rm a,b,d,e}$, which implies that the cancellation is not exact, but rather enriches the physics. 

\subsection{On the multiplet splitting}  \label{sec:HH-multi}

\begin{figure}
    \centering
    \includegraphics[width=\linewidth]{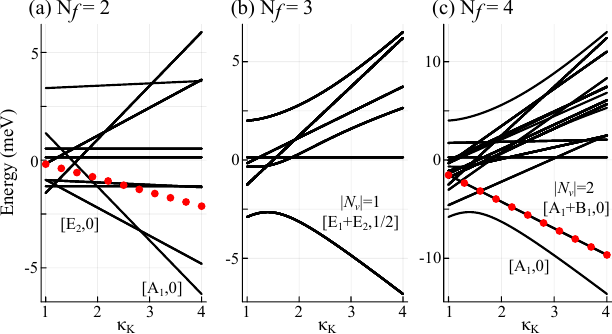}
    \caption{\label{fig:split-9} (a-c) Multiplet splittings induced by $\hat{H}_{\rm A} + \hat{H}_{\rm H}$ in $N_f=2,3,4$ subspaces, with $\kappa_\Gamma=1$ and $U_0=9\mrm{eV}$. 
    (a-c) Phonon-induced multiplet splittings in $N_f=2,3,4$ subspaces, with $\kappa_\Gamma=1$ fixed. 
    Each black line indicates a multiplet level, with symmetry labels tabulated in \cref{tab:2e,tab:3e,tab:4e}, respectively. 
    The symmetry irreps of several lowest-energy states are explicitly marked. 
    In (a) and (c), the energy expectation value of a $\rm TIVC \times QAH/QSH$ order and a $\rm TIVC$ order are also plotted as red dots for comparison, respectively.
    These Hartree-Fock orders are superpositions of energy eigenstates. }
\end{figure}

As an immediate implication, $\hat{H}_{\rm H}$ also participates in the multiplet splitting. 
In particular, among all the two-electron states tabulated in \cref{tab:2e}, while the energy of $[A_1, 0]$ is lifted by an amount of $0.544\times 10^{-3}U_0$ by $\hat{H}_{\rm H}$, the energy of the $[E_2, 0]$ channel is only slightly lifted by $0.010 \times 10^{-3} U_0$. 
Therefore, with the realistic range of $U_0$ and $\kappa_K$ taken into consideration, there can be a region where the two-electron ground states are $[E_2, 0]$ instead of $[A_1, 0]$, which can be seen from \cref{fig:split-9}(a). On the other hand, the three- and four-electron ground states are scarcely affected by $\hat{H}_{\rm H}$, as seen from \cref{fig:split-9}(b,c). They remain in the same symmetry sectors. In particular, the four-electron ground state remains a non-degenerate singlet. 

\subsection{On the symmetry-breaking orders}

$\hat{H}_{\rm H}$ also greatly enriches the self-consistent Hartree-Fock mean-field phase diagram at even-integer fillings. We take $U_0 = 3,6,9$eV, and plot the results in \cref{fig:hund-phd}. 

For $\nu=0$, due to the Hund's nature, $\hat{H}_{\rm H}$ brings the $\rm SP$ order to the ground state at $\kappa_\Gamma = \kappa_K = 0$, and the area occupied by $\rm SP$ gradually expands with increasing $U_0$. 
For intermediate $\kappa_\Gamma$ and $\kappa_K$, since $|\frac{K_{\rm a,e}}{J_{\rm a,e}}| > |\frac{K_{\rm b,d}}{J_{\rm b,d}}|$, with growing $U_0$, the phases stabilized by $J_{\rm a}$ and $J_{\rm e}$ will become disfavored first. As a consequence, the $\rm OP$ order stabilized by $J_{\rm a}$ prominently shrinks in the phase diagram at $3$eV, and becomes completely absent at $U_0=6$eV. 
On the contrary, a new orbital-spin polarized order ($\rm OSP$) with $Q = \sigma^z s^\mu$ appears in the phase diagram of $U_0 = 9$eV. In this region, $\kappa_\Gamma J_{\rm a} + K_{\rm a}$ and $\kappa_K J_{\rm e} + K_{\rm e}$ are negative, while $\kappa_\Gamma J_{\rm b} + K_{\rm b}$ and $\kappa_K J_{\rm d} + K_{\rm d}$ stay positive. By the perturbative calculation tabulated in \cref{tab:0_energy}, $\rm OSP$ thus saves more energy through $\hat{H}_{\rm A} + \hat{H}_{\rm H}$ than both $\rm OP$ and $\rm SP$. 
We also note that, for the parameter $\kappa_\Gamma = 1$ and $\kappa_K = 3.2$, the ground state remains $\rm TIVC$ throughout. 

For $\nu=-2$, similarly, $\hat{H}_{\rm H}$ brings the spin-polarized $\rm KIVC\times SP$ and $\rm VP\times SP$ to the ground state at small $\kappa_\Gamma$ and $\kappa_K$. The two orders are disfavored by $K$-phonons and $\Gamma$-phonons, respectively, hence competing to form a phase boundary with slope $\frac{\Delta \kappa_\Gamma}{\Delta \kappa_K} \sim 0.7$, which extends slightly beneath $\kappa_\Gamma = \kappa_K = 0$. The area occupied by $\rm (KIVC\times SP)_{tw}$ shrinks accordingly. Moreover, the $\rm OP\times QAH/QSH$ order observed at $U_0=0$ gives way to $\rm OP \times SP$, because in that region, $\kappa_K$ is small, and $\kappa_K J_{\rm e} + K_{\rm e}$ will be negative to favor a spin-polarized component. 
The area occupied by $\rm NIOC \times NIVC $ greatly diminishes, and disappears for $U_0 = 6,9$eV. In the vicinity of $\kappa_\Gamma=1$ and $\kappa_K=3.2$, $\rm TIVC \times QAH/QSH$ prevails.

\section{Conclusion and discussion} \label{sec:discuss}

In this work, we have performed a systematic study on the EPC to the emergent local $f$-orbitals in MATBG, in order to shed light on how phonons participate in the strongly correlated physics. 
We identified and analyzed the phonon modes that have a significant EPC, and established that phonons mainly induce an on-site anti-Hund's interaction $\hat{H}_{\rm A}$ on each $f$-orbital. 
How the carbon atomic Hubbard $U_0$ contributes to a Hund's interaction $\hat{H}_{\rm H}$ that partly counteracts $\hat{H}_{\rm A}$ is also discussed. 
As one of the immediate applications, we explored the close competition between various symmetry-breaking orders at even-integer fillings in the presence of $\hat{H}_{\rm A}$ and $\hat{H}_{\rm H}$. 

Nevertheless, whether phonons are functioning behind more experimental facts is yet to be unveiled. 
Now, we briefly discuss the possibility of realizing a light symmetric semi-metal formed by only $c$-electrons at CNP. 

\subsection{Mott semi-metal at charge-neutrality}

In this state, each $f$-site forms the non-degenerate $[A_1, 0]$ singlet under $\hat{H}_{\rm A}+\hat{H}_{\rm H}$ (see \cref{fig:split-0}(c) and \cref{fig:split-9}(c)), and opens up a featureless Mott gap in the low-energy end. $c$-electrons are decoupled from $f$-sites and form the Dirac Fermi liquid. We dub such a state as Mott-SM. We note that, in traditional heavy fermion materials, such $f$-singlets are typically realized by strong crystalline electric field (CEF) effect on the atomic $f$-orbitals, hence dubbed as CEF singlets \cite{Satoshi_2002_nFL}. In other theoretical proposals, models with a non-degenerate ground state that forms a non-trivial irrep under crystalline symmetries are also designed \cite{Yao_2010_fragileMott}. Those states cannot be adiabatically connected to any band insulator in the presence of time-reversal, hence dubbed as fragile Mott states. The $N_f=4$ ground state, by forming the $A_1$ irrep, does not belong to that category. 

By the large $U$, the charge fluctuation at each $f$-site is asymptotically frozen, so the total $f$-electron number approaches some integer $N_f$. At CNP, $N_f=4$. On doping away from CNP, particles/holes first populate on the light $c$-electron bands, and re-shuffles to the $f$-orbital only when the next integer filling is approached \cite{Zhou_2024_Kondo,Hu_2023_Kondo,Rai_2023_DMFT}. Consequently, we roughly take $N_f \approx \lfloor \nu \rfloor + 4$ for $\nu>0$ and $N_f \approx \lfloor \nu \rfloor + 5$ for $\nu<0$, where $\lfloor \nu \rfloor$ takes the largest integer that does not exceed $\nu$. 
The Hamiltonian can be formally written as $\hat{H}_f + \hat{H}_c + \hat{H}_{fc}$. 
In the $N_f = 4$ subspace, $\hat{H}_f$ only contains the multiplet splittings induced by (anti-)Hund's $\hat{H}_{\rm A,H}$. 
$\hat{H}_c$ contains the kinetic term $\hat{H}^{(c)}_0$ and the Coulomb interaction $\hat{H}_V$. Since $c$-electrons are light and un-correlated, it suffices to treat the repulsive $\hat{H}_V$ at the Hartree-Fock level. 
$\hat{H}_{fc}$ describes the mutual couplings, which in the full Hilbert space with un-restricted $N_f$, contains a kinetic hybridization $\hat{H}^{(cf)}_{0}$, and two Coulomb terms - $\hat{H}_W$ and $\hat{H}_J$. 
Particularly, $\hat{H}_W$ acts as a density-density interaction, hence simply contributes a chemical potential to the $c$-electrons proportional to $N_f=4$, and can be absorbed into $\hat{H}^{(c)}_0$. 
We only need to keep $\hat{H}_J$. 

In the absence of multiplet splittings $\hat{H}_{\rm A,H}$, as $\hat{H}^{(cf)}_0$ induces high-energy virtual hoppings in and out of each $f$-site, integrating out such processes leads to a Kondo coupling that takes the form $\hat{H}_K \sim -\frac{\hat{H}^{(cf)}_0 \hat{H}^{(cf)\dagger}_0}{U}$ \cite{Zhou_2024_Kondo,Hu_2023_Kondo}. We first discuss the scenario without $\hat{H}_J$, which defines a standard Kondo lattice problem, with an array of U(8) $f$-moments interacting anti-ferromagnetically with the non-interacting $c$-bath \cite{Chou_2023_Kondo, Zhou_2024_Kondo, Hu_2023_Kondo, Datta_2023_heavy, Rai_2023_DMFT, Chou_2023_scaling, Lau_2023_topological}. Two different phases compete in a Kondo lattice. On the one hand, $c$-electrons tend to screen the $f$-moments through $\hat{H}_K$ directly, and the screening energy scale is characterized by the Kondo temperature $T_{\rm K}$. 
On the other hand, the $c$-bath also mediates an indirect RKKY interaction between $f$-moments at different sites, which is characterized by $J_{\rm RKKY}$. 
Recovering $\hat{H}_J$ will be insignificant to the Kondo screening, because ferromagnetic couplings flow irrelevantly in the low-energy end. Nevertheless, $\hat{H}_J$ serves as an equally important source of RKKY interactions besides $\hat{H}_K$ \cite{Hu_2023_Kondo}. 
At CNP, it is found that $T_{\rm K}$ is vanishingly small due to the Dirac nature of $c$-electrons \cite{Zhou_2024_Kondo}, while $J_{\rm RKKY} \lesssim 1$meV and is of the ferromagnetic type \cite{Hu_2023_Kondo}. Therefore, a symmetry-breaking order is expected. 

With the multiplet splittings $\hat{H}_{\rm A,H}$ turned on, however, the physics can significantly change. 
In particular, when the splittings far exceed both $T_{\rm K}$ and $J_{\rm RKKY}$, the low-energy physics should only involve the ground state multiplet \cite{nozieres_kondo_1980}, hence the Hilbert space for the $f$-moment is further restricted. 
In our case, as discussed in \cref{sec:multi,sec:HH-multi}, the ground state on the $f$-site forms the non-degenerate $[A_1, 0]$ irrep, hence any $fc$-couplings reduce to an identity on the $f$-moment, rendering the $f$-impurity and the $c$-electrons effectively decoupled. Correspondingly, neither Kondo screening nor symmetry-breaking could take place. 
Put on the lattice, the many-body ground state corresponds to an array of $f$-singlets inactively immersed in the light Dirac semi-metal formed by $c$-electrons. 

Since Kondo screening is not competitive at CNP even when compared with RKKY \cite{Zhou_2024_Kondo}, it suffices to discuss the stability of Mott-SM against the competition of symmetry-breaking orders. Note that, any symmetry breaking order must have a higher energy under $\hat{H}_{\rm A} + \hat{H}_{\rm H}$ than the non-degenerate ground state multiplet. For example, the energy of $\rm TIVC$ has been plotted in \cref{fig:split-0}(c) and \cref{fig:split-9}(c), which is the most favored order at $\kappa_K=3.2$, and the gap to the ground state multiplet is around $\Delta E_{\rm AH} \sim 3$meV. Therefore, $J_{\rm RKKY}$ must overcome the energy gap $\Delta E_{\rm AH}$, so as to compensate the on-site energy penalty to involve higher multiplet levels, and incorporate symmetry-breaking. At CNP, $J_{\rm RKKY} \lesssim 1$meV estimated in the scenario without multiplet splittings taken into consideration is smaller than $E_{\rm AH}$, hence realizing a Mott-SM is not impossible. Nevertheless, the actual physics can be sensitive to details such as the form of the RKKY interaction. Therefore, to determine the actual physics at CNP, further researches are called for. 

We comment on the experimental implication of the Mott-SM state. 
Owing to the fast Dirac velocity corresponding to the $c$-electrons $v_\star = 430\mrm{meV \cdot nm}$ \cite{Song_2022}, which is the same order as the SLG Dirac Fermi velocity $v_F$, Mott-SM is sharply distinct from the uncorrelated semi-metallic MATBG bands with almost vanishing Dirac velocity. 
Nevertheless, the semi-metallic $\rm NIOC$ discussed in \cref{sec:mf-0} also hosts a large Dirac velocity, and for both $\rm NIOC$ and Mott-SM, the Dirac points are located in the vicinity of $\GGamma_M$. We hence propose to distinguish the two phases from the local density of states of the occupied bands. Specifically, Mott-SM populates an irrep that transforms trivially under $C_{3z}$, and the strain-induced $C_{3z}$-breaking enters only at the perturbative level. On the contrary, the $\rm NIOC$ semi-metal populates orbitals that strongly violates the $C_{3z}$ symmetry. For example, if $Q=\sigma^x$, $f^\dagger_{\beta\eta s} + f^\dagger_{\ovl{\beta}\eta s}$ for all $\eta$ and $s$ is populated, which exhibits a strong $C_{3z}$-breaking. 

To close this article, we also briefly discuss whether the Dirac $c$-electrons alone can support an SC state. 
For this sake, we assume that all the degrees of freedom on $f$-impurities are frozen, \textit{e.g.}, by forming the non-degenerate $[A_1, 0]$ singlet at $N_f=2$, and doping to $\nu \in (-2.5, -2]$ mainly amounts to doping the $c$-electron bands with $\nu_c \in (-0.5, 0]$. 
Since there are two valence $c$-electron bands per spin per valley, each valence band is doped with filling $\frac{\nu_c}{8}$, which is $\lesssim 0.06$, and corresponds to a Fermi vector $k_F \lesssim 0.22 k_\theta$. 
Since $c$-electrons are very light, we estimate the dimensionless EPC constant $\lambda = g D(k_F)$ for a conventional BCS superconductor, and assume the absence of Coulomb repulsion. 
Here, $g$ is the zero-frequency phonon-mediated interaction strength among $c$-electrons, and can be estimated as the same order as the $f$-electrons, $1\sim 4$meV. At $k_F=0.22 k_\theta$, $D(k_F)$ on the Dirac bands becomes the largest (\cref{fig:MBZ}(c)), with $D(k_F) = \frac{\Omega_M}{2\pi} \frac{k_F^2}{v_\star k_F} = 4\times 10^{-3}\mrm{meV}^{-1}$, which is still very small due to the fast Dirac velocity. One can also compare this result with the BCS theory carried out on the uncorrelated flat bands with almost vanishing Dirac velocity \cite{Lian_2019_SCac}, where the density of states is on the order $1\mrm{meV}^{-1}$. 
Consequently, $\lambda$ is on the order of $10^{-2}$, which is too weak to generate superconductivity on the experimentally probed temperature, as the BCS transition temperature is exponential in $-\frac{1}{\lambda}$. 



\begin{acknowledgements}
We are grateful to Seung-Sup B. Lee for useful discussions. 
Z.-D. S., Y.-J. W. and G.-D. Z. were supported by National Natural Science Foundation of China (General Program No. 12274005), National Key Research and Development Program of China (No. 2021YFA1401900), and Innovation Program for Quantum Science and Technology (No. 2021ZD0302403). 
B. L. is supported by the National Science Foundation under award DMR-2141966, and the National Science Foundation through Princeton University's Materials Research Science and Engineering Center DMR-2011750.
\end{acknowledgements}

\appendix
\onecolumngrid

\newpage

\tableofcontents

\newpage

\section{Symmetry actions on the phonon fields} \label{app:sym}

We analyze how crystalline symmetries act on the microscopic and continuum phonon fields. Let the graphene lattice vector $\RR^{(l)}$ of layer $l=1,2$ spanned by $\mbf{a}^{(l)}_{1,2}$, 
\begin{align}
    \mbf{a}^{(l)}_{1} = R_{(-1)^{l-1}\frac{\theta}{2}} \circ  (\frac{1}{2}, \frac{\sqrt{3}}{2}) a_0 \ , \qquad \mbf{a}^{(l)}_{2} = R_{(-1)^{l-1}\frac{\theta}{2}} \circ  (-\frac{1}{2}, \frac{\sqrt{3}}{2}) a_0 \ , 
\end{align}
where $R_{\frac{\theta}{2}} \circ$ denotes an anti-counterclockwise rotation by $\frac{\theta}{2}$. 
Then, $\RR^{(l)} = N_1 \mbf{a}_1^{(l)} + N_2 \mbf{a}_2^{(l)}$ ($N_{1,2} \in \mbb{Z}$). 
The carbon atoms are located at $\RR^{(l, \alpha)} = \RR^{(l)} + \ttau^{(l, \alpha)}$, where $\ttau^{(l, \alpha)}$ denotes the sublattice location within each graphene unit cell. We will choose 
\begin{align}
    \ttau^{(l, A)} = \frac{1}{3} \left( \mbf{a}^{(l)}_1 + \mbf{a}^{(l)}_2  \right) ~, \qquad \ttau^{(l, B)} = -\frac{1}{3} \left( \mbf{a}^{(l)}_1 + \mbf{a}^{(l)}_2  \right)
\end{align}

We define the spatial rotation actions on the phonon fields as the following, so as to stay compatible with the actions on the electron fields (\cref{eq:Dg}), 
\begin{align}
    C_{nz} \begin{pmatrix}
        \uu_{x}(\RR^{(l, \alpha)}) \\
        \uu_{y}(\RR^{(l, \alpha)}) \\
        \uu_{z}(\RR^{(l, \alpha)}) \\
    \end{pmatrix} C_{nz}^{-1} 
    =& \begin{pmatrix}
        \cos\frac{2\pi}{n} & \sin\frac{2\pi}{n} & 0 \\ 
        -\sin\frac{2\pi}{n} & \cos\frac{2\pi}{n} & 0 \\
        0 & 0 & 1 \\
    \end{pmatrix} \begin{pmatrix}
        \uu_{x}(C_{nz} \RR^{(l, \alpha)}) \\
        \uu_{y}(C_{nz} \RR^{(l, \alpha)}) \\
        \uu_{z}(C_{nz} \RR^{(l, \alpha)}) \\
    \end{pmatrix} , \quad n = 2,3,6 \\
    C_{2x} \begin{pmatrix}
        \uu_{x}(\RR^{(l, \alpha)}) \\
        \uu_{y}(\RR^{(l, \alpha)}) \\
        \uu_{z}(\RR^{(l, \alpha)}) \\
    \end{pmatrix} C_{2x}^{-1} 
    =& \begin{pmatrix}
        1 & 0 & 0 \\ 
        0 & - 1 & 0 \\
        0 & 0 & - 1 \\
    \end{pmatrix} \begin{pmatrix}
        \uu_{x}(C_{2x} \RR^{(l, \alpha)}) \\
        \uu_{y}(C_{2x} \RR^{(l, \alpha)}) \\
        \uu_{z}(C_{2x} \RR^{(l, \alpha)}) \\
    \end{pmatrix}
\end{align}
where $C_{3z} \RR^{(l, \alpha)}$ belongs to sublattice $\RR^{(l, \alpha)}$, $C_{6z}\RR^{(l, \alpha)}$ and $C_{2z}\RR^{(l, \alpha)}$ belong to sublattice $\RR^{(l, \ovl{\alpha})}$, and $C_{2x} \RR^{(l, \alpha)}$ belongs to sublattice $\RR^{(\ovl{l}, \ovl{\alpha})}$. Furthermore, since $\uu_i(\RR^{(l, \alpha)})$ are real numbers measuring the displacement of each atom, it is natural to define the time-reversal action $T$ on phonon fields as
\begin{align}
    T \uu_i(\RR^{(l, \alpha)}) T^{-1} = \uu_i(\RR^{(l, \alpha)})
\end{align}

Applying the Fourier transformation \cref{eq:def_uq}, for $\qq \in \gBZ^{(l)}$ there is, 
\begin{align}
    C_{6z} \begin{pmatrix}
        \uu_{\qq, x}^{(l, \alpha)} \\
        \uu_{\qq, y}^{(l, \alpha)} \\
        \uu_{\qq, z}^{(l, \alpha)} \\
    \end{pmatrix} C_{6z}^{-1} 
    =& \begin{pmatrix}
        \frac{1}{2} & \frac{\sqrt{3}}{2} & 0 \\ 
        - \frac{\sqrt{3}}{2} & \frac{1}{2} & 0 \\
        0 & 0 & 1 \\
    \end{pmatrix} \begin{pmatrix}
        \uu_{C_{6z}\qq, x}^{(l, \ovl{\alpha})} \\
        \uu_{C_{6z}\qq, y}^{(l, \ovl{\alpha})} \\
        \uu_{C_{6z}\qq, z}^{(l, \ovl{\alpha})} \\
    \end{pmatrix} \\ 
    C_{2x} \begin{pmatrix}
        \uu_{\qq, x}^{(l, \alpha)} \\
        \uu_{\qq, y}^{(l, \alpha)} \\
        \uu_{\qq, z}^{(l, \alpha)} \\
    \end{pmatrix} C_{2x}^{-1} 
    =& \begin{pmatrix}
        1 & 0 & 0 \\ 
        0 & - 1 & 0 \\
        0 & 0 & - 1 \\
    \end{pmatrix} \begin{pmatrix}
        \uu_{C_{2x} \qq, x}^{(\ovl{l}, \ovl{\alpha})} \\
        \uu_{C_{2x} \qq, y}^{(\ovl{l}, \ovl{\alpha})} \\
        \uu_{C_{2x} \qq, z}^{(\ovl{l}, \ovl{\alpha})} \\
    \end{pmatrix} \\
    T  \uu_{\qq, i}^{(l, \alpha)}  T^{-1} =& ~ 
    \uu_{- \qq, i}^{(l, \alpha)} 
\end{align}

Consequently, for the continuum phonons with $\chi@\Gamma = (r, \XA/\YA)$, by definition \cref{eq:uG_q}, they transform as
\begin{align}   \label{eq:C6z_rXAYA}
    C_{6z} \begin{pmatrix}
        u_{\qq,r,\XA} \\ 
        u_{\qq,r,\YA} \\
    \end{pmatrix} C_{6z}^{-1} =& \begin{pmatrix}
        \frac{1}{2} & \frac{\sqrt{3}}{2} \\
        - \frac{\sqrt{3}}{2} & \frac{1}{2} \\
    \end{pmatrix} \begin{pmatrix}
        u_{C_{6z}\qq,r,\XA} \\ 
        u_{C_{6z}\qq,r,\YA} \\
    \end{pmatrix}  \\ 
    \label{eq:C2x_rXAYA}
    C_{2x} \begin{pmatrix}
        u_{\qq,r,\XA} \\ 
        u_{\qq,r,\YA} \\
    \end{pmatrix} C_{2x}^{-1} =& \begin{pmatrix}
        -1 & 0 \\
        0 & 1 \\
    \end{pmatrix} \begin{pmatrix}
        u_{C_{2x}\qq,r,\XA} \\ 
        u_{C_{2x}\qq,r,\YA} \\
    \end{pmatrix} 
\end{align}
which forms an $E_1$ irrep of the $D_6$ group. For $\chi@\Gamma = (r, \XO/\YO)$, 
\begin{align}   \label{eq:C6z_rXOYO}
    C_{6z} \begin{pmatrix}
        u_{\qq,r,\XO} \\ 
        u_{\qq,r,\YO} \\
    \end{pmatrix} C_{6z}^{-1} =& (-1) \begin{pmatrix}
        \frac{1}{2} & \frac{\sqrt{3}}{2} \\
        -\frac{\sqrt{3}}{2} & \frac{1}{2} \\
    \end{pmatrix} \begin{pmatrix}
        u_{C_{6z}\qq,r,\XO} \\ 
        u_{C_{6z}\qq,r,\YO} \\
    \end{pmatrix}  \\ \label{eq:C2x_rXOYO}
    C_{2x} \begin{pmatrix}
        u_{\qq,r,\XO} \\ 
        u_{\qq,r,\YO} \\
    \end{pmatrix} C_{2x}^{-1} =& \begin{pmatrix}
        1 & 0 \\
        0 & -1 \\
    \end{pmatrix} \begin{pmatrix}
        u_{C_{2x}\qq,r,\XO} \\ 
        u_{C_{2x}\qq,r,\YO} \\
    \end{pmatrix} 
\end{align}
which forms an $E_2$ irrep of the $D_6$ group. For $\chi@\Gamma = (r, \ZA)$, 
\begin{align}   \label{eq:C6z_rZA}
    C_{6z} u_{\qq,r,\ZA} C_{6z}^{-1} =& u_{C_{6z}\qq,r,\ZA} \\ \label{eq:C2x_rZA}
    C_{2x} u_{\qq,r,\ZA} C_{2x}^{-1} =& u_{C_{2x}\qq,r,\ZA} 
\end{align}
which forms an $A_1$ irrep of the $D_6$ group. 

For $K$-phonons, as explained in \cref{sec:epc-phn}, it is more convenient to work with the layer basis $l=1,2$. Therefore, we only discuss the representations of $C_{nz}$ actions, which transform within each layer. Under $C_{2z}$, there is 
\begin{align}   \label{eq:C2z_A1}
    C_{2z} u_{\qq,l,\Aone} C_{2z}^{-1} =& + u_{-\qq,l,\Aone} \\ 
    \label{eq:C2z_B1}
    C_{2z} u_{\qq,l,\Bone} C_{2z}^{-1} =& - u_{-\qq,l,\Bone} 
\end{align}
which reflects their nature as the $A_1$ or $B_1$ irreps of the $D_6$ group of each graphene layer. 
Nevertheless, by definition Eq. (\ref{eq:uK_q}), they do not transform in a simple form under $C_{3z}$. To exploit the $C_{3z}$ symmetry, we have to focus on each valley component (\cref{eq:uK_eta,eq:uK_q}), 
\begin{align}   \label{eq:uK_eta_app}
    u_{\qq, l, \eta K} = \frac{u_{\qq+\eta\KK_l, l, \Aone} - i \eta u_{\qq+\eta\KK_l, l, \Bone}}{\sqrt{2}} = \frac{-i\eta\left( \uu^{(l, A)}_{\qq+\eta\KK^{(l)}, x} + \uu^{(l, B)}_{\qq+\eta\KK^{(l)}, x} \right) - \left( \uu^{(l, A)}_{\qq+\eta\KK^{(l)}, y} - \uu^{(l, B)}_{\qq+\eta\KK^{(l)}, y} \right)}{2}
\end{align}
By the following identity, 
\begin{align}
    \uu^{(l, \alpha)}_{C_{3z}(\qq'+\eta\KK^{(l)}) , i} =& \frac{1}{\sqrt{N_g}} \sum_{\RR^{(l, \alpha)}} e^{-i (C_{3z}\qq'+ C_{3z}\eta\KK^{(l)}) \cdot \RR^{(l, \alpha)}} \uu_i(\RR^{(l, \alpha)}) \\\nonumber
    =& \frac{1}{\sqrt{N_g}} \sum_{\RR^{(l, \alpha)}} e^{-i (C_{3z}\qq'+ \eta\KK^{(l)}) \cdot \RR^{(l, \alpha)}} e^{-i \eta \left( C_{3z}\KK^{(l)} - \KK^{(l)} \right) \cdot \RR^{(l, \alpha)}} \uu_i(\RR^{(l, \alpha)}) \\\nonumber
    =& [e^{-i\eta \frac{2\pi}{3} \sigma_z}]_{\alpha, \alpha} \cdot \uu^{(l, \alpha)}_{C_{3z}\qq'+\eta\KK^{(l)} , i }
\end{align}
where we have applied $e^{-i \eta \left( C_{3z}\KK^{(l)} - \KK^{(l)} \right) \cdot \RR^{(l, \alpha)}} = e^{-i \eta \left( C_{3z}\KK^{(l)} - \KK^{(l)} \right) \cdot \ttau^{(l, \alpha)}} = [e^{-i\eta \frac{2\pi}{3} \sigma_z}]_{\alpha, \alpha}$, it can be verified that, 
\begin{align}   \label{eq:C3z_eta}
    C_{3z} u_{\qq, l, \eta K} C_{3z}^{-1} = u_{(C_{3z}\qq), l, \eta K} 
\end{align}
In comparison, by \cref{eq:uK_q}, we also compute how $u_{\qq,l,\Aone}$ transforms under $C_{3z}$, 
\begin{align}
    C_{3z} u_{\qq, l, \Aone} C_{3z}^{-1} &= C_{3z} \frac{u_{\qq-\KK_l, l, +K} + u_{\qq+\KK_l, l, -K}}{\sqrt{2}} C_{3z}^{-1} = \frac{u_{C_{3z}\qq-C_{3z}\KK_l, l, +K} + u_{C_{3z}\qq+C_{3z}\KK_l, l, -K}}{\sqrt{2}} \\\nonumber
    &= \frac{u_{C_{3z}\qq+(\KK_l-C_{3z}\KK_l), l, \Aone} - i u_{C_{3z}\qq+(\KK_l-C_{3z}\KK_l), l, \Bone} + u_{C_{3z}\qq-(\KK_l-C_{3z}\KK_l), l, \Aone} + i u_{C_{3z}\qq-(\KK_l-C_{3z}\KK_l), l, \Bone}}{2}
\end{align}
which hybrids two different momenta $C_{3z}\qq \pm (\KK_l + C_{3z}\KK_l)$ and complicates analysis. 

Finally, for arbitrary $\chi@\Gamma$ and $\chi@K$, there is 
\begin{align}   \label{eq:T_u}
    T u_{\qq, \chi} T^{-1} = u_{-\qq, \chi} \ .
\end{align}

\section{The inter-layer EPC of \texorpdfstring{$(r,\XA/\YA)$}{(r,XA/YA)} modes} \label{app:divergence}

We now discuss the inter-layer term in the EPC vertex of $(r, \XA/\YA)$ phonons (\cref{eq:C_r_XA}) in detail. 
In the main-text, we have found that this term remains finite even when $\qq=0$, bringing about an ostensible divergence. 
Now, we will show that in the energy band basis, the intra-energy-band elements actually vanish linearly with $\qq$, and only inter-energy-band elements remain finite. 
We then argue the finite value of inter-energy-band elements in the $\qq\to0$ limit simply results from the overall drift of Bloch wave functions when the two layers are rigidly shifted, hence should be deducted if properly imposing the Born-Oppenheimer (adiabatic) approximation. 

For this purpose, we focus the on the $(r,\XA)$ mode, and single out the inter-layer term (denoted by an extra superscript ``2''), 
\begin{align}   \label{eq:C_r_XA_2}
    C^{(r,\XA),2}(\qq;\rr) = - \frac{w_1}{\theta} \partial_y T(\rr) 
\end{align}
The analysis for $(r,\YA)$ modes follows analogously. The Bloch states on the $n$-th energy band with MBZ momentum $\kk$ and energy $E_{\kk,n}$ reads,  
\begin{align} 
    \Psi^\dagger_{\kk, n, \eta, s} = \sum_{l\alpha} \int\mrm{d}^2\rr ~ e^{i\kk\cdot\rr} ~ \varpi^{(\eta)}_{l\alpha, n\kk}(\rr) \psi^\dagger_{l, \alpha, \eta, s}(\rr)
\end{align}
The Bloch functions are periodic in the momentum space,  
\begin{align} \label{eq:varpi_trans_G}
    \Psi^\dagger_{\kk+\GG,n,\eta,s} = \Psi^\dagger_{\kk,n,\eta,s}, \quad \textrm{so that} \quad 
    \varpi^{(\eta)}_{l\alpha,n\kk}(\rr) = e^{i\GG\cdot\rr} \varpi^{(\eta)}_{l\alpha,n\kk+\GG}(\rr)
\end{align}
for an arbitrary moir\'e reciprocal lattice vector $\GG$ (spanned by $\mbf{g}_{M1,M2}$ as depicted in \cref{fig:MBZ}(a)). 
The Bloch functions also necessarily enjoy the moir\'e translation symmetry, and we choose the representation as, 
\begin{align} \label{eq:varpi_trans_R}
    T_{\Delta\RR} \Psi^\dagger_{\kk,n,\eta,s} T_{\Delta\RR}^{-1} = e^{-i\kk\cdot\Delta\RR} \Psi^\dagger_{\kk,n,\eta,s}, \quad \textrm{so that} \quad 
    \varpi^{(\eta)}_{l\alpha,n\kk}(\rr) = e^{i\eta\KK_l\cdot\Delta\RR} \varpi^{(\eta)}_{l\alpha,n\kk}(\rr+\Delta\RR)
\end{align}
where the layer and valley dependent $e^{i\eta\KK_l\cdot\Delta\RR}$ factor in the second equation follows from \cref{eq:psi_TR}.

The normalization of the Bloch function obeys
\begin{align}
    1 = \sum_{l\alpha} \int \mrm{d}^2\rr ~\varpi^{(\eta)*}_{l\alpha,n\kk}(\rr) \varpi^{(\eta)}_{l\alpha,n\kk}(\rr) = N_M \sum_{l\alpha} \int_{\rm MUC} \mrm{d}^2\rr 
    ~ \varpi^{(\eta)*}_{l\alpha,n\kk}(\rr) \varpi^{(\eta)}_{l\alpha,n\kk}(\rr)
\end{align}
where $N_M \sim \theta^2 N_g$ denotes the total number of MUC's. 

In terms of these Bloch states, the EPC vertex induced by $(r,\XA)$ phonons reads,  
\begin{align} 
    \hat{H}_{\rm epc} &= \frac{1}{\sqrt{N_g}} \sum_{s,\eta} \sum_{\kk,\kk'\in\MBZ} \sum_{|\qq| \lesssim \Lambda}  u_{\qq,r,\XA} \Psi^\dagger_{\kk,n,\eta,s} L^{(r,\XA),2}_{n\eta, n'\eta}(\qq;\kk,\kk') \Psi_{\kk',n',\eta,s} 
\end{align}
with 
\begin{align} \label{eq:L}
    L^{(r,\XA),2}_{n\eta, n'\eta}(\qq;\kk,\kk') &= \sum_{l\alpha,l'\alpha'} \int \mrm{d}^2\rr ~ e^{-i\kk\cdot\rr} \varpi^{(\eta)*}_{l\alpha,n\kk}(\rr) \left[ C^{(r,\XA),2}_{l\alpha\eta,l'\alpha'\eta}(\qq;\rr) \right] \varpi^{(\eta)}_{l'\alpha',n'\kk'}(\rr) e^{i(\kk'+\qq) \cdot \rr} 
\end{align}
We remark that: 
1) $(r, \XA)$ phonons preserve the independent $\rm U(1)^2 \times SU(2)^2$ symmetries; 
2) by virtue of \cref{eq:varpi_trans_G}, the $L$-matrix (\cref{eq:L}) depends on $\kk$ and $\kk'$ only modulo MBZ - namely, changing either $\kk\to\kk+\GG$ or $\kk'\to\kk'+\GG$ does not affect the integral value (but changing $\qq\to\qq+\GG$ does); 
3) by the moir\'e translation symmetry \cref{eq:varpi_trans_R}, the $L$-matrix (\cref{eq:L}) should be non-vanishing only when $\kk=\kk'+\qq$ mod MBZ. 
Equivalently, one can then verify by \cref{eq:varpi_trans_R} and the fact $[T(\rr+\Delta\RR)]_{1\alpha\eta,2\alpha'\eta} = e^{i\eta\qq_1\cdot\Delta\RR} [T(\rr)]_{1\alpha\eta,2\alpha'\eta}$ that the integrand of \cref{eq:L} is periodic in $\rr$ with MUC periodicity. 

To evaluate \cref{eq:L}, we apply the following trick, 
\begin{align}   \label{eq:C_r_XA_equi}
    C^{(r,\XA),2}(\qq;\rr) = - \frac{1}{\theta} [\vec{\partial}_y,  H(\rr)]
\end{align}
where by using the symbol $\vec{\partial}_y$, we emphasize that this derivative operates on the product of all objects that are functions of $\rr$ after it. $H(\rr) = v_F \left(-i \vec{\partial}_{\rr} \cdot \mbs{\sigma}^{(\eta)} \right) \rho_0 + w_1 T(\rr)$ denotes the full first-quantized matrix in $\hat{H}_0$ (\cref{eq:H0}). 
For the term containing $\vec{\partial}_y H(\rr)$, 
\begin{align}
    & \int\mrm{d}^2\rr \cdots \vec{\partial}_y \sum_{l'\alpha'} \left( [H(\rr) ]_{l\alpha\eta,l'\alpha'\eta'} \left[ \varpi^{(\eta)}_{l'\alpha',n'\kk'}(\rr) e^{i\kk'\cdot\rr} \right] e^{i\qq\cdot\rr} \right) \\\nonumber
    &= \int\mrm{d}^2\rr \cdots \vec{\partial}_y \sum_{l'\alpha'} \left(  e^{i\qq\cdot\rr} [v_F \qq\cdot\mbs{\sigma^{(\eta)}} ]_{l\alpha\eta,l'\alpha'\eta'} \left[ \varpi^{(\eta)}_{l'\alpha',n'\kk'}(\rr) e^{i\kk'\cdot\rr} \right] + e^{i\qq\cdot\rr} [H(\rr)]_{l\alpha\eta,l'\alpha'\eta'} \left[ \varpi^{(\eta)}_{l'\alpha',n'\kk'}(\rr) e^{i\kk'\cdot\rr} \right] \right)  \\\nonumber
    &= \int\mrm{d}^2\rr \cdots \vec{\partial}_y \left(  e^{i\qq\cdot\rr} \sum_{l'\alpha'} [v_F \qq\cdot\mbs{\sigma^{(\eta)}} ]_{l\alpha\eta,l'\alpha'\eta'} \left[\varpi^{(\eta)}_{l'\alpha',n'\kk'}(\rr) e^{i\kk'\cdot\rr} \right] + e^{i\qq\cdot\rr} E_{\kk',n'} \left[ \varpi^{(\eta)}_{l\alpha,n'\kk'}(\rr) e^{i\kk'\cdot\rr} \right] \right) 
\end{align}
where $\cdots$ omitted factors before $  \vec{\partial}_y H(\rr)$. For the term containing $ H(\rr) \vec{\partial}_y$, 
\begin{align}
    \int\mrm{d}^2\rr  \sum_{l\alpha} \left[ e^{-i\kk\cdot\rr} \varpi^{(\eta)*}_{l\alpha,n\kk}(\rr) \right] [H(\rr)]_{l\alpha\eta,l'\alpha'\eta'} \vec{\partial}_y \cdots = \int\mrm{d}^2\rr  E_{\kk,n} \left[ e^{-i\kk\cdot\rr} \varpi^{(\eta)*}_{l'\alpha',n\kk}(\rr) \right] \vec{\partial}_y \cdots
\end{align}
where $\cdots$ omitted factors after $ H(\rr) \vec{\partial}_y$. An integration-by-part has been implicitly carried out, where the periodicity of the integrand is exploited. 
Combined together (recall that $L$-matrix is non-vanishing only for $\kk=\kk'+\qq$ mod MBZ), we obtain
\begin{align}
    L_{n\eta,n'\eta}^{(r,\XA),2}(\qq;\kk'+\qq,\kk') &= -\frac{v_F}{\theta} \qq \cdot \int \mrm{d}^2\rr \sum_{l\alpha,l'\alpha'} 
    e^{-i(\kk'+\qq)\cdot\rr} \varpi^{(\eta)*}_{l\alpha,n\kk'+\qq}(\rr) [\mbs{\sigma}^{(\eta)}]_{l\alpha\eta,l'\alpha'\eta'} \vec{\partial}_y \varpi^{(\eta)}_{l'\alpha',n'\kk'}(\rr) e^{i(\kk'+\qq) \cdot \rr} \\\nonumber 
    &+ \frac{1}{\theta} (E_{\kk'+\qq,n} - E_{\kk',n'}) \int \mrm{d}^2\rr \sum_{l\alpha} e^{-i(\kk'+\qq)\cdot\rr} \varpi^{(\eta)*}_{l\alpha,n\kk'+\qq}(\rr) \vec{\partial}_y \varpi^{(\eta)}_{l\alpha,n'\kk'}(\rr) e^{i(\kk'+\qq) \cdot \rr}
\end{align}
It can thus be seen that, for the first term, the EPC strengths vanish linearly when $\qq\to0$ for all band indices $n,n'$, while for the second term, the intra-energy-band EPC strengths vanishes as $E_{\kk'+\qq,n} - E_{\kk',n} \to 0$ when $\qq\to0$, but the inter-energy-band EPC strength can remain finite. 

To understand the inter-energy-band EPC strength at $\qq=0$, we consider a rigid relative glide $\uu$ between the two layers, which will move the moir\'e potential as $T(\rr) \to T(\rr-\td{\uu})$, where $\td{\uu} = \frac{1}{\theta} \hat{\mbf{z}} \times \uu$. Consequently, a Bloch state with momentum $\kk$ and Bloch function $\varpi^{(\eta)}_{l\alpha,n\kk}(\rr)$ should move along with $\td{\uu}$, too. Its energy $E_{\kk,n}$ will be un-changed, and we denote the new wave-function by $\varpi^{(\eta)'}_{l\alpha,n\kk}(\rr) = \varpi^{(\eta)}_{l\alpha,n\kk}(\rr - \td{\uu})$. 
The overlap between $\varpi^{(\eta)'}_{l\alpha,n\kk}(\rr)$ and $\varpi^{(\eta)}_{l\alpha,n\kk}(\rr)$ necessarily diminishes, hence $\varpi^{(\eta)'}_{l\alpha,n\kk}(\rr)$ must also contain $\varpi^{(\eta)}_{l\alpha,n'\kk}(\rr)$ with general $n'\not=n$, which span a complete basis for Bloch functions, to preserve the norm. 
Such a process has a finite amplitude proportional to $\uu$ at the leading order, even if $\qq=0$.
Nevertheless, they do not contribute to EPC after imposing the Born-Oppenheimer approximation, just as in stoichiometric materials. 

\section{Projected EPC on \texorpdfstring{$f$}{f}-orbitals} \label{app:vertex}

In this section, we derive the general symmetry-constrained form of $M^{(\chi@\Gamma)}_{+,+}(\qq)$, for $\chi@\Gamma = (r, \XA/\YA)$, $(r, \XO/\YO)$, and $M^{(\chi@K)}_{+,-}(\qq)$ for $\chi@K = (l, \Aone/\Bone)$ ($l=1,2$), which are shown in the main-text to be non-negligible for the $f$-orbitals. The parametrizing functions for $M^{(\chi@\Gamma)}_{+,+}(\qq)$ will be classified using irreps of the $D_3$ group, which is the point group obeyed by electrons in the $\eta=+$ valley. 
The parametrizing functions for $M^{(l, \Aone)}_{+,+}(\qq)$ will be constrained by the $C_{3z}$ rotation symmetry. Finally, we also explain technique details in deriving the Gaussian formulae of the projected EPC vertices in \cref{tab:rXAYA_irrep,tab:rXOYO_irrep,tab:lA1_irrep}. 

\subsection{\texorpdfstring{$(r, \XA/\YA)$}{(r, XA/YA)}} \label{app:vertex-rXAYA}

For the parametrization \cref{eq:M_para_G}, we have shown in Sec. \ref{sec:vertex-rXAYA} that, $C_{2z}T$ and the Hermiticity jointly dictate only $M^{(r, \XA/\YA)}_{z, Re}(\qq)$ ($\qq$-even) and $M^{(r, \XA/\YA)}_{0/x/y, Im}(\qq)$ ($\qq$-odd) can survive. Now we show that the emergent particle-hole $P$ further forbids $M^{(r, \XA/\YA)}_{0,Im}(\qq)$. 
It can be checked that $D(P) C^{(r, \XA/\YA)}(\qq; \rr) D^\dagger(P) = C^{(r, \XA/\YA)}(-\qq; -\rr)$ from \cref{eq:DP,eq:C_r_XA,eq:C_r_YA}, which implies
\begin{align}
    \langle w^{(+)}_{\beta}(\rr) | C^{(\chi)} (\qq;\rr) | w^{(+)}_{\beta'}(\rr) \rangle =& \langle w^{(+)}_{\beta}(\rr) | D^\dagger(P) D(P) C^{(\chi)} (\qq;\rr) D^\dagger(P) D(P) | w^{(+)}_{\beta'}(\rr) \rangle  \\\nonumber
    =& (-1)^{\beta+\beta'} \langle w^{(+)}_{\beta}(-\rr) | C^{(\chi)} (-\qq;-\rr) | w^{(+)}_{\beta'}(-\rr) \rangle , \qquad \chi = (r, \XA/\YA) 
\end{align}
In the above bra-ket notation, only layer ($l$) and sublattice ($\alpha$) indices are summed over, while $\rr$ is \textit{not} integrated. The first equation has inserted $1 = D^\dagger(P) D(P)$, and the second equation follows from \cref{eq:DP_eig}. Inserting this relation into the definition of $M^{(r, \XA/\YA)}_{\beta+, \beta'+}(\qq)$ (Eq. (\ref{eq:M})), we will conclude that $M^{(r, \XA/\YA)}_{\beta+, \beta'+}(\qq) = (-1)^{\beta+\beta'} M^{(r, \XA/\YA)}_{\beta+, \beta'+}(-\qq)$, namely, $\beta$-diagonal elements must depend evenly on $\qq$, while $\beta$-off-diagonal elements must depend oddly on $\qq$. Combining with the parity requirements imposed by Hermiticity, only $M^{(r, \XA/\YA)}_{z, Re}(\qq)$ (even) and $M^{(r, \XA/\YA)}_{x/y, Im}(\qq)$ (odd) can survive.

We now impose $C_{3z}$ and $C_{2x}$, which span a $D_3$ point group at $\GGamma$, to solve their general $\qq$ dependence. 
To begin with, by squaring \cref{eq:C6z_rXAYA}, $C_{3z}$ acts as
\begin{align} \label{eq:C3z_rXAYA}
    C_{3z} \begin{pmatrix}
        u_{\qq,r,\XA} \\
        u_{\qq,r,\YA} \\
    \end{pmatrix} C_{3z}^{-1} = \begin{pmatrix}
        -\frac{1}{2} & \frac{\sqrt{3}}{2} \\
        -\frac{\sqrt{3}}{2} & -\frac{1}{2} \\
    \end{pmatrix} \begin{pmatrix}
        u_{C_{3z}\qq,r,\XA} \\
        u_{C_{3z}\qq,r,\YA} \\
    \end{pmatrix} 
\end{align}
By conjugating $\hat{H}_{\rm epc}$ with $C_{3z}$, the invariance of the EPC vertex dictates 
\begin{align} 
    e^{i\frac{2\pi}{3} \sigma_z} \left[ -\frac{1}{2} M^{(r, \XA)}_{+,+}(\qq) - \frac{\sqrt{3}}{2} M^{(r, \YA)}_{+,+}(\qq) \right] e^{-i\frac{2\pi}{3} \sigma_z} &= M^{(r, \XA)}(C_{3z}\qq) \\ 
    e^{i\frac{2\pi}{3} \sigma_z} \left[ \frac{\sqrt{3}}{2} M^{(r, \XA)}_{+,+}(\qq) - \frac{1}{2} M^{(r, \YA)}_{+,+}(\qq) \right] e^{-i\frac{2\pi}{3} \sigma_z} &= M^{(r, \YA)}(C_{3z}\qq)
\end{align}
where \cref{eq:DfC3z,eq:C3z_rXAYA} are used. On the other hand, the invariance under $C_{2x}$ dictates (using \cref{eq:DfC2x,eq:C2x_rXAYA}), 
\begin{align} 
    \sigma_x \left[ -  M^{(r, \XA)}_{+,+}(\qq) \right] \sigma_x &= M^{(r, \XA)}_{+,+}(C_{2x}\qq) \\ 
    \sigma_x \left[  M^{(r, \YA)}_{+,+}(\qq) \right] \sigma_x &= M^{(r, \YA)}_{+,+}(C_{2x}\qq)
\end{align}

Decomposed with the surviving parameters, the diagonal elements parametrized by $M^{(r, \XA/\YA)}_{z, Re}(\qq)$ must obey the following restrictions, 
\begin{align}
    \begin{pmatrix}
        -\frac{1}{2} & -\frac{\sqrt{3}}{2} \\
        \frac{\sqrt{3}}{2} & -\frac{1}{2} \\
    \end{pmatrix} \begin{pmatrix}
        M^{(r, \XA)}_{z, Re}(\qq) \\
        M^{(r, \YA)}_{z, Re}(\qq) \\
    \end{pmatrix} = \begin{pmatrix}
        M^{(r, \XA)}_{z, Re}(C_{3z}\qq) \\
        M^{(r, \YA)}_{z, Re}(C_{3z}\qq) \\
    \end{pmatrix}  , \quad
    \begin{pmatrix}
        1 & 0 \\
        0 & -1 \\
    \end{pmatrix} \begin{pmatrix}
        M^{(r, \XA)}_{z, Re}(\qq) \\
        M^{(r, \YA)}_{z, Re}(\qq)
    \end{pmatrix} = \begin{pmatrix}
        M^{(r, \XA)}_{z, Re}(C_{2x}\qq) \\
        M^{(r, \YA)}_{z, Re}(C_{2x}\qq)
    \end{pmatrix}
\end{align}
which form an $E$ irrep of the $D_3$ group. The lowest order of real functions that are even in $\qq$ and can furnish such an irrep must be of $O(|\qq|^2)$, and take the following form 
\begin{align}  \label{eq:appC_M_eE}
    M^{(r, \XA)}_{z, Re}(\qq) \sim \qq_x^2 - \qq_y^2  , \qquad 
    M^{(r, \YA)}_{z, Re}(\qq) \sim -2\qq_x\qq_y , \qquad \boxed{ \textrm{even}~E }
\end{align}

For the off-diagonal elements, by definition of the parametrization functions \cref{eq:M_para_G}, 
\begin{align} \label{eq:appC7}
    M^{(\chi)}_{2+, 1+}(\qq) &= i\left( M^{(\chi)}_{x,Im}(\qq) + i M^{(\chi)}_{y,Im}(\qq) \right) , \\ \nonumber
    M^{(\chi)}_{1+, 2+}(\qq) &= i\left( M^{(\chi)}_{x,Im}(\qq) - i M^{(\chi)}_{y,Im}(\qq) \right) , 
\end{align}
hence $M^{(\chi)}_{2+, 1+}(\qq) = - M^{(\chi)*}_{1+, 2+}(\qq)$, and it suffices to track one of them, \textit{e.g.}, $M^{(\chi)}_{2+, 1+}(\qq)$. The restrictions of $C_{3z}$ and $C_{2x}$ symmetries read,  
\begin{align}   \label{eq:M_rXAYA_B8}
    e^{-i\frac{4\pi}{3}} \begin{pmatrix}
        - \frac{1}{2} & - \frac{\sqrt{3}}{2} \\
        \frac{\sqrt{3}}{2} & - \frac{1}{2} \\
    \end{pmatrix}
    \begin{pmatrix}
        M^{(r, \XA)}_{2+, 1+}(\qq) \\
        M^{(r, \YA)}_{2+, 1+}(\qq) \\
    \end{pmatrix} = \begin{pmatrix}
        M^{(r, \XA)}_{2+, 1+}(C_{3z}\qq) \\
        M^{(r, \YA)}_{2+, 1+}(C_{3z}\qq) \\
    \end{pmatrix}, \quad \begin{pmatrix}
        -1 & 0 \\
        0 & 1 \\
    \end{pmatrix} \begin{pmatrix}
        M^{(r, \XA)}_{2+, 1+}(\qq) \\
        M^{(r, \YA)}_{2+, 1+}(\qq) \\
    \end{pmatrix} = - \begin{pmatrix}
        M^{(r, \XA)}_{2+, 1+}(C_{2x}\qq) \\
        M^{(r, \YA)}_{2+, 1+}(C_{2x}\qq) \\ 
    \end{pmatrix}^*
\end{align}
where the second relation also exploited the relation $M^{(\chi)}_{2+, 1+}(\qq) = - M^{(\chi)*}_{1+, 2+}(\qq)$. 
To simplify the problem to irreps of real functions, we left-multiply $\begin{pmatrix}
    1 & i \\
    1 & -i \\
\end{pmatrix}$
to \cref{eq:M_rXAYA_B8}, and introduce the following complex functions
\begin{align}  \label{eq:appC8}
    M^{(\pm)}(\qq) = \frac{M^{(r, \XA)}_{2+, 1+}(\qq) \pm i M^{(r, \YA)}_{2+, 1+}(\qq)}{2} 
\end{align}
Now, \cref{eq:M_rXAYA_B8} will be block-diagonalized as, 
\begin{align}   \label{eq:appC9}
    e^{-i\frac{2\pi}{3}} M^{(+)}(\qq) =&  M^{(+)}(C_{3z} \qq), \quad M^{(+)}(\qq) = M^{(+)*}(C_{2x}\qq) \\   \label{eq:appC10}
    M^{(-)}(\qq) =& M^{(-)}(C_{3z} \qq) , \quad  M^{(-)}(\qq) = M^{(-)*}(C_{2x}\qq) 
\end{align}

For $M^{(+)}(\qq)$, since it is odd in $\qq$, one can directly conclude from \cref{eq:appC9} that the leading $\qq$-dependence should be proportional to $M^{(+)}(\qq) \sim \qq_x - i\qq_y$ with a real coefficient. By substituting \cref{eq:appC8} and \cref{eq:appC7} sequentially, there will be
\begin{align}   \label{eq:appC_M_oE}
    \mrm{Re}M^{(+)}(\qq) = \frac{M^{(r,\XA)}_{y,Im}(\qq) + M^{(r,\YA)}_{x,Im}(\qq)}{2} \sim \qq_x , \quad \mrm{Im}M^{(+)}(\qq) = \frac{M^{(r,\XA)}_{x,Im}(\qq) - M^{(r,\YA)}_{y,Im}(\qq)}{2} \sim \qq_y, \quad \boxed{\textrm{odd}~E}
\end{align}
They span a $\qq$-odd $E$ irrep of $D_3$ group. 

On the other hand, by \cref{eq:appC10}, the real and imaginary parts of $M^{(-)}(\qq)$ can transform separately under $C_{3z}$ and $C_{2x}$, 
\begin{align}   \label{eq:M_rXOYO_m}
    \begin{pmatrix}
        \mrm{Re} M^{(-)}(\qq) \\
        \mrm{Im} M^{(-)}(\qq) \\
    \end{pmatrix} = \begin{pmatrix}
        \mrm{Re} M^{(-)}(C_{3z} \qq) \\
        \mrm{Im} M^{(-)}(C_{3z} \qq) \\
    \end{pmatrix}, \quad \begin{pmatrix}
        1 & 0 \\
        0 & -1 \\ 
    \end{pmatrix} \begin{pmatrix}
        \mrm{Re} M^{(-)}(\qq) \\
        \mrm{Im} M^{(-)}(\qq) \\
    \end{pmatrix} = \begin{pmatrix}
        \mrm{Re} M^{(-)}(C_{2x} \qq) \\
        \mrm{Im} M^{(-)}(C_{2x} \qq) \\
    \end{pmatrix} .
\end{align}
which correspond to an $A_1$ and $A_2$ irrep of $D_3$, respectively. 
By inserting \cref{eq:appC8} and \cref{eq:appC7}, 
\begin{align}  \label{eq:appC_M_A1}
    - \mrm{Re} M^{(-)}(\qq) &= \frac{M^{(r,\XA)}_{y, Im}(\qq) - M^{(r, \YA)}_{x,Im}(\qq)}{2} \sim \mrm{Re} (\qq_x+i\qq_y)^3 \sim \qq_x (\qq_x - \sqrt{3}\qq_y) (\qq_x + \sqrt{3}\qq_y) , \quad \boxed{\textrm{odd}~A_1} \\  \label{eq:appC_M_A2}
    \mrm{Im} M^{(-)}(\qq) &= \frac{M^{(r,\XA)}_{x, Im}(\qq) + M^{(r, \YA)}_{y,Im}(\qq)}{2} \sim \mrm{Im} (\qq_x+i\qq_y)^3 \sim \qq_y (\qq_y - \sqrt{3}\qq_x) (\qq_y + \sqrt{3}\qq_x) , \quad \boxed{\textrm{odd}~A_2}
\end{align}

Now we obtain the Gaussian formulae listed in \cref{tab:rXAYA_irrep}. Put in other words, we will find the coefficients in \cref{eq:appC_M_eE,eq:appC_M_oE,eq:appC_M_A1,eq:appC_M_A2}. We will exemplify with $(r,\XA)$ modes, and the $(r,\YA)$ modes follow similarly. 

First, we separate the $C$ matrix \cref{eq:C_r_XA} to an intra-layer term (superscript ``1'') and an inter-layer term (superscript ``2'')
\begin{align} 
    C^{(r,\XA),1}(\qq;\rr) &= \frac{v_F}{4r_0} (-iq_x \rho_z \sigma_x \tau_0 +iq_y \rho_z \sigma_y \tau_z) \\
    C^{(r,\XA),2}(\qq;\rr) &=  - \frac{w_1}{\theta} \partial_y T(\rr) 
\end{align}
The intra-layer term (``1'') is independent of $\rr$, and evaluating the projected vertex (integral \cref{eq:M}) will be equivalent to finding the Fourier component of a Guassian function at wave vector $\qq$. To be concrete, we evaluate the following diagonal element, 
{\small \begin{align}
    M^{(r,\XA),1}_{1+,1+}(\qq) &= (-i)\frac{v_F}{4r_0} \int \mrm{d}^2\rr ~ e^{i\qq\cdot\rr} ~ \sum_{l} (-1)^{l-1} \left[ w^{(+)*}_{lA,1}(\rr) (q_x+iq_y) w^{(+)}_{lB,1}(\rr) +  w^{(+)*}_{lB,1}(\rr) (q_x-iq_y) w^{(+)}_{lA,1}(\rr) \right] \\\nonumber
    &= (-i)\frac{v_F}{4r_0} \int \mrm{d}^2\rr ~ e^{i\qq\cdot\rr} \cdot 2 \cdot \left[ (- \frac{\alpha_1 \alpha_2}{2\pi}) \frac{x+iy}{\lambda_1 \lambda_2^2} e^{-\left( \frac{1}{2\lambda_1^2} + \frac{1}{2\lambda_2^2} \right)\rr^2} (q_x+iq_y) + (-\frac{\alpha_1\alpha_2}{2\pi}) \frac{x-iy}{\lambda_1 \lambda_2^2} e^{-\left( \frac{1}{2\lambda_1^2} + \frac{1}{2\lambda_2^2} \right)\rr^2} (q_x-iq_y) \right] \\\nonumber
    &= i \frac{v_F}{4\pi r_0} \frac{\alpha_1\alpha_2 }{\lambda_1 \lambda_2^2} \left[ q_x \int\mrm{d}^2\rr e^{i\qq\cdot\rr} (2x) e^{-\frac{\rr^2}{{\lambda'}^2}} - q_y \int\mrm{d}^2\rr e^{i\qq\cdot\rr} (2y) e^{-\frac{\rr^2}{{\lambda'}^2}} \right]
\end{align}}%
In the second line, summation over $l$ contributes to the factor $2$, and to write the Gaussian factor in a more compact form, we have defined $\frac{2}{{\lambda'}^2} = \frac{1}{\lambda_1^2} + \frac{1}{\lambda_2^2}$ in the third line. We note that, $\int\mrm{d}^2\rr e^{i\qq\cdot\rr} e^{-\frac{\rr^2}{{\lambda'}^2}} = \pi {\lambda'}^2 e^{-\frac{1}{4}\qq^2{\lambda'}^2}$, and adding extra factor $x$ or $y$ in the integrand on the left hand side is equivalent to differentiating with $-i\partial_{q_x}$ or $-i\partial_{q_y}$. The results read, 
\begin{align}  \label{eq:C_XA_11_1}
    M^{(r,\XA),1}_{1+,1+}(\qq) = -\frac{v_F}{4 r_0} \frac{\alpha_1\alpha_2 {\lambda'}^4}{\lambda_1 \lambda_2^2}  \left(q_x^2 - q_y^2 \right) e^{-\frac{1}{4} \qq^2{\lambda'}^2 } 
\end{align}

We also evaluate an off-diagonal element, 
{\small \begin{align}
    M^{(r,\XA),1}_{2+,1+}(\qq) &= (-i)\frac{v_F}{4r_0} \int \mrm{d}^2\rr ~ e^{i\qq\cdot\rr} ~ \sum_{l} (-1)^{l-1} \left[ w^{(+)*}_{lA,2}(\rr) (q_x+iq_y) w^{(+)}_{lB,1}(\rr) +  w^{(+)*}_{lB,2}(\rr) (q_x-iq_y) w^{(+)}_{lA,1}(\rr) \right] \\\nonumber
    &= (-i)\frac{v_F}{4r_0} \int \mrm{d}^2\rr ~ e^{i\qq\cdot\rr} ~ \sum_{l} (-1)^{l-1} \left[ w^{(+)}_{lB,1}(-\rr) (q_x+iq_y) w^{(+)}_{lB,1}(\rr) +  w^{(+)}_{lA,1}(-\rr) (q_x-iq_y) w^{(+)}_{lA,1}(\rr) \right] \\\nonumber 
    &= (-i)\frac{v_F}{4r_0} \int \mrm{d}^2\rr ~ e^{i\qq\cdot\rr} \cdot 2i \cdot \left[ \frac{\alpha_2^2}{2\pi} \frac{-(x+iy)^2}{\lambda_2^4} e^{-\frac{\rr^2}{\lambda_2^2}} (q_x+iq_y) + \frac{\alpha_1^2}{2\pi} \frac{1}{\lambda_1^2} e^{-\frac{\rr^2}{\lambda_1^2}} (q_x-iq_y) \right]
\end{align}}%
where the second line used \cref{eq:w_Gauss_2}. In deriving the third line, the phase factors in the wave-functions produce a factor $(e^{i\frac{\pi}{4}\zeta_l})^2 = i\zeta_l = i (-1)^{l-1}$ in \cref{eq:w_Gauss_B}, and the summation over $l$ produces the factor $2$. The final results read, 
\begin{align}  \label{eq:C_XA_21_1}
    M^{(r,\XA),1}_{2+,1+}(\qq) &=  \frac{v_F}{4 r_0} \alpha_1^2 e^{-\frac{1}{4} \qq^2 \lambda_1^2} (q_x-iq_y) + \frac{v_F}{16 r_0} \alpha_2^2 \lambda_2^2  e^{-\frac{1}{4} \qq^2 \lambda_2^2} (q_x+iq_y)^3
\end{align}

The inter-layer (``2'') element relies on $\rr$. We explicitly write out the matrix elements, with basis sorted as $(l\alpha) = (1A),(1B),(2A),(2B)$, 
{\small \begin{align}  \label{eq:appC19}
    C^{(r,\XA), 2}(\qq;\rr)  &= \left( -\frac{w_1}{\theta} \right) \sum_{j=1}^{3} (i\qq_{j,y}) \left[ e^{i\qq_j \cdot \rr} 
    \begin{pmatrix}
        0 & 0 & u_0 & e^{-i\frac{2\pi}{3}(j-1)} \\
        0 & 0 & e^{i\frac{2\pi}{3}(j-1)} & u_0 \\
        0 & 0 & 0 & 0 \\
        0 & 0 & 0 & 0 \\
    \end{pmatrix} - e^{-i\qq_j \cdot \rr} \begin{pmatrix}
        0 & 0 & 0 & 0 \\
        0 & 0 & 0 & 0 \\
        u_0 & e^{-i\frac{2\pi}{3}(j-1)} & 0 & 0 \\
        e^{i\frac{2\pi}{3}(j-1)} & u_0 & 0 & 0 \\
    \end{pmatrix} \right]
\end{align}}%
It can thus be observed that, if inserting \cref{eq:appC19} into the integral \cref{eq:M}, for each $j=1,2,3$, there will be two terms corresponding to the Fourier component of Gaussian functions at wave vector $\qq\pm\qq_j$, respectively, producing factors $e^{-\frac{1}{4}(\qq\pm\qq_j)^2 \lambda_{1,2}^2}$. An exact analytic evaluation can be long and provides little extra physical insights. 
We thus exploit the highly localized nature of $f$-orbitals, and expand \cref{eq:appC19} in powers of $(\qq_j \cdot \rr)$. We will comment on the error of this approximation in the end of the calculation. To the linear order of $(\qq_j \cdot \rr)$, 
\begin{align}   \label{eq:appC20}
    C^{(r,\XA), 2}(\qq;\rr)  &\approx \left( -\frac{w_1 k_\theta}{\theta} \right)  
    \begin{pmatrix}
        0 & 0 & -u_0 \frac{3}{2} k_\theta y & -i\frac{3}{2} - k_\theta \frac{3}{4} (-ix+y) \\
        0 & 0 & -i\frac{3}{2} - k_\theta \frac{3}{4} (ix+y) & -u_0 \frac{3}{2} k_\theta y \\
        -u_0 \frac{3}{2} k_\theta y & i\frac{3}{2} - k_\theta \frac{3}{4} (-ix+y) & 0 & 0 \\
        i\frac{3}{2} - k_\theta \frac{3}{4} (ix+y) & -u_0 \frac{3}{2} k_\theta y  & 0 & 0 \\
    \end{pmatrix} \\\nonumber
    &= \left( \frac{w_1 k_\theta}{\theta} \right) \left[u_0 \frac{3}{2} (k_\theta y) \rho_x - \frac{3}{2}i \rho_y \sigma_x - \frac{3}{4} (k_\theta x) \rho_x \sigma_y - \frac{3}{4} (k_\theta y) \rho_x \sigma_x \right]
\end{align}
Crucially, the first two terms always anti-commute with the chiral operator $\td{S} = \rho_y \sigma_z$ defined in \cref{sec:vertex-G-1}, hence the integral will vanish. Based on \cref{eq:appC20}, we evaluate the diagonal element, 
\begin{align}
    M^{(r,\XA),2}_{1+,1+}(\qq) &= \left(\frac{3}{4} \frac{w_1 k_\theta^2}{\theta} \right) \int\mrm{d}^2\rr e^{i\qq\cdot\rr} \left[ w^{(+)*}_{1A,1}(\rr) w^{(+)}_{2B,1}(\rr) + w^{(+)*}_{2A,1}(\rr) w^{(+)}_{1B,1}(\rr) \right] (-ix+y) \\\nonumber
    & \qquad  + \left( \frac{3}{4} \frac{w_1 k_\theta^2}{\theta} \right) \int\mrm{d}^2\rr e^{i\qq\cdot\rr} \left[ w^{(+)*}_{1B,1}(\rr) w^{(+)}_{2A,1}(\rr) + w^{(+)*}_{2B,1}(\rr) w^{(+)}_{1A,1}(\rr) \right] (ix+y) \\\nonumber
    &= \left(\frac{3}{4} \frac{w_1 k_\theta^2}{\theta} \right) \int\mrm{d}^2\rr e^{i\qq\cdot\rr} \left[\sum_l e^{-i\frac{\pi}{4} \zeta_{\ovl{l}}} (-\zeta_l) e^{i\frac{\pi}{4} \zeta_l} \right] \left[ \frac{\alpha_1 \alpha_2}{2\pi} \frac{x+iy}{\lambda_1 \lambda_2^2}  e^{-\frac{\rr^2}{{\lambda'}^2}} \right] (-ix+y) \\\nonumber
    & \qquad + \left(\frac{3}{4} \frac{w_1 k_\theta^2}{\theta} \right) \int\mrm{d}^2\rr e^{i\qq\cdot\rr} \left[\sum_l (-\zeta_{\ovl{l}}) e^{-i\frac{\pi}{4} \zeta_{\ovl{l}}} e^{i\frac{\pi}{4} \zeta_l} \right] \left[ \frac{\alpha_1 \alpha_2}{2\pi} \frac{x-iy}{\lambda_1 \lambda_2^2}  e^{-\frac{\rr^2}{{\lambda'}^2}} \right] (ix+y) \\\nonumber
    &= \left( -\frac{3}{2} \frac{w_1 k_\theta^2}{\theta} \right) \frac{\alpha_1 \alpha_2}{2\pi} \frac{1}{\lambda_1 \lambda_2^2}  \int\mrm{d}^2\rr e^{i\qq\cdot\rr} \left((x+iy)^2 + (x-iy)^2 \right) e^{-\frac{\rr^2}{{\lambda'}^2}} \\\nonumber
    &= \left( -3 \frac{w_1 k_\theta^2}{\theta} \right) \frac{\alpha_1 \alpha_2}{2\pi} \frac{1}{\lambda_1 \lambda_2^2}  \int\mrm{d}^2\rr e^{i\qq\cdot\rr} \left(x^2 - y^2\right) e^{-\frac{\rr^2}{{\lambda'}^2}} 
\end{align}
Finally we arrive at, 
\begin{align}  \label{eq:C_XA_11_2}
    M^{(r,\XA),2}_{1+,1+}(\qq) = \left( \frac{3}{8} \frac{w_1 k_\theta^2 {\lambda'}^2}{\theta} \right) \alpha_1 \alpha_2 \frac{{\lambda'}^4}{\lambda_1 \lambda_2^2} \left( q_x^2 - q_y^2 \right) e^{-\frac{1}{4} \qq^2 {\lambda'}^2}
\end{align}
We also evaluate the off-diagonal element, 
\begin{align}
    M^{(r,\XA),2}_{2+,1+}(\qq) &= \left(\frac{3}{4} \frac{w_1 k_\theta^2}{\theta} \right) \int\mrm{d}^2\rr e^{i\qq\cdot\rr} \left[ w^{(+)*}_{1A,2}(\rr) w^{(+)}_{2B,1}(\rr) + w^{(+)*}_{2A,2}(\rr) w^{(+)}_{1B,1}(\rr) \right] (-ix+y) \\\nonumber
    & \qquad  + \left( \frac{3}{4} \frac{w_1 k_\theta^2}{\theta} \right) \int\mrm{d}^2\rr e^{i\qq\cdot\rr} \left[ w^{(+)*}_{1B,2}(\rr) w^{(+)}_{2A,1}(\rr) + w^{(+)*}_{2B,2}(\rr) w^{(+)}_{1A,1}(\rr) \right] (ix+y) \\\nonumber
    &= \left(\frac{3}{4} \frac{w_1 k_\theta^2}{\theta} \right) \int\mrm{d}^2\rr e^{i\qq\cdot\rr} \left[ w^{(+)}_{1B,1}(-\rr) w^{(+)}_{2B,1}(\rr) + w^{(+)}_{2B,1}(-\rr) w^{(+)}_{1B,1}(\rr) \right] (-ix+y) \\\nonumber
    & \qquad  + \left( \frac{3}{4} \frac{w_1 k_\theta^2}{\theta} \right) \int\mrm{d}^2\rr e^{i\qq\cdot\rr} \left[ w^{(+)}_{1A,1}(-\rr) w^{(+)}_{2A,1}(\rr) + w^{(+)}_{2A,1}(-\rr) w^{(+)}_{1A,1}(\rr) \right] (ix+y) \\\nonumber
    &=  \left(\frac{3}{4} \frac{w_1 k_\theta^2}{\theta} \right) \int\mrm{d}^2\rr e^{i\qq\cdot\rr} \cdot 2 \cdot \left( \left[ \frac{\alpha_2^2}{2\pi} \frac{(x+iy)^2}{\lambda_2^4} e^{-\frac{\rr^2}{\lambda_2^2}} \right] (-ix+y) +  \left[ \frac{\alpha_1^2}{2\pi} \frac{1}{\lambda_1^2} e^{-\frac{\rr^2}{\lambda_1^2}} \right] (ix+y) \right) 
\end{align}
where again \cref{eq:w_Gauss_2} is exploited in the second equation, and we arrive at
\begin{align} \label{eq:C_XA_21_2}
    M^{(r,\XA),2}_{2+,1+}(\qq) =\left( - \frac{3 w_1 k_\theta^2}{8 \theta} \alpha_1^2 \lambda_1^2 \right) e^{-\frac{1}{4}\qq^2 \lambda_1^2} (q_x - iq_y) +  \left( - \frac{3 w_1 k_\theta^2}{32 \theta} \alpha_2^2 \lambda_2^4 \right) e^{-\frac{1}{4}\qq^2 \lambda_2^2} (q_x + iq_y)^3
\end{align}

Referring to the parametrization \cref{eq:M_para_G}, the diagonal element is expressed as
\begin{align}  \label{eq:appC29}
    M^{(r, \XA)}_{1+,1+}(\qq) = M^{(r,\XA)}_{z,Re}(\qq) 
\end{align}
For \cref{eq:appC29}, its left hand side is given by summing up \cref{eq:C_XA_11_1,eq:C_XA_11_2}, while the right hand side must conform to the symmetry constraint \cref{eq:appC_M_eE}. By comparing the two sides, the 4th row in \cref{tab:rXAYA_irrep} can be immediately obtained. 

On the other hand, the off-diagonal element is expressed as
\begin{align} \label{eq:appC30}
    M^{(r, \XA)}_{2+,1+}(\qq) &= i \left( M^{(r,\XA)}_{x,Im}(\qq) + i M^{(r,\XA)}_{y,Im}(\qq) \right) \\\nonumber
    &= i \left( \frac{M^{(r,\XA)}_{x,Im}(\qq) + M^{(r,\YA)}_{y,Im}(\qq)}{2} + \frac{M^{(r,\XA)}_{x,Im}(\qq) - M^{(r,\YA)}_{y,Im}(\qq)}{2} \right) \\\nonumber
    & \qquad  - \left( \frac{M^{(r,\XA)}_{y,Im}(\qq) + M^{(r,\YA)}_{x,Im}(\qq)}{2} + \frac{M^{(r,\XA)}_{y,Im}(\qq) - M^{(r,\YA)}_{x,Im}(\qq)}{2} \right) 
\end{align}
where in the second equation we find the decomposition using the irrep functions \cref{eq:appC_M_A1,eq:appC_M_A2,eq:appC_M_oE}. The left hand side of \cref{eq:appC30} is given by summing up \cref{eq:C_XA_21_1,eq:C_XA_21_2}. By comparing the $O(|\qq|)$ terms with \cref{eq:appC_M_oE}, the 3rd row of \cref{tab:rXAYA_irrep} can be obtained. Comparing the real and imaginary parts of the $O(|\qq|^3)$ terms with \cref{eq:appC_M_A1,eq:appC_M_A2}, the 1st and 2nd rows of \cref{tab:rXAYA_irrep} are obtained. 

In general, including higher order expansions $(\qq_j \cdot \rr)^N$ not only brings about the higher order $\qq$-dependence (from $O(\qq^{N})$ to $O(\qq^{N+2})$), but also adds corrections to the lower order coefficients. For example, $\int\mrm{d}^2\rr e^{i\qq\cdot\rr} \rr^2 e^{-\frac{\rr^2}{\lambda^2}} = \pi \lambda^4 e^{-\frac{1}{4}\qq^2\lambda^2} (1 - \frac{\qq^2\lambda^2}{4})$. Therefore, expanding to the linear order of $(\qq_j\cdot\rr)$ does not necessarily produce accurate $O(|\qq|)$ and $O(|\qq|^2)$ coefficients. 
However, by dimension analysis, after integral over $\rr$, these corrections contain a factor proportional to $(k_\theta \lambda_{1,2})^N \sim (\frac{\lambda_{1,2}}{a_M})^N$, hence for the same $O(|\qq|^M)$ ($M < N$) order coefficient, the larger $N$ corrections should be un-important when $\frac{\lambda_{1,2}}{a_M} \ll 1$. Therefore, we argue the above approximation is valid because the $f$-orbitals are indeed well localized. The statement can be verified by comparing the approximate Gaussian results with numeric results in \cref{tab:rXAYA_irrep}.

\subsection{\texorpdfstring{$(r, \XO/\YO)$}{(r, XO/YO)}}    \label{app:vertex-rXOYO}

Since the optical modes transform under $C_{2z}T$ as $(C_{2z}T) u_{\qq,\chi} (C_{2z}T)^{-1} = u_{\qq,\chi}$ by \cref{eq:C6z_rXOYO,eq:T_u}, without the minus sign as opposed to the acoustic modes, $M^{(r, \XO/\YO)}_{+, +}(\qq)$ should commute with $\sigma^x K$, namely, only $M^{(r, \XO/\YO)}_{z, Im}(\qq)$ and $M^{(r, \XO/\YO)}_{0/x/y, Re}(\qq)$ can survive for arbitrary $\qq$. 
By the Hermiticity requirement explained below \cref{eq:M_para_G}, $M^{(r, \XO/\YO)}_{z, Im}(\qq)$ is odd in $\qq$, while $M^{(r, \XO/\YO)}_{0/x/y, Re}(\qq)$ are even. Moreover, there is $D(P) C^{(r, \XO/\YO)}(\qq; \rr) D^\dagger(P) = - C^{(r, \XO/\YO)}(-\qq; -\rr)$ by \cref{eq:DP,eq:C_r_XO,eq:C_r_YO}. Consequently, 
\begin{align}
    \langle w^{(+)}_{\beta}(\rr) | C^{(\chi)} (\qq;\rr) | w^{(+)}_{\beta'}(\rr) \rangle 
    = (-1)^{\beta+\beta'-1} \langle w^{(+)}_{\beta}(-\rr) | C^{(\chi)} (-\qq;-\rr) | w^{(+)}_{\beta'}(-\rr) \rangle , \qquad \chi = (r, \XO/\YO)
\end{align}
which thus dictates $\beta$-diagonal elements and $\beta$-off-diagonal elements to be odd and even in $\qq$, respectively. Combined with the parity constraints, only $M^{(r, \XO/\YO)}_{z, Im}(\qq)$ and $M^{(r, \XO/\YO)}_{x/y, Re}(\qq)$ can survive. 

By squaring \cref{eq:C6z_rXOYO}, $C_{3z}$ acts as
\begin{align} \label{eq:C3z_rXOYO}
    C_{3z} \begin{pmatrix}
        u_{\qq,r,\XO} \\
        u_{\qq,r,\YO} \\
    \end{pmatrix} C_{3z}^{-1} = \begin{pmatrix}
        -\frac{1}{2} & \frac{\sqrt{3}}{2} \\
        -\frac{\sqrt{3}}{2} & -\frac{1}{2} \\
    \end{pmatrix} \begin{pmatrix}
        u_{C_{3z}\qq,r,\XO} \\
        u_{C_{3z}\qq,r,\YO} \\
    \end{pmatrix} 
\end{align}
By conjugating $\hat{H}_{\rm epc}$ with $C_{3z}$, the invariance dictates (using \cref{eq:DfC3z,eq:C3z_rXOYO})
\begin{align} \label{eq:appC34}
    e^{i\frac{2\pi}{3} \sigma_z} \left[ -\frac{1}{2} M^{(r, \XO)}(\qq) - \frac{\sqrt{3}}{2} M^{(r, \YO)}(\qq) \right] e^{-i\frac{2\pi}{3} \sigma_z} &= M^{(r, \XO)}(C_{3z}\qq) \\ 
    e^{i\frac{2\pi}{3} \sigma_z} \left[ \frac{\sqrt{3}}{2} M^{(r, \XO)}(\qq) - \frac{1}{2} M^{(r, \YO)}(\qq) \right] e^{-i\frac{2\pi}{3} \sigma_z} &= M^{(r, \YO)}(C_{3z}\qq) 
\end{align}
and with $C_{2x}$, (using \cref{eq:DfC2x,eq:C2x_rXOYO})
\begin{align} 
    \sigma_x \left[  M^{(r, \XO)}(\qq) \right] \sigma_x &= M^{(r, \XO)}(C_{2x}\qq) \\ \label{eq:appC37}
    \sigma_x \left[ - M^{(r, \YO)}(\qq) \right] \sigma_x &= M^{(r, \YO)}(C_{2x}\qq)
\end{align}

The $\beta$-diagonal elements thus need to observe
\begin{align}
    \begin{pmatrix}
        -\frac{1}{2} & -\frac{\sqrt{3}}{2} \\
        \frac{\sqrt{3}}{2} & -\frac{1}{2} \\
    \end{pmatrix} \begin{pmatrix}
        M^{(r, \XO)}_{z, Im}(\qq) \\
        M^{(r, \YO)}_{z, Im}(\qq) \\
    \end{pmatrix} = \begin{pmatrix}
        M^{(r, \XO)}_{z, Im}(C_{3z}\qq) \\
        M^{(r, \YO)}_{z, Im}(C_{3z}\qq) \\
    \end{pmatrix}  , \quad
    \begin{pmatrix}
        -1 & 0 \\
        0 & 1 \\
    \end{pmatrix} \begin{pmatrix}
        M^{(r, \XO)}_{z, Im}(\qq) \\
        M^{(r, \YO)}_{z, Im}(\qq)
    \end{pmatrix} = \begin{pmatrix}
        M^{(r, \XO)}_{z, Im}(C_{2x}\qq) \\
        M^{(r, \YO)}_{z, Im}(C_{2x}\qq)
    \end{pmatrix}
\end{align}
which span an $E$ irrep of $D_3$. As odd functions, the leading $\qq$-dependence can be of order $O(\qq)$, and proportional to
\begin{align}
    M^{(r, \XO)}_{z, Im}(\qq) \sim \qq_y, \qquad M^{(r, \YO)}_{z, Im}(\qq) \sim -\qq_x, \qquad \boxed{\textrm{odd}~E}
\end{align} 

For $\beta$-off-diagonal elements, by definition of the parametrization \cref{eq:M_para_G},
\begin{align}
    M^{(\chi)}_{2+,1+}(\qq) &= \left( M^{(\chi)}_{x, Re}(\qq) + i M^{(\chi)}_{y, Re}(\qq) \right) \\
    M^{(\chi)}_{1+,2+}(\qq) &= \left( M^{(\chi)}_{x, Re}(\qq) - i M^{(\chi)}_{y, Re}(\qq) \right)
\end{align}
hence $M^{(\chi)}_{1+,2+}(\qq) = M^{(\chi)*}_{2+,1+}(\qq)$, and we focus on $M^{(\chi)}_{1+,2+}(\qq)$. \cref{eq:appC34} to \cref{eq:appC37} are now transcribed as 
\begin{align} 
    e^{-i\frac{4\pi}{3}} \begin{pmatrix}
        - \frac{1}{2} & - \frac{\sqrt{3}}{2} \\
        \frac{\sqrt{3}}{2} & - \frac{1}{2} \\
    \end{pmatrix}
    \begin{pmatrix}
        M^{(r, \XO)}_{2+, 1+}(\qq) \\
        M^{(r, \YO)}_{2+, 1+}(\qq) \\
    \end{pmatrix} = \begin{pmatrix}
        M^{(r, \XO)}_{2+, 1+}(C_{3z}\qq) \\
        M^{(r, \YO)}_{2+, 1+}(C_{3z}\qq) \\
    \end{pmatrix}, \quad \begin{pmatrix}
        1 & 0 \\
        0 & -1 \\
    \end{pmatrix} \begin{pmatrix}
        M^{(r, \XO)}_{2+, 1+}(\qq) \\
        M^{(r, \YO)}_{2+, 1+}(\qq) \\
    \end{pmatrix} = \begin{pmatrix}
        M^{(r, \XO)}_{2+, 1+}(C_{2x}\qq) \\
        M^{(r, \YO)}_{2+, 1+}(C_{2x}\qq) \\ 
    \end{pmatrix}^*
\end{align}
which is identical to \cref{eq:M_rXAYA_B8}. Consequently, by defining $M^{(\pm)}(\qq) = \frac{M^{(r, \XO)}_{2+,1+}(\qq) \pm i M^{(r, \YO)}_{2+,1+}(\qq)}{2}$, $M^{(\pm)}(\qq)$ must obey the identical equations as \cref{eq:appC9}, respectively. 
The only difference is that, here, $M^{\pm}(\qq)$ should be even in $\qq$. Therefore, $M^{(+)}(\qq) \sim (\qq_x+i\qq_y)^2$, namely, its real and imaginary parts span an $E$ irrep with $O(|\qq|^2)$ dependence at the leading order, 
\begin{align}
    \mrm{Re} M^{(+)}(\qq) = \frac{M^{(r,\XO)}_{x,Re}(\qq) - M^{(r,\YO)}_{y,Re}(\qq)}{2} \sim (\qq_x^2 - \qq_y^2), \quad \mrm{Im} M^{(+)}(\qq) = \frac{M^{(r,\XO)}_{y,Re}(\qq) + M^{(r,\YO)}_{x,Re}(\qq)}{2} \sim (2\qq_x \qq_y), \quad \boxed{\textrm{even}~E}
\end{align}
On the other hand, $\mrm{Re}M^{(-)}(\qq)$ will form a $\qq$-even $A_1$ irrep, 
\begin{align}
    \mrm{Re}M^{(-)}(\qq) = \frac{M^{(r,\XO)}_{x,Re}(\qq) + M^{(r,\YO)}_{y,Re}(\qq)}{2} \sim 1 , \qquad \boxed{\textrm{even}~A_1}
\end{align}
while $\mrm{Im}M^{(-)}(\qq)$ will form a $\qq$-even $A_2$ irrep. It can be verified that the leading $\qq$-dependence of such irrep starts at $O(|\qq|^6)$, as $q_x (q_x-\sqrt{3}q_y) (q_x+\sqrt{3}q_y) q_y (q_y-\sqrt{3}q_x) (q_y+\sqrt{3}q_x)$. 

Evaluating the Gaussian formulae for $M^{(r,\XO/\YO)}(\qq)$ highly resembles the procedures in the last subsection, and we leave out calculation details. Here, since $C^{(r,\XO/\YO)}$ does not contain $\rr$-dependence, the results can be obtained without applying approximations. They are summarized in \cref{tab:rXOYO_irrep}.

\subsection{\texorpdfstring{$(l, \Aone/\Bone)$}{(l, A1/B1)}}   \label{app:vertex-K}

The parametrization for $M^{(\chi@K)}_{+, -}(\qq)$ takes the form of Eq. (\ref{eq:M_para_K}), and as stated in the main-text, $M^{(l, \Bone)}_{+, -}(\qq) = i M^{(l, \Aone)}_{+, -}(\qq)$, while dictated by $C_{2z}T$, only $M^{(l,\Aone)}_{0/x, Re}(\qq)$ and $M^{(l,\Aone)}_{z, Im}(\qq)$ can survive for generic $\qq$. 

To apply the $C_{3z}$ symmetry, we re-write the terms in Eq. (\ref{eq:Hepc}) involving the $K$-phonons as
\begin{align}   \label{eq:Hepc_2}
    \hat{H}_{\rm epc} = \frac{1}{\sqrt{N_g}} \sum_{\RR,\eta,s} \sum_{l} \sum_{|\qq| \lesssim \Lambda} e^{i\qq\cdot\RR} \times  \sqrt{2} u_{\qq+\KK_l,l,-K} \times f^\dagger_{\RR\beta+ s} M^{(l, \Aone)}_{\beta+,\beta'-}(\qq) f_{\RR\beta'- s} + \mrm{H.c.}
\end{align}
where \cref{eq:uK_eta,eq:uK_eta_app} are exploited. We note again that $u_{\qq,l,-K}$ transforms according to \cref{eq:C3z_eta} under $C_{3z}$ rotation. Therefore, conjugating \cref{eq:Hepc_2} with $C_{3z}$, the symmetry requires 
\begin{align}
    e^{i\frac{2\pi}{3} \sigma_z} M_{+,-}^{(l, \Aone)}(\qq) e^{i\frac{2\pi}{3} \sigma_z} = M_{+,-}^{(l, \Aone)}(C_{3z}\qq + C_{3z}\KK_l - \KK_l) 
\end{align}
which dictates $M^{(l, \Aone)}_{+,-}(\qq)$ to be $C_{3z}$-symmetric around $-\KK_l$. 
Decomposed with the surviving parameters, for the $\beta$-diagonal elements, the independent restrictions read, 
\begin{align}
    e^{i\frac{4\pi}{3}} M^{(l,\Aone)}_{1+,1-}(\qq) = M^{(l,\Aone)}_{1+,1-}(C_{3z}\qq+C_{3z}\KK_l-\KK_l)
\end{align}
hence $M^{(l,\Aone)}_{1+,1-}(\qq) \sim (\qq+\KK_l)_x - i(\qq+\KK_l)_y$ with generically a \textit{complex} coefficient. Note that $M^{(l,\Aone)}_{1+,1-}(\qq) = M^{(l,\Aone)}_{0,Re}(\qq) + i M^{(l,\Aone)}_{z,Im}(\qq)$ by definition of \cref{eq:M_para_K}. 

As for the $\beta$-off-diagonal elements, there is $M_{2+,1-}^{(l,\Aone)}(\qq) = M_{x,Re}^{(l,\Aone)}(\qq)$ by definition of \cref{eq:M_para_K}, and
\begin{align}
    M^{(l, \Aone)}_{x, Re}(\qq) = M^{(l, \Aone)}_{x, Re}(C_{3z}\qq + C_{3z}\KK_l - \KK_l) 
\end{align}
and $M^{(l, \Aone)}_{x, Re}(\qq)$ can approach a finite constant around $\qq\to -\KK_l$, in contrast with $M^{(l,\Aone)}_{0,Re}(\qq)$ and $M^{(l,\Aone)}_{z,Im}(\qq)$. 

Evaluating the Gaussian formulae for $M_{+,-}^{(l,\Aone)}(\qq)$ also highly resembles that for $M^{(r,\XA/\YA),1}(\qq)$ and $M^{(r,\XO/\YO)}(\qq)$. The major difference comes from the extra phase factor $e^{i\KK_l \cdot \rr}$ in \cref{eq:C_l_A1}. This factor will always combine with the $e^{i\qq\cdot\rr}$ phase, so that the final results always rely on $\qq+\KK_l$ as a whole. The results are summarized in \cref{tab:lA1_irrep}.

\section{Inter-\texorpdfstring{$f$}{f}-site interactions}  \label{app:inter}  

\begin{figure}[tb]
    \centering
    \includegraphics[width = 0.7\linewidth]{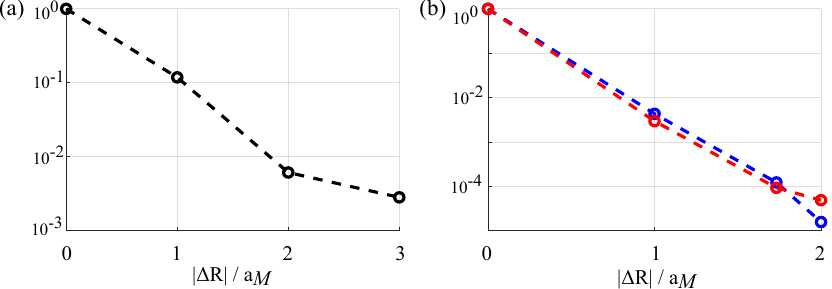}
    \caption{\label{fig:intersite} (a) The decay of the inter-site EPC element with $\Delta\RR$, $| M^{(r,\YO)}_{1+,2+}(0;\Delta\RR) | / | M^{(r,\YO)}_{1+,2+}(0;0) |$, calculated from the numeric Wannier functions, where $\Delta\RR$ is chosen to lie in the $y$-direction, parallel to $(\mbf{a}_1 + \mbf{a}_2)$ (see \cref{fig:MBZ}(c)). (b) The decay of the phonon-mediated inter-site interactions with $\Delta\RR$, $|| \td{V}^\eta(\Delta\RR) || / ||\td{V}^\eta(0)||$ (blue) and $||{V}^{\eta\eta}(\Delta\RR) || / ||{V}^{\eta\eta}(0)||$ (red), calculated from the numeric Wannier functions. }
\end{figure}

\subsection{Inter-site EPC vertex}  \label{app:inter-vertex}

The general form of the EPC vertex on the local orbitals reads, 
\begin{align}
    \hat{H}_{\rm epc} &= \frac{1}{\sqrt{N_g}} \sum_{\RR,\RR' } \sum_{\eta, \eta',\beta, \beta'} \sum_{s} \sum_{|\qq| \lesssim \Lambda} \sum_{\chi} e^{i\qq\cdot(\RR-\RR')} \times  u_{\qq, \chi} \times f^\dagger_{\RR \beta\eta s} M^{(\chi)}_{\beta\eta, \beta'\eta'}(\qq;\RR-\RR') f_{\RR'\beta'\eta's} 
\end{align}
where 
\begin{align} \label{eq:M_RR2}
    M^{(\chi)}_{\beta\eta, \beta'\eta'}(\qq; \RR-\RR') &= \int \mrm{d}^2 \rr ~ e^{i\qq\cdot\rr} ~ w^{(\eta)*}_{l\alpha,\beta}(\rr-\RR) \left[ C^{(\chi)}(\qq;\rr) \right]_{l\alpha\eta,l'\alpha'\eta'} w^{(\eta')}_{l'\alpha',\beta'}(\rr-\RR') 
\end{align}
only depends on the relative position $\RR - \RR'$. 
In the main text, we have focused on the $\RR=\RR'$ (on-site) vertices. Now we show that $\RR\not=\RR'$ (inter-site) vertices are indeed negligible. 

For all phonon modes $\chi$, $C^{(\chi)}(\qq;\rr)$ is either independent of $\rr$, or only varies slowly on the moir\'e scale. The size of the integral is thus controlled by the overlap of two Wannier functions located at $\RR$ and $\RR'$, which decays exponentially with $|\RR-\RR'|$. 
For concreteness, let us consider a phonon mode where $C^{(\chi)}(\qq;\rr)$ does not depend on $\rr$, \textit{e.g.} $(r,\XO/\YO)$, and two Guassian Wannier functions (\cref{eq:w_Gauss_A,eq:w_Gauss_B}) located at $0$ and ${\Delta\RR}$. If roughly treating $\lambda_1 = \lambda_2 = \lambda'$, and focusing on the Gaussian factors, there will be 
\begin{align} \label{eq:vertex_Gauss_decay}
    e^{-\frac{\rr^2}{2\lambda^{\prime 2}}} e^{-\frac{(\rr-\Delta\RR)^2}{2\lambda^{\prime 2}}} = e^{-\frac{(\rr-\Delta\RR/2)^2}{\lambda^{\prime 2}}} e^{-\frac{\Delta\RR^2}{4\lambda^{\prime 2}}} 
\end{align}
The first factor simply describes the integrand peaks at $\Delta\RR/2$, while the second factor $e^{-\frac{\Delta\RR^2}{4\lambda^{\prime 2}}}$ describes the Gaussian decay of \cref{eq:M_RR2} with $|\Delta\RR|$. For the nearest-neighboring $f$-orbitals, it can be estiamted that $e^{-\frac{a_M^2}{4{\lambda'}^2}} < 10^{-3}$, which is extremely small. 
With the polynomials before the Guassian factors taken into consideration, by dimension analysis, extra factors proportional to $\left( \frac{a_M}{\lambda'} \right)^n$ can arise, which narrows the difference. 
We directly calculate the size of element $M^{(r,\YO)}_{1+,2+}(0;\Delta\RR)$ in \cref{fig:intersite}(a) with the numeric Wannier functions, and find the nearest-neighbor EPC vertex is typically $10^{-1}$ of the on-site size. 
Since $M^{(r,\YO)}_{1+,2+}(0;\Delta\RR)$ scatters an electron from site $\RR'$ to $\RR'+\Delta\RR$, for such an EPC vertex to affect the low-energy subspace where the $f$-occupation number at each site $N_f$ is fixed, it has to be combined with a vertex $M^{(r,\YO)}_{1+,2+}(0;-\Delta\RR)$, which can scatter another electron from $\RR'+\Delta\RR$ back to $\RR'$. Such phonon-mediated pair swapping interactions between nearest-neighbor sites are thus of the order $10^{-2}$ of the on-site interactions, which is negligible for the experimental probed temperature, \textit{e.g.}, 1K$\sim$0.1meV. 

\subsection{Inter-site interaction}   \label{app:inter-int}

After showing that inter-site EPC vertices are negligibly weak, now we examine the inter-site interactions brought by two on-site EPC vertices and a phonon line that connects two sites, namely, the terms with $\RR\not=\RR'$ in \cref{eq:Sint}. 
It suffices to focus on the zero-frequency part, given by 
\begin{align}   \label{eq:Vee_inter}
    V^{\eta'\eta}_{{\beta'}{\alpha'}, {\beta}{\alpha}}(\Delta\RR) = - \frac{1}{N_g m_0} \sum_{\chi@\Gamma} \sum_{|\qq|<\Lambda} e^{i\qq\cdot\Delta\RR} \frac{M^{(\chi@\Gamma)}_{\beta'\eta',\beta\eta'}(-\qq) M^{(\chi@\Gamma)}_{\alpha'\eta,\alpha\eta}(\qq)} {\omega^2_{\qq,\chi}}
\end{align}
\begin{align}   \label{eq:Ve_inter}
    \td{V}^{\eta}_{{\beta'}{\alpha'},{\beta}{\alpha}}(\Delta\RR) = - \frac{1}{N_g m_0} \sum_{\chi@K} \sum_{|\qq|<\Lambda} e^{i\qq\cdot\Delta\RR} \frac{M^{(\chi@K)}_{\beta'\eta,\beta\ovl{\eta}}(-\qq) M^{(\chi@K)}_{\alpha'\ovl{\eta},\alpha\eta}(\qq)} {\omega^2_{\qq,\chi}}
\end{align}
which are generalizations for the on-site expressions \cref{eq:Vee,eq:Ve}. 
As discussed in \cref{sec:vertex}, the function $M^{(\chi)}_{\beta_1\eta_1,\beta'_1\eta_1'}(\qq) D^{(\chi)}(\qq,0) M^{(\chi)}_{\beta_2\eta_2,\beta_2'\eta_2'}(\qq)$ can be modeled by Gaussian packets with width $e^{-\frac{1}{2} |\qq|^2 {\lambda'}^2}$. Summing over $\qq$ in \cref{eq:Vee_inter,eq:Ve_inter} is equivalent to finding the Fourier transformation of such Gaussian packets, which decays quickly as $e^{-\frac{1}{2} (\frac{|\Delta\RR|}{\lambda'})^2}$. 
We calculate $||V^{\eta\eta}(\Delta\RR)||$ and $||\td{V}^{\eta}(\Delta\RR)||$ with the numeric Wannier functions, where $||\cdots||$ stands for Frobenius norm, and display the results in \cref{fig:intersite}(b). The nearest-neighboring interactions are also $10^{-2}$ of the on-site values. 

\section{Three and four electron states of \texorpdfstring{$\hat{H}_{\rm A}$}{HA}}   \label{app:multi}

\begin{table*}[tb]
    \centering
    \begin{tabular}{c|c|c|c|c|c}
    \hline\hline
        $|N_v|$ & $|L|$ & $[\rho, j]$ ($\times$ fold) & rep. w.f. & degeneracy & energy \\
    \hline
        3 & 1 & $[E_1 + E_2, \frac{1}{2}]$ & $f^\dagger_{1+\up} f^\dagger_{1+\down} f^\dagger_{2+\up}$ & 8 & $J_{\rm a} + J_{\rm b}$ \\
    \hline
        1 & 3 & $[A_1 + A_2 + B_1 + B_2, \frac{1}{2}]$ & $f^\dagger_{1+\up} f^\dagger_{1+\down} f^\dagger_{2-\up}$ & 8 & $J_{\rm a} + J_{\rm d}$ \\
    \hline
        \multirow{2}{*}{1} & \multirow{2}{*}{1} & $[E_1 + E_2, \frac{1}{2}]$ ($\times 3$) & \cref{eq:3e_gs} & 8 ($\times 3$) & \cref{eq:3e_H} \\
    \cline{3-6}
         &  & $[E_1 + E_2, \frac{3}{2}]$ & $f^\dagger_{1+\up} f^\dagger_{1-\up} f^\dagger_{2+\up}$ & 16 &  $J_{\rm a} + J_{\rm b} + J_{\rm d} + J_{\rm e}$ \\
    \hline\hline
    \end{tabular}
    \caption{\label{tab:3e} Three-electron states of $\hat{H}_{\rm A}$. ``$\times$ fold'' indicates how many times the same $[\rho, j]$ irrep appears. }

    \centering
    \begin{tabular}{c|c|c|c|c|c}
    \hline\hline
        $|N_v|$ & $|L|$ & $[\rho, j]$ ($\times$ fold) & rep. w.f. & degeneracy & energy \\
    \hline
        4 & 0 & $[A_1 + B_1, 0]$ & $f^\dagger_{1+\up} f^\dagger_{1+\down} f^\dagger_{2+\up} f^\dagger_{2+\down}$ & 2 & $2(J_{\rm a} + J_{\rm b})$ \\
    \hline
        \multirow{2}{*}{2} & \multirow{2}{*}{2} & $[E_1 + E_2, 0]$ & $\sqrt{\frac{1}{2}} f^\dagger_{1+\up} f^\dagger_{1+\down} (f^\dagger_{2+\up} f^\dagger_{2-\down} - f^\dagger_{2+\down} f^\dagger_{2-\up})$ & 4 & $2J_{\rm a} + J_{\rm b} + J_{\rm d} - J_{\rm e}$ \\
    \cline{3-6}
         &  & $[E_1 + E_2, 1]$ & $f^\dagger_{1+\up} f^\dagger_{1+\down} f^\dagger_{2+\up} f^\dagger_{2-\up}$ & 12 & $2J_{\rm a} + J_{\rm b} + J_{\rm d} + J_{\rm e}$ \\
    \hline
        \multirow{4}{*}{2} & \multirow{4}{*}{0} & $[A_1 + B_1, 0]$ & $\frac{1}{2} f^\dagger_{1+\up} f^\dagger_{1+\down} (f^\dagger_{1-\up} f^\dagger_{2+\down} - f^\dagger_{1-\down} f^\dagger_{2+\up}) + [C_{2x}]$ & 2 & $2J_{\rm b} - 3J_{\rm d} + J_{\rm e}$ \\
    \cline{3-6}
         &  & $[A_2 + B_2, 0]$ & $\frac{1}{2} f^\dagger_{1+\up} f^\dagger_{1+\down} (f^\dagger_{1-\up} f^\dagger_{2+\down} - f^\dagger_{1-\down} f^\dagger_{2+\up}) - [C_{2x}]$ & 2 & $J_{\rm d} + J_{\rm e}$ \\
    \cline{3-6}
         &  & $[A_1 + B_1, 1]$ & $\sqrt{\frac{1}{2}} f^\dagger_{1+\up} f^\dagger_{1+\down} f^\dagger_{1-\up} f^\dagger_{2+\up} + [C_{2x}]$ & 6 & $2 J_{\rm b} + J_{\rm d} + J_{\rm e}$ \\
    \cline{3-6}
         &  & $[A_2 + B_2, 1]$ & $\sqrt{\frac{1}{2}} f^\dagger_{1+\up} f^\dagger_{1+\down} f^\dagger_{1-\up} f^\dagger_{2+\up} - [C_{2x}]$ & 6 & $J_{\rm d} + J_{\rm e}$ \\
    \hline
        0 & 4 & $[E_2, 0]$ & $f^\dagger_{1+\up} f^\dagger_{1+\down} f^\dagger_{2-\up} f^\dagger_{2-\down}$ & 2 & $2(J_{\rm a} + J_{\rm d})$  \\
    \hline
        \multirow{4}{*}{0} & \multirow{4}{*}{2} & $[E_2, 0]$ & $\frac{1}{2} f^\dagger_{1+\up} f^\dagger_{1+\down} (f^\dagger_{1-\up} f^\dagger_{2-\down} - f^\dagger_{1-\down} f^\dagger_{2-\up}) + [C_{2z}]$ & 2 & $-3J_{\rm b}+2J_{\rm d}+J_{\rm e}$ \\
    \cline{3-6}
         &  & $[E_1, 0]$ & $\frac{1}{2} f^\dagger_{1+\up} f^\dagger_{1+\down} (f^\dagger_{1-\up} f^\dagger_{2-\down} - f^\dagger_{1-\down} f^\dagger_{2-\up}) - [C_{2z}]$ & 2 & $J_{\rm b} + J_{\rm e}$ \\
    \cline{3-6}
         &  & $[E_2, 1]$ & $\sqrt{\frac{1}{2}} f^\dagger_{1+\up} f^\dagger_{1+\down} f^\dagger_{1-\up} f^\dagger_{2-\up} + [C_{2z}]$ & 6 & $J_{\rm b} + 2J_{\rm d}+ J_{\rm e}$ \\
    \cline{3-6}
         &  & $[E_1, 1]$ & $\sqrt{\frac{1}{2}} f^\dagger_{1+\up} f^\dagger_{1+\down} f^\dagger_{1-\up} f^\dagger_{2-\up} - [C_{2z}]$ & 6 & $J_{\rm b} + J_{\rm e}$ \\
    \hline
        \multirow{5}{*}{0} & \multirow{5}{*}{0} & $[A_1, 0]$ ($\times 3$) & \cref{eq:4e_gs} & 1 ($\times 3$) & \cref{eq:4e_H} \\
    \cline{3-6}
         &  & $[B_2, 0]$ & $\sqrt{\frac{1}{2}} f^\dagger_{1+\up} f^\dagger_{1+\down} f^\dagger_{1-\up} f^\dagger_{1-\down} - [C_{2x}]$ & 1 & $-6 J_{\rm a} + 2J_{\rm e}$ \\
    \cline{3-6}
         &  & $[A_2, 1]$ & $\frac{1}{2} f^\dagger_{1+\up} f^\dagger_{1-\up} (f^\dagger_{2+\up} f^\dagger_{2-\down} - f^\dagger_{2+\down} f^\dagger_{2-\up}) - [C_{2x}]$ & 3 & $2 (J_{\rm a} + J_{\rm d})$ \\
    \cline{3-6}
         &  & $[B_1, 1]$ & $\frac{1}{2} f^\dagger_{1+\up} f^\dagger_{1-\up} (f^\dagger_{2+\up} f^\dagger_{2-\down} - f^\dagger_{2+\down} f^\dagger_{2-\up}) + [C_{2x}]$ & 3 & $2 (J_{\rm a} + J_{\rm b})$ \\
    \cline{3-6}
         &  & $[B_2, 1]$ & $\frac{1}{2} f^\dagger_{1+\up} f^\dagger_{1-\up} (f^\dagger_{2+\up} f^\dagger_{2-\down} + f^\dagger_{2+\down} f^\dagger_{2-\up}) - [C_{2x}]$ & 3 & $2 (J_{\rm a} + J_{\rm e})$ \\
    \cline{3-6}
         &  & $[A_1, 2]$ & $f^\dagger_{1+\up} f^\dagger_{1-\up} f^\dagger_{2+\up} f^\dagger_{2-\up}$ & 5 & $2(J_{\rm a} + J_{\rm b} + J_{\rm d} + J_{\rm e})$ \\
    \hline\hline
    \end{tabular}
    \caption{\label{tab:4e} Four-electron states of $\hat{H}_{\rm A}$. $[C_{2x}]$ or $[C_{2z}]$ indicates a term that differs by a conjugation of $C_{2x}$ or $C_{2z}$ than its former term. }
\end{table*}

\subsection{Three-electron states}  \label{app:multi-3e}

We have tabulated the three-electron eigenstates in \cref{tab:3e}. Specially, there are three $[E_1 + E_2, \frac{1}{2}]$ irreps in total. In each irrep, we choose the state with the highest $N_v$, $L$, and $j_z$ as the representative state, as given by 
\begin{align}  \label{eq:3e_gs}
    |1\rangle_3 &= f^\dagger_{1+\up} f^\dagger_{1+\down} f^\dagger_{1-\up} |0\rangle \\ \nonumber
    |2\rangle_3 &= \sqrt{\frac{1}{2}}(f^\dagger_{1+\up} f^\dagger_{2+\down} - f^\dagger_{1+\down} f^\dagger_{2+\up}) f^\dagger_{2-\up} |0\rangle \\ \nonumber
    |3\rangle_3 &= \sqrt{\frac{1}{6}}(f^\dagger_{1+\up} f^\dagger_{2+\down} + f^\dagger_{1+\down} f^\dagger_{2+\up}) f^\dagger_{2-\up}  - \sqrt{\frac{2}{3}} f^\dagger_{1+\up} f^\dagger_{2+\up} f^\dagger_{2-\down} |0\rangle
\end{align}
By symmetry, each of the representative states $|\Phi\rangle_3$ ($\Phi=1,2,3$) is allowed to couple with one another, but they cannot couple to other states in the three irreps by carrying different quantum number ($N_v$, $L$, or the spin-$z$ component $j_z$). The coupling matrix is given by
\begin{align}  \label{eq:3e_H}
    _3\langle \Phi | \hat{H}_{\rm A} | \Phi' \rangle_3 =  \begin{pmatrix}
            -3 J_{\rm a}+J_{\rm e} & -\sqrt{2} J_{\rm b} + \sqrt{\frac{1}{2}}J_{\rm d} & -\sqrt{\frac{3}{2}} J_{\rm d} \\
            \cdots & J_{\rm a}-J_{\rm b}+\frac{1}{2} J_{\rm d}+\frac{1}{2} J_{\rm e}  & \frac{\sqrt{3}}{2}J_{\rm d} - \frac{\sqrt{3}}{2} J_{\rm e} \\ 
            \cdots & \cdots & J_{\rm a}+J_{\rm b}- \frac{1}{2} J_{\rm d} - \frac{1}{2} J_{\rm e}  \\
        \end{pmatrix}
\end{align}
The ellipsis can be completed by Hermitian conjugation. The spectrum within this subspace can be obtained by diagonalizing \cref{eq:3e_H}, but does not adopt a simple expression. In general, the ground state will be a superposition of different $|\Phi\rangle_3$ with coefficients relying on specific parameters.

\subsection{Four-electron states}   \label{app:multi-4e}

The four-electron states are tabulated in \cref{tab:4e}. There are three $[A_1, 0]$ irreps that are allowed to couple with one another, and their wave-functions are given by $|\Phi\rangle_4$ with $\Phi=1,2,3$, 
\begin{align}  \label{eq:4e_gs}
    |1\rangle_4 &= \frac{1}{\sqrt{2}} \left[ f^\dagger_{1+\up} f^\dagger_{1+\down} f^\dagger_{1-\up} f^\dagger_{1-\down} + f^\dagger_{2+\up} f^\dagger_{2+\down} f^\dagger_{2-\up} f^\dagger_{2-\down} \right] |0\rangle \\\nonumber
    |2\rangle_4 &= \frac{1}{2} \left[ f^\dagger_{1+\up} f^\dagger_{1-\up} f^\dagger_{2+\down} f^\dagger_{2-\down} + f^\dagger_{1+\down} f^\dagger_{1-\down} f^\dagger_{2+\up} f^\dagger_{2-\up}  - f^\dagger_{1+\up} f^\dagger_{1-\down} f^\dagger_{2+\up} f^\dagger_{2-\down} - f^\dagger_{1+\down} f^\dagger_{1-\up} f^\dagger_{2+\down} f^\dagger_{2-\up} \right] |0\rangle \\\nonumber
    |3\rangle_4 &= \frac{1}{2\sqrt{3}} \left[ f^\dagger_{1+\up} f^\dagger_{1-\down} f^\dagger_{2+\up} f^\dagger_{2-\down} + f^\dagger_{1+\down} f^\dagger_{1-\up} f^\dagger_{2+\down} f^\dagger_{2-\up} + f^\dagger_{1+\up} f^\dagger_{1-\down} f^\dagger_{2+\up} f^\dagger_{2-\down} + f^\dagger_{1+\down} f^\dagger_{1-\up} f^\dagger_{2+\down} f^\dagger_{2-\up} \right. \\\nonumber
    & \quad \left.- 2 f^\dagger_{1+\up} f^\dagger_{1-\down} f^\dagger_{2+\down} f^\dagger_{2-\up} -2 f^\dagger_{1+\down} f^\dagger_{1-\up} f^\dagger_{2+\up} f^\dagger_{2-\down}  \right] |0\rangle 
\end{align}  
and the coupling matrix reads
\begin{align} \label{eq:4e_H}
    & _4\langle \Phi | \hat{H}_{\rm A} | \Phi' \rangle_4 =  \begin{pmatrix}
            -6J_{\rm a} + 2J_{\rm e} &  \sqrt{2}J_{\rm b} - 2\sqrt{2}J_{\rm d}  & \sqrt{6} J_{\rm b}   \\
            \cdots & 2J_{\rm a} + J_{\rm b} - 2J_{\rm d} + J_{\rm e} & -\sqrt{3} J_{\rm b} + \sqrt{3} J_{\rm e} \\
            \cdots & \cdots & 2J_{\rm a} - J_{\rm b} + 2J_{\rm d} - J_{\rm e} \\
    \end{pmatrix}
\end{align}
Specially, we also notice that $|2\rangle_4$ is also the product of the two states in the two-electron $[E_2, 0]$ doublet (see \cref{tab:2e}). There is no simple expression for the eigenvalues of \cref{eq:4e_H} either, and generically, the four-electron ground state is a superposition of $|\Phi\rangle_4$ ($\Phi=1,2,3$) with coefficients depending on parameters. 

\section{Hartree-Fock decomposition of \texorpdfstring{$\hat{H}_{\rm A}$}{HA}} \label{app:HF}

We calculate the expectation value of $\hat{H}_{\rm A}$ under an arbitrary Hartree-Fock order $O$ (\cref{eq:O}), $\langle \hat{H}_{\rm A} \rangle_O$. Recall that we have focused on the traceless part hence set $J_{\rm c} = 0$. For the four terms proportional to $J_{\rm \xi}$ with $\xi = {\rm a,b,d,e}$, 
\begin{align}
    E_{\rm a} &= \left( - \frac{J_{\rm a}}{2} \right) \sum_{\beta,\beta',\eta,\eta',s,s'} (-1)^{\beta+\beta'} \left\langle f^\dagger_{\beta\eta s} f^\dagger_{\beta'\eta' s'} f_{\beta'\eta's'} f_{\beta\eta s} \right\rangle_{O} \\\nonumber
    &= \left( - \frac{J_{\rm a}}{2} \right) \sum_{\beta,\beta',\eta,\eta',s,s'} (-1)^{\beta+\beta'} O_{\beta\eta s, \beta\eta s} O_{\beta'\eta' s', \beta'\eta' s'} - \zeta_{\beta} \zeta_{\beta'}  O_{\beta\eta s, \beta'\eta' s'} O_{\beta'\eta' s', \beta\eta s} \\\nonumber
    &= \left( - \frac{J_{\rm a}}{2} \right) \left( \mrm{Tr}^2 [O\cdot \sigma_z] - \mrm{Tr} [(O\cdot\sigma_z)^2] \right) 
\end{align}

\begin{align}
    E_{\rm b} &= \left( - \frac{J_{\rm b}}{2} \right) \sum_{\beta,\eta,s,s'} \left\langle f^\dagger_{\beta\eta s} f^\dagger_{\ovl{\beta}\eta s'} f_{\beta \eta s'} f_{\ovl{\beta}\eta s} + f^\dagger_{\beta\eta s} f^\dagger_{\beta \ovl{\eta} s'} f_{\ovl{\beta}\ovl{\eta} s'} f_{\ovl{\beta} \eta s} \right\rangle_{O} \\\nonumber
    &= \left( - \frac{J_{\rm b}}{2} \right) \sum_{\beta,\eta,s,s'} O_{\beta\eta s, \ovl{\beta}\eta s} O_{\ovl{\beta}\eta s', \beta \eta s'} + O_{\beta\eta s, \ovl{\beta} \eta s} O_{\beta \ovl{\eta} s', \ovl{\beta}\ovl{\eta} s'} - O_{\beta\eta s, \beta \eta s'} O_{\ovl{\beta}\eta s', \ovl{\beta}\eta s} - O_{\beta\eta s, \ovl{\beta}\ovl{\eta} s'} O_{\beta \ovl{\eta}s', \ovl{\beta} \eta s} \\\nonumber
    &= \left( - \frac{J_{\rm b}}{2} \right) \left( \frac{\mrm{Tr}^2 [O \cdot \sigma_x] + \mrm{Tr}^2[O \cdot \sigma_y \tau_z]}{2} - \frac{\mrm{Tr}[(O \cdot \sigma_x)^2] + \mrm{Tr}[(O \cdot \sigma_y \tau_z)^2]}{2} \right)
\end{align}
Note that, in deriving the third line above, we have used the following tricks, 
\begin{align}
    (\mrm{Tr}[O\cdot \sigma_x])^2 + (\mrm{Tr}[O\cdot \sigma_y\tau_z])^2 &= \left( \sum_{\beta\eta s} O_{\beta\eta s,\ovl{\beta}\eta s} \right)^2 + \left( \sum_{\beta\eta s} i (-1)^{\beta-1} \eta O_{\beta\eta s,\ovl{\beta}\eta s} \right)^2 \\\nonumber
    &= \sum_{\beta,\beta'} \sum_{\eta,\eta'} \sum_{s,s'} O_{\beta\eta s,\ovl{\beta}\eta s}  O_{\beta'\eta's',\ovl{\beta'}\eta's'} - \sum_{\beta,\beta'} \sum_{\eta,\eta'} \sum_{s,s'} (-1)^{\beta+\beta'} (\eta\eta') O_{\beta\eta s,\ovl{\beta}\eta s}  O_{\beta'\eta's',\ovl{\beta'}\eta's'} \\\nonumber
    &= \sum_{\beta,\beta'} \sum_{\eta,\eta'} \sum_{s,s'} \left( 1 - (-1)^{\beta+\beta'}(\eta\eta') \right) O_{\beta\eta s,\ovl{\beta}\eta s}  O_{\beta'\eta's',\ovl{\beta'}\eta's'} \\\nonumber
    &= 2 \sum_{\beta,\beta'} \sum_{\eta,\eta'} \sum_{s,s'} \left( \delta_{\beta,\ovl{\beta'}} \delta_{\eta,\eta'} + \delta_{\beta,\beta'} \delta_{\eta,\ovl{\eta'}} \right) O_{\beta\eta s,\ovl{\beta}\eta s}  O_{\beta'\eta's',\ovl{\beta'}\eta's'}
\end{align}
and 
\begin{align}
    \mrm{Tr}[(O\cdot \sigma_x)^2]) + \mrm{Tr}[(O\cdot \sigma_y\tau_z)^2]) &= \sum_{\beta,\beta'} \sum_{\eta,\eta'} \sum_{s,s'} O_{\beta\eta s,\beta'\eta's'} O_{\ovl{\beta'}\eta's',\ovl{\beta}\eta s} + O_{\beta\eta s,\beta'\eta's'} (i(-1)^{\beta'} \eta') O_{\ovl{\beta'}\eta's',\ovl{\beta}\eta s} (i(-1)^{\ovl{\beta}} \eta) \\\nonumber
    &= \sum_{\beta,\beta'} \sum_{\eta,\eta'} \sum_{s,s'} \left( 1 + (-1)^{\beta+\beta'} (\eta\eta') \right) O_{\beta\eta s,\beta'\eta's'} O_{\ovl{\beta'}\eta's',\ovl{\beta}\eta s} \\\nonumber
    &= 2 \sum_{\beta,\beta'} \sum_{\eta,\eta'} \sum_{s,s'} \left( \delta_{\beta,\beta'} \delta_{\eta,\eta'} + \delta_{\beta,\ovl{\beta'}} \delta_{\eta,\ovl{\eta'}} \right) O_{\beta\eta s,\beta'\eta's'} O_{\ovl{\beta'}\eta's',\ovl{\beta}\eta s} 
\end{align}
Similar tricks will also be applied in deriving $E_{\rm d,e}$. 
\begin{align}
    E_{\rm d} &= \left( - \frac{J_{\rm d}}{2} \right) \sum_{\beta,\eta,s,s'} \left\langle f^\dagger_{\beta \eta s} f^\dagger_{\beta \ovl{\eta} s'} f_{\ovl{\beta} \eta s'} f_{\ovl{\beta} \ovl{\eta} s} + f^\dagger_{\beta \eta s} f^\dagger_{\ovl{\beta} \ovl{\eta} s'} f_{\beta \eta s'} f_{\ovl{\beta} \ovl{\eta} s} \right\rangle_{O} \\\nonumber
    &= \left( - \frac{J_{\rm d}}{2} \right) \sum_{\beta,\eta,s,s'} O_{\beta\eta s, \ovl{\beta} \ovl{\eta} s} O_{\beta \ovl{\eta} s', \ovl{\beta} \eta s'} + O_{\beta\eta s, \ovl{\beta} \ovl{\eta} s} O_{\ovl{\beta}\ovl{\eta} s', \beta \eta s'} - O_{\beta\eta s, \ovl{\beta} \eta s'} O_{\beta \ovl{\eta} s', \ovl{\beta} \ovl{\eta} s} - O_{\beta\eta s, \beta \eta s'} O_{\ovl{\beta} \ovl{\eta}s', \ovl{\beta} \ovl{\eta} s} \\\nonumber
    &= \left( - \frac{J_{\rm d}}{2} \right) \left( \frac{\mrm{Tr}^2 [O \cdot \sigma_x \tau_x] + \mrm{Tr}^2[O \cdot \sigma_x \tau_y] }{2} -  \frac{\mrm{Tr}[O \cdot \sigma_x \tau_x]^2 + \mrm{Tr}[O \cdot \sigma_x \tau_y]^2 }{2} \right)
\end{align}

{\small \begin{align}
    E_{\rm e} &= \left( - \frac{J_{\rm e}}{2} \right) \sum_{\beta,\eta,s,s'} \left\langle f^\dagger_{\beta \eta s} f^\dagger_{\beta \ovl{\eta} s'} f_{\beta \eta s'} f_{\beta \ovl{\eta} s} \right\rangle_{O} \\\nonumber
    &= \left( - \frac{J_{\rm e}}{2} \right) \sum_{\beta,\eta,s,s'} O_{\beta\eta s, \beta \ovl{\eta} s} O_{\beta \ovl{\eta} s', \beta \eta s'} - O_{\beta\eta s, \beta\eta s'} O_{\beta \ovl{\eta} s', \beta \ovl{\eta} s}  \\\nonumber
    &= \left( - \frac{J_{\rm e}}{2} \right) \left( \frac{\mrm{Tr}^2 [O \cdot \tau_x] + \mrm{Tr}^2[O \cdot \tau_y] + \mrm{Tr}^2 [O \cdot \sigma_z\tau_x] + \mrm{Tr}^2[O \cdot \sigma_z \tau_y]}{4} -  \frac{\mrm{Tr}[O \cdot \tau_x]^2 + \mrm{Tr}[O \cdot \tau_y]^2 + \mrm{Tr}[O \cdot \sigma_z \tau_x]^2 + \mrm{Tr}[O \cdot \sigma_z \tau_y]^2}{4} \right)
\end{align}}

In sum, all the interactions can be written in the following form, 
\begin{align}
    E_{\xi} = \frac{J_\xi}{2} \frac{1}{|\mcl{P}_{\xi}|} \sum_{P_\xi \in \mcl{P}_{\xi}} \left[ - \left(\mrm{Tr} [O \cdot P_\xi]\right)^2 + \mrm{Tr} \left[ (O \cdot P_\xi)^2 \right] \right]
\end{align}
with the set of Pauli matrices $\mcl{P}_{\rm \xi}$ tabulated in \cref{tab:Pxi}. 
By defining the traceless part in $O$ as $Q$ in Eq. (\ref{eq:Q}), and identifying that $\mrm{Tr} [P_{\xi}] = \mrm{Tr} [Q] = \mrm{Tr} [P_{\xi} Q P_{\xi}] = 0$, one concludes that $\mrm{Tr}[O \cdot P_{\xi}] = \frac{\nu+4}{8} \mrm{Tr}[Q \cdot P_{\xi}]$, and $\mrm{Tr}[(O \cdot P_{\xi})^2] = (\frac{\nu+4}{8})^2 \mrm{Tr}[1 + (Q \cdot P_{\xi})^2]$, from which Eq. (\ref{eq:Hxi}) can be derived. 

\section{Details on the self-consistent Hartree-Fock calculation} \label{app:HFcalc}

\begin{table}[tb]
    \centering
    \begin{tabular}{l|c|ccc|cc}
    \hline\hline
        Orders & $B[O^f]$ & $B[O^{c,12}]$ & $B[O^{c,34}]$ & $B[O^{c,1234}]$ & $B[O^{cf,12}]$ & $B[O^{cf,34}]$ \\
    \hline
        $\rm KIVC$ & 1.280 & 0.093 & 0.224 & 0.005 & 0.001 & 0.044 \\
        $\rm TIVC$ & 1.283 & 0.082 & 0.148 & 0.006 & 0.001 & 0.045 \\
        $\rm NIOC_{0}$ & 1.290 & 0.032 & 0 & 0.006 & 0.001 & 0.044 \\
        $\rm OP$ & 1.282 & 0.083 & 0.147 & 0.006 & 0.001 & 0.046 \\
    \hline
        $\rm KIVC \times SP$ & 1.061 & 0.061 & 0.222 & 0.008 & 0.099 & 0.054 \\
        $\rm TIVC \times QAH$ & 1.057 & 0.056 & 0.168 & 0.009 & 0.105 & 0.056 \\
    \hline\hline
    \end{tabular}
    \caption{\label{tab:BO} Self-consistent order paramter symmetry-breaking strengths, calculated at $\kappa_\Gamma = \kappa_K = 0$ and $U_0 = 0$. 
    The projector ansatz at $\nu=0$ and $\nu=-2$ have $B[O^f] = \sqrt{2} = 1.414$ and $B[O^f] = \sqrt{\frac{3}{2}} = 1.225$, respectively, and the symmetry-breaking strengths in $O^c$ and $O^{cf}$ are assumed to be subsidiary and negligibly weak. 
    The self-consistent values verify this picture. }
\end{table}

The procedures of our self-consistent Hartree-Fock calculation closely follow Ref.~\cite{Song_2022}. The only modification is that $\hat{H}_{\rm A,H}$ are added. We briefly summarize the main steps and numeric parameters. 

The Hamiltonian used in the numeric calculations includes the kinetic energy $\hat{H}_0$ (\cref{eq:Hf0}), the full Coulomb interactions $\hat{H}_{\rm C} = \hat{H}_U + \hat{H}_W + \hat{H}_J + \hat{H}_V$ (\cref{eq:HfC,eq:HU,eq:HW,eq:HJ,eq:HV}), and the on-$f$-site splittings $\hat{H}_{\rm A,H}$, which are induced by phonons and the atomic Hubbard. The physical parameters we adopt have been presented in the main text. 
Interactions are decomposed with the following mean-field order parameters, 
\begin{align}  \label{eq:Of}
    O^{f}_{\beta\eta s, \beta'\eta's'} &= \frac{1}{N_M} \sum_{\kk\in\MBZ} \langle f^\dagger_{\kk\beta\eta s} f_{\kk\beta'\eta's'} \rangle \\ \label{eq:Oc}
    O^{c}_{b\eta s, b'\eta's'} &= \frac{1}{N_M} \sum_{\kk\in\MBZ} \sum_{\GG} \langle c^\dagger_{\kk+\GG b\eta s} c_{\kk+\GG b'\eta's'} \rangle - \frac{1}{2} \delta_{bb'} \delta_{\eta\eta'} \delta_{ss'} \\ \label{eq:Ocf} 
    O^{cf}_{b\eta s, \beta'\eta's'} &= \frac{1}{N_M} \sum_{\kk\in\MBZ} \sum_{\GG} \langle c^\dagger_{\kk+\GG b\eta s} f_{\kk\beta'\eta's'} \rangle
\end{align}
Here, we have divided $\kk'$ that runs over the whole momentum space satisfying $|\kk'| \lesssim \Lambda_c$ into $\kk' = \kk+\GG$, where $\kk\in\MBZ$, and $|\GG|\lesssim\Lambda_c$ is a moir\'e reciprocal lattice vector. \cref{eq:Of} is equivalent to \cref{eq:O}, in the absence of translation symmetry breaking. The $\kk$-dependence is averaged, which is validated by the absence of $\kk$-dependence in the interacting terms $\hat{H}_{\rm C}$ and $\hat{H}_{\rm A,H}$.

In the self-consistent loops, we have only kept the $|\GG|=0$ shell, because $c$-electrons with large $\GG$ quickly decouple from the low-energy theory. The convergence is tested by further including the $|\GG| \leq \sqrt{3} k_\theta$ shells. 
When plotting the mean-field bands in \cref{fig:epc-phd}, the $|\GG| \leq \sqrt{3} k_\theta$ shell is exploited to capture the correct periodicity in the moir\'e reciprocal space. 
A $13\times 13$ $\kk$ mesh in the MBZ is taken during the self-consistent loops. 

In the main text, we have assumed that the symmetry-breaking effect is mainly carried by $O^f$, so that the phonon-favored orders are mainly selected by the on-$f$-site $\hat{H}_{\rm H,A}$, by perturbatively lowering the corresponding total energies. 
Now we quantitatively verify this statement. 
First, we define the $b, b'=1,2$ and $b,b'=3,4$ blocks in $O^c$ as $O^{c,12}$ and $O^{c,34}$, respectively, and define the $b=1,2$, $b'=3,4$ block as $O^{c,1234}$. We also define the $b=1,2$ and $b=3,4$ blocks in $O^{cf}$ as $O^{cf,12}$ and $O^{cf,34}$, respectively. 
Without spontaneously breaking the crytalline $C_{2z}T$, $C_{3z}$, $C_{2x}$, the time-reversal $T$, and valley U(1) and the spin SU(2), the above order parameters, which are all $8\times8$ matrices, can only contain the following components
\begin{align}
    O^f |_{\rm sym}  &\propto \sigma^0 \tau^0 s^0 \\
    O^{c,12} |_{\rm sym}  &\propto \sigma^0 \tau^0 s^0 , \qquad 
    O^{c,34} |_{\rm sym}  \propto \sigma^{0,x} \tau^0 s^0 , \qquad  
    O^{c,1234} |_{\rm sym}  = 0  \\
    O^{cf,12} |_{\rm sym} &\propto \sigma^0 \tau^0 s^0 , \qquad O^{cf,34} |_{\rm sym}  = 0
\end{align}
We therefore quantify the symmetry-breaking strength of an $8\times 8$ matrix $O$ through
\begin{align}
    B[O] = \bigg| \bigg| O - O|_{\rm sym} \bigg| \bigg| 
\end{align}
where $||\cdots||$ is the Frobenius norm. We tabulate the typical symmetry-breaking strengths found in the self-consistent calculations in \cref{tab:BO}. It can be seen that, the $O^f$ sector carries the major symmetry-breaking effect, and the self-consistent strength is close to the projector ansatz proposed in the main text. The symmetry-breaking in $O^c$ and $O^{cf}$ sectors, on the other hand, are much weaker. The results thus validate the perturbative analysis of the phonon-favored order presented in the main text using the projector ansatz of $O^f$. 

In general, phonons also induce $cf$ and $cc$ interactions, which have not been incorporated in our current calculations. Since the Bloch function spread of $c$-electrons are $0.22\sim0.33 a_M$ \cite{calugaru_twisted_2023}, similar to or even wider than the $f$-electron Wannier orbitals ($0.18\sim0.19 a_M$, see \cref{sec:ele-THF}), the corresponding $cf$ and $cc$ strength will be of the same order, or weaker than, the $f$-on-site interactions $\hat{H}_{\rm A,H}$. 
However, due to the much weaker symmetry-breaking strength in the $O^c$ and $O^{cf}$ parameters, the $cf$ and $cc$ interaction will contribute much less to the total Hartree-Fock energy than the on-site $\hat{H}_{\rm A,H}$. 
To be concrete, the Hartree-Fock energy correction due to $\hat{H}_{\rm A}$ is proportional to $B[O^f]^2$, which is of the $O(1)$ order. However, the correction due to $c^\dagger f^\dagger f c$ interactions will be proportional to $B[O^f] B[O^c]$ and $B[O^{cf}]^2$, and the correction due to $c^\dagger c^\dagger c c$ will be proportional to $B[O^c]^2$, which are both much smaller than $O(1)$. 
Therefore, further incorporating these effects should not qualitatively affect the results presented in the main text.

\section{EPC vertices from the microscopic tight-binding lattice} \label{app:cont}

In this section, we will start from the microscopic lattice model of MATBG to derive the continuum model, including the static electron Hamiltonian $\hat{H}_0$ (summarized in \cref{sec:ele-cont}), and the EPC vertices $\hat{H}_{\rm epc}$ (summarized as the $C^{(\chi)}(\qq)$ matrices in \cref{sec:epc-epc}). The hopping amplitudes between two carbon $p_z$-orbitals that are spatially separated by $\rr$ is modeled by the standard Slater-Koster form given in Eq. (\ref{eq:SKhop}). In general, depending on whether the electron hops within one layer or across two layers, we divide both $\hat{H}_0$ and $\hat{H}_{\rm epc}$ into the single-layer part (subscript SLG) and the inter-layer part (subscript moir\'e). 

The physical results will be presented in the coordinate system where $\KK^{(2)} - \KK^{(1)}$ lies exactly along the $y$-axis (which we term as the MATBG coordinates, exemplified in \cref{fig:MBZ}) in the end. Nevertheless, while deriving the SLG hoppings, it is more convenient to start with a coordinate system where $\KK^{(l)}$ lies exactly along the $x'$-axis (which we term as the SLG coordinates, and differs by an angle of $\pm\frac{\theta}{2}$ from the MATBG coordinates), and then rotate the results back to the MATBG coordinates. We will thus adopt this convention, and when the SLG coordinates are used, the layer index $l$ will be kept implicit, too. 

\subsection{Single-layer hoppings} \label{app:cont-SLG}

In the SLG coordinates, we denote the directions of the three nearest neighbor bonds (pointing from $A$ sublattice to $B$ sublattice) as $\dd_{1,2,3}$,  
\begin{align}
    \dd_1 = \frac{a_0}{\sqrt{3}}(0, 1, 0), ~~\dd_2 = \frac{a_0}{\sqrt{3}}(\frac{\sqrt{3}}{2}, -\frac{1}{2}, 0), ~~\dd_3 = \frac{a_0}{\sqrt{3}}(-\frac{\sqrt{3}}{2}, -\frac{1}{2}, 0)
\end{align}
Correspondingly, the graphene $\KK$-valley is located at $\KK=\frac{4\pi}{3a_0}(1, 0, 0)$. 

As the hopping amplitude decays exponentially with the atom spacing, it suffices to keep the nearest-neighbor hoppings. We check the nearest-neighbor hopping between the $B$ sublattice and $A$, which we expand to the linear order of lattice vibration $\uu_i$, 
\begin{align}
    t(\dd_j+\uu(\RR^{(A)}+\dd_j)-\uu(\RR^{(A)})) \approx t(\dd_j) + \left( \uu(\RR^{(A)}+\dd_j) - \uu(\RR^{(A)}) \right) \cdot \nabla t(\dd_j) 
\end{align}
By observing that 
\begin{align}
    t(\dd_j) = V_\pi, ~~~~\nabla t(\dd_j) = -\frac{V_\pi}{r_0} \hat{\dd}_j
\end{align}
and carrying out the inverse Fourier transformation of \cref{eq:def_uq}, 
\begin{align}
    \uu(\RR^{(l,\alpha)}) = \frac{1}{\sqrt{N_g}} \sum_{\qq\in\gBZ} \uu_{\qq}^{(l,\alpha)} e^{i\qq\cdot\RR^{(l,\alpha)}} 
\end{align}
we obtain
\begin{align}   \label{eq:SLG_hop}
    t(\dd_j+\uu(\RR^{(A)}+\dd_j)-\uu(\RR^{(A)})) 
    \approx V_\pi + \left( -\frac{V_\pi}{r_0} \right) \frac{1}{\sqrt{N_g}} \sum_{\qq\in\gBZ} e^{i\qq\cdot\RR^{(A)}} \left( \uu^{(B)}_\qq e^{i\qq\cdot\dd_j} - \uu^{(A)}_\qq  \right) \cdot \hat{\dd_j} 
\end{align}
It can be thus seen that the out-of-plane modes will not enter the SLG EPC vertex, as all $\dd_j$ lie in-plane. In terms of the second-quantized operators, the electrons hop within the layer as, 
\begin{align}
    \hat{H}_{\rm SLG} = \sum_{s} \sum_{\dd_j} \sum_{\RR^{(A)}} 
    \psi^\dagger_s(\RR^{(A)}) ~ t(\dd_j+\uu(\RR^{(A)}+\dd_j)-\uu(\RR^{(A)})) ~ \psi_s(\RR^{(A)} + \dd_j) + \mrm{H.c.}
\end{align}

Substituting the inverse Fourier transformation of Eq. (\ref{eq:psi_k}), 
\begin{align} 
    \psi^\dagger_{s}(\RR^{(l, \alpha)}) = \frac{1}{\sqrt{N_g}} \sum_{\eta} \sum_{|\kk| \lesssim \Lambda} e^{-i(\kk + \eta\KK^{(l)}) \cdot \RR^{(l, \alpha)}} \psi^\dagger_{\kk, l, \alpha, \eta, s}
\end{align}
and Eq. (\ref{eq:SLG_hop}) into $\hat{H}_{\rm SLG}$, we obtain
\begin{align}
    \hat{H}_{\rm SLG} = & \frac{1}{N_g} \sum_s \sum_{{\RR}^{(A)}} \sum_{\eta, \eta'} \sum_{|\kk|, |\kk'| \lesssim \Lambda} e^{-i(\kk+\eta\KK)\cdot\RR^{(A)}} \psi^\dagger_{\kk, A, \eta, s} \times \\\nonumber
    & \sum_{\dd_j} \left[ V_\pi + \left( -\frac{V_\pi}{r_0} \right) \frac{1}{\sqrt{N_g}} \left[ \sum_{\qq \in \gBZ} e^{i\qq \cdot \RR^{(A)}} \left( \uu^{(B)}_{\qq} e^{i\qq\cdot\dd_j} - \uu_{\qq}^{(A)} \right)\cdot\hat{\dd}_j \right]  \right] \psi_{\kk', B, \eta', s} e^{i(\kk'+\eta'\KK)\cdot(\RR^{(A)}+\dd_j)} + \mrm{H.c.}
\end{align}
Depending on containing $\uu_i$ or not, we separate $\hat{H}_{\rm SLG} = \hat{H}_{\rm 0, SLG} + \hat{H}_{\rm epc, SLG}$. For $\hat{H}_{\rm 0, SLG}$, summation over $\RR^{(A)}$ imposes momentum conservation,  $\kk = \kk'$ and $\eta=\eta'$, therefore, 
\begin{align}   \label{eq:H0SLG_prim}
    \hat{H}_{\rm 0, SLG} = & V_\pi \sum_s \sum_{\eta} \sum_{|\kk| \lesssim \Lambda} \psi^\dagger_{\kk, A, \eta, s}  \left[ \sum_{\dd_j} e^{i \kk \cdot \dd_j} e^{i\eta\frac{2\pi}{3}(j-1)} \right] \psi_{\kk, B, \eta, s} + \mrm{H.c.} \\\nonumber
    = & \left( -\frac{\sqrt{3}}{2} a_0 V_\pi \right) \sum_s \sum_{\eta} \sum_{|\kk| \lesssim \Lambda} \psi^\dagger_{\kk, A, \eta, s}  \left[ \eta k_x - i k_y \right] \psi_{\kk, B, \eta, s} + \mrm{H.c.}
\end{align}
where we have used that $e^{i\KK \cdot \dd_j} = e^{i \frac{2\pi}{3}(j-1)}$, and then expanded $e^{i\kk\cdot\dd_j}$ to the first order in $\kk\cdot\dd_j$. For $\hat{H}_{\rm epc, SLG}$, summation over $\RR^{(A)}$ requires $\qq = \kk - \kk' + (\eta-\eta')\KK$, 
\begin{align}   \label{eq:HepcSLG_prim}
    \hat{H}_{\rm epc, SLG} = & \frac{1}{\sqrt{N_g}} \left( -\frac{V_\pi}{r_0} \right) \sum_s \sum_{\eta, \eta'} \sum_{|\kk|, |\kk'| \lesssim \Lambda} \sum_{\qq \in \gBZ} \delta_{\qq,\ \kk-\kk'+(\eta-\eta')\KK} \\\nonumber
    ~& \times \psi^\dagger_{\kk, A, \eta, s} 
     \left[ \sum_{\dd_j} e^{i\kk'\cdot\dd_j} e^{i\eta'\frac{2\pi}{3}(j-1)} \left( \uu^{(B)}_{\qq} e^{i\qq\cdot\dd_j} - \uu_{\qq}^{(A)} \right)\cdot\hat{\dd}_j \right] \psi_{\kk', B, \eta', s} + \mrm{H.c.}
\end{align}
Depending on whether the phonon is $\Gamma$-phonon or the $K$-phonon, the momentum conservation will impose $\eta=\eta$ and $\ovl{\eta}=  \eta'$, respectively. We will thus further separate $\hat{H}_{\rm epc, SLG} = \hat{H}_{\rm epc, SLG, \Gamma} + \hat{H}_{\rm epc, SLG, K}$.

\subsubsection{\texorpdfstring{$\hat{H}_{\rm 0, SLG}$}{H0,SLG}} \label{app:cont-SLG-H0}

For $\hat{H}_{\rm 0, SLG}$, Eq. (\ref{eq:H0SLG_prim}), by defining $v_F = -\frac{\sqrt{3}}{2}V_\pi a_0$, and adding up the two parts that are Hermitian conjugate to each other, we obtain
\begin{align} 
    \hat{H}_{0,\mrm{SLG}} = v_F \sum_s \sum_{|\kk|<\Lambda} \sum_{\eta} \psi^\dagger_{\kk,\alpha',\eta', s} \left[  k_x \sigma_x \tau_z + k_y \sigma_y \tau_0 \right]_{\alpha'\eta',\alpha\eta} \psi_{\kk,\alpha,\eta, s}
\end{align}
where we have introduced the Pauli matrices $[\sigma_{\mu}]_{\alpha',\alpha}$ in the sublattice index, and $[\tau_\mu]_{\eta',\eta}$ in the valley index, with $\mu=0,x,y,z$. 

Now we recover the layer index and rotate back to the MATBG coordinates. In that way, we should replace $(k_x, k_y)$ with $(\cos\frac{\theta}{2}k_x + (-1)^l \sin\frac{\theta}{2}k_y, \sin\frac{\theta}{2}k_x + (-1)^{l-1} \cos\frac{\theta}{2}k_y)$ for layer $l=1,2$, respectively, so that the Dirac equations from the two layers are slightly rotated relative to each other.  
Nevertheless, it has been shown that dropping all the $O(\theta)$ terms have little effect on the electronic properties \cite{TBG2_2021}. Therefore, we will adopt this approximation, writing $\hat{H}_{0,\mrm{SLG}}$ as 
\begin{align} 
\label{eq:H0SLG_k}
    \hat{H}_{0,\mrm{SLG}} = v_F \sum_{|\kk| \lesssim \Lambda} \sum_{\eta, l, s} \psi^\dagger_{\kk,l,\alpha',\eta',s} \left[ k_x \sigma_x \tau_z + k_y \sigma_y \tau_0 \right]_{\alpha'\eta',\alpha\eta} \psi_{\kk,l,\alpha,\eta,s}
\end{align}

Written in terms of the continuum fields Eq. (\ref{eq:psi_r}), Eq. (\ref{eq:H0SLG_k}) reads, 
\begin{align}
\label{eq:H0SLG_r}
    \hat{H}_{0,\mrm{SLG}} = v_F \sum_{\eta, l, s} \int\mrm{d}^2\rr ~  \psi^\dagger_{l,\alpha',\eta',s}(\rr) \left[ -i \partial_x \sigma_x\tau_z - i\partial_y \sigma_y \tau_0 \right]_{\alpha'\eta',\alpha\eta} \psi_{l,\alpha,\eta,s}(\rr) 
\end{align}

\subsubsection{\texorpdfstring{$\hat{H}_{\mrm{epc, SLG},\Gamma}$}{Hepc,SLG,G}} \label{app:cont-SLG-G}

For the $\Gamma$-phonon term in Eq. (\ref{eq:HepcSLG_prim}), we focus on those $\uu^{(\alpha)}_{\qq}$ with $\qq\cdot\dd_j\ll 1$ which impose $\eta=\eta'$. Thus we obtain
\begin{align}
\label{eq:HepcSLG_1}
    \hat{H}_{\mrm{epc},\mrm{SLG},\Gamma} = \frac{1}{\sqrt{N_g}} \left(-\frac{V_\pi}{r_0}\right) \sum_{\eta, s} \sum_{|\kk|, |\qq| \lesssim \Lambda} \psi^\dagger_{\kk+\qq,A,\eta,s} \left[ \sum_{j=1}^{3} e^{i\kk\cdot\dd_j} e^{i\eta\frac{2\pi}{3}(j-1)} \left( \uu^{(B)}_{\qq} e^{i\qq\cdot\dd_j} - \uu_{\qq}^{(A)} \right)\cdot\hat{\dd}_j \right] \psi_{\kk,B,\eta,s} + \mrm{H.c.} 
\end{align}
First, expanding $e^{i\qq\cdot\dd_j}$ to the linear order of $\qq\cdot\dd_j$, we have
\begin{align}
    \left( \uu^{(B)}_{\qq} e^{i\qq\cdot\dd_j} - \uu_{\qq}^{(A)} \right) \approx \left( \left(\uu_{\qq}^{(B)} - \uu_{\qq}^{(A)} \right) + (i\qq\cdot\dd_j) \uu_{\qq}^{(B)} \right)
\end{align}
Moreover, here it suffices to approximate $e^{i\kk\cdot\dd_j}\approx1$, because the zeroth order will not get canceled in any phonon channel, leaving all higher orders negligible. 
Referring to the definition of the $\Gamma$-phonon modes, Eq. (\ref{eq:uG_q}), the square bracket in Eq. (\ref{eq:HepcSLG_1}) will now read, 
\begin{align} \label{eq:HepcSLG_Gbracket}
    ~& \left[ \sum_{j=1}^{3} e^{i\eta\frac{2\pi}{3}(j-1)} \left( \left(\uu_{\qq}^{(B)} - \uu_{\qq}^{(A)} \right) + (i\qq\cdot\dd_j) \uu_{\qq}^{(B)}\right)\cdot\hat{\dd}_j \right] \\\nonumber 
    =&  \left( -\frac{3}{\sqrt{2}} \right) \left[ i\eta u_{\qq,\XO} + u_{\qq,\YO} \right] + \left(\frac{\sqrt3 a_0}{4}\right) (iq_x -\eta q_y) ( \uu^{(B)}_{\qq,x} + i\eta \uu^{(B)}_{\qq,y}) \\\nonumber 
    \approx& \left( -\frac{3}{\sqrt{2}} \right) \left[ i\eta u_{\qq,\XO} + u_{\qq,\YO} \right] + \left( \frac{\sqrt{3}a_0}{4\sqrt{2}} \right) \left( -iq_x + \eta q_y  \right) \left(u_{\qq,\XA} + i\eta u_{\qq,\YA} \right)
\end{align}
Note that in deriving the last line, the EPC strength of the second term is linear in $|\qq|$. Consequently, if decomposing $\uu_{\qq}^{(B)}$ into acoustic and optical components, only the acoustic component will contribute significantly, because the acoustic phonon frequency is also linear in $|\qq|$. 
The optical component can be simply added to the first term, where the linear-in-$|\qq|$ strength will be negligible. 
By inserting Eq. (\ref{eq:HepcSLG_Gbracket}) into the square brackets of Eq. (\ref{eq:HepcSLG_1}), and writing down the Hermitian conjugate terms explicitly, $\hat{H}_{\rm epc, SLG, \Gamma}$ reads
\begin{align}
    \hat{H}_{\mrm{epc},\mrm{SLG},\Gamma} &= \frac{1}{\sqrt{N_g}} \left( -\frac{V_\pi}{r_0} \right) \times \sum_{\eta, s} \sum_{|\kk|, |\qq| \lesssim \Lambda} \\\nonumber
    ~& \psi^\dagger_{\kk+\qq,A,\eta,s} \left[ \left( -\frac{3}{\sqrt{2}} \right) \left[ i\eta u_{\qq,\XO} + u_{\qq,\YO} \right] + \left( \frac{\sqrt{3}a_0}{4\sqrt{2}} \right) \left( -iq_x + \eta q_y  \right) \left(u_{\qq,\XA} + i\eta u_{\qq,\YA} \right)  \right] \psi_{\kk,B,\eta,s} \\\nonumber 
    +& \psi^\dagger_{\kk,B,\eta,s} \left[ \left( -\frac{3}{\sqrt{2}} \right) \left[ -i\eta u_{-\qq,\XO} + u_{-\qq,\YO} \right] + \left( \frac{\sqrt{3}a_0}{4\sqrt{2}} \right) \left( iq_x + \eta q_y  \right) \left(u_{-\qq,\XA} - i\eta u_{-\qq,\YA} \right)  \right] \psi_{\kk+\qq,A,\eta,s}
\end{align}
where we have used $u_{\qq,\chi}^\dagger = u_{-\qq,\chi}$. Next, by first changing the summed variables as $\kk+\qq \to \kk$ and then $-\qq\to \qq$, 
\begin{align} 
    \hat{H}_{\mrm{epc},\mrm{SLG},\Gamma} &= \frac{1}{\sqrt{N_g}} \left( -\frac{V_\pi}{r_0} \right) \times \sum_{\eta,s} \sum_{|\kk|, |\qq| < \Lambda}  \\\nonumber
    & \psi^\dagger_{\kk+\qq,A,\eta,s} \left[ \left( -\frac{3}{\sqrt{2}} \right) \left[ i\eta u_{\qq,\XO} + u_{\qq,\YO} \right] + \left( \frac{\sqrt{3}a_0}{4\sqrt{2}} \right) \left( -iq_x + \eta q_y  \right) \left(u_{\qq,\XA} + i\eta u_{\qq,\YA} \right)  \right] \psi_{\kk,B,\eta,s} \\\nonumber 
    +& \psi^\dagger_{\kk+\qq,B,\eta,s} \left[ \left( -\frac{3}{\sqrt{2}} \right) \left[ -i\eta u_{\qq,\XO} + u_{\qq,\YO} \right] + \left( \frac{\sqrt{3}a_0}{4\sqrt{2}} \right) \left( -iq_x - \eta q_y  \right) \left(u_{\qq,\XA} - i\eta u_{\qq,\YA} \right)  \right] \psi_{\kk,A,\eta,s}
\end{align}
Using the Pauli matrix $[\sigma_{\mu}]_{\alpha',\alpha}$ and $[\tau_\mu]_{\eta',\eta}$, $\hat{H}_{\mrm{epc,SLG},\GGamma}$ further simplifies to  
\begin{align}  \label{eq:HepcSLG_G}
    \hat{H}_{\mrm{epc},\mrm{SLG},\Gamma} =& \frac{1}{\sqrt{N_g}} \left( -\frac{V_\pi}{r_0} \right) \left(\frac{\sqrt{3}a_0}{4\sqrt{2}}\right) \sum_{\eta, \eta', s} \psi^\dagger_{\kk+\qq,\alpha',\eta',s} \left[ u_{\qq,\XA} \left[ (iq_x) (-\sigma_x\tau_0) + (iq_y)( \sigma_y\tau_z) \right] \right. \\\nonumber 
    ~& ~~~~~~~~~~~~~ + \left. u_{\qq,\YA} \left[ (iq_x)(\sigma_y\tau_z) + (iq_y)(\sigma_x\tau_0) \right] \right]_{\alpha'\eta',\alpha\eta} \psi_{\kk,\alpha,\eta,s} \\\nonumber 
    +& \frac{1}{\sqrt{N_g}} \left( -\frac{V_\pi}{r_0} \right)\left( \frac{3}{\sqrt{2}} \right) \sum_{\eta, \eta', s} \psi^\dagger_{\kk+\qq,\alpha',\eta', s} \left[ u_{\qq,\XO} \left( \sigma_y \tau_z \right) - u_{\qq,\YO} \left( \sigma_x \tau_0 \right) \right]_{\alpha'\eta',\alpha\eta} \psi_{\kk,\alpha,\eta, s}  
\end{align}

Now we recover the layer index $l$, and rotate Eq. (\ref{eq:HepcSLG_G}) back to the MATBG coordinates. In doing so, not only $\kk$ and $\qq$ need to be rotated by an angle of $\pm\frac{\theta}{2}$, the $\XA$ and $\XO$ modes will also mix with $\YA$ and $\YO$ by an amount of $\sim\frac{\theta}{2}$. 
Nonetheless, the extra correction terms to appear are all of the order $O(\theta)$, and hence negligible. 
Fouriered to the continuum fields by exploiting Eq. (\ref{eq:psi_r}) and (\ref{eq:u_r}), and by replacing $v_F = -\frac{\sqrt{3}}{2} a_0 V_\pi$, 
\begin{align}  \label{eqn:Hepc_SLG_G_r}
    \hat{H}_{\mrm{epc, SLG},\Gamma} =& \left( \frac{1}{2\sqrt{2}} \frac{v_F}{r_0} \right) \sum_{\eta, l, s} \int\mrm{d}^2\rr~ \psi^\dagger_{l,\alpha',\eta',s}(\rr) \left[  \left(\partial_y u_{l,\YA}(\rr) - \partial_x u_{l,\XA}(\rr) \right) (\sigma_x\tau_0)   + \right. \\\nonumber 
    ~& ~~~~~~~~~~~~ \left. \left(\partial_y u_{l,\XA}(\rr) + \partial_x u_{l,\YA}(\rr) \right) (\sigma_y\tau_z) \right]_{\alpha'\eta',\alpha\eta}  \psi_{l,\alpha,\eta,s}(\rr) \\\nonumber 
    +& \left( \sqrt{6} \frac{v_F}{a_0r_0} \right) \sum_{\eta,l,s} \int \mrm{d}^2\rr~  \psi^\dagger_{l,\alpha',\eta',s}(\rr) \left[ u_{l,\XO}(\rr) \left( \sigma_y \tau_z \right) - u_{l,\YO}(\rr) \left( \sigma_x \tau_0 \right) \right]_{\alpha'\eta',\alpha\eta} \psi_{l,\alpha,\eta,s}(\rr)
\end{align}
\par 

Note that, if defining
\begin{align}
    \left[\AA^{(l)}(\rr)\right]_{\eta',\eta} = \frac{1}{2\sqrt{2}} \frac{1}{r_0} \left( \partial_x u_{l,\XA}(\rr) - \partial_{y} u_{l,\YA}(\rr) , - \partial_{y} u_{l,\XA}(\rr) - \partial_{x} u_{l,\YA}(\rr) \right) \otimes [\tau_z]_{\eta',\eta}
\end{align}
then the acoustic part in $\hat{H}_{\rm epc, SLG, \Gamma}$ modifies $\hat{H}_{0,\mrm{SLG}}$ into the form of a Dirac fermion coupled to a pseudo gauge field,   
\begin{align}
    \hat{H}_{0,\mrm{SLG}} + \hat{H}^{(ac)}_{\mrm{epc,SLG},\GGamma} 
    = v_F \int \mrm{d}^2\rr ~ \psi^\dagger_{l,\alpha',\eta',s}(\rr) \left[ \left( -i\partial_{x} - \AA^{(l)}_x(\rr) \right) \sigma_x\tau_z + \left( -i\partial_y - \AA^{(l)}_y(\rr) \right) \sigma_y\tau_0 \right]_{\alpha'\eta',\alpha\eta} \psi_{l,\alpha,\eta,s}(\rr)
\end{align}
which takes the form as derived in Ref. \cite{Suzuura_2002, Koshino_epc_2020}. On the other hand, if identifying that $\sqrt{6}\frac{v_F}{a_0 r_0} = \frac{3}{\sqrt{2}} \frac{|V_\pi|}{r_0} \approx 145\mrm{eV\cdot nm^{-1}}$, then the optical EPC vertex along with the corresponding EPC strengths (the second term in Eq. (\ref{eqn:Hepc_SLG_G_r}) involving the $\XO$ and $\YO$ branches) will take the same form as that used in Ref. \cite{Basko_2008, Wu_2018_SCop}. \par

\subsubsection{\texorpdfstring{$\hat{H}_{{\rm epc, SLG}, K}$}{Hepc,SLG,K}} \label{app:cont-SLG-K}

We then investigate the $K$-phonons involved in Eq. (\ref{eq:HepcSLG_prim}). For this sake, we focus on those $\uu^{(\alpha)}_{\qq}$ with $\qq\approx\eta\KK$, which impose $\eta'=\ovl{\eta}$. To begin with, we exploit the SLG coordinates and re-define $\qq\to\qq+\eta\GGamma'$, so that for the re-defined $\qq$, there is $\qq\cdot\dd_j\ll1$. 
\begin{align}
\label{eq:Hepc_SLG_K_1}
    \hat{H}_{\mrm{epc,SLG},K} =& \frac{1}{\sqrt{N_g}} \left( -\frac{V_\pi}{r_0} \right) \sum_{\eta, s} \sum_{|\kk|, |\qq| \lesssim \Lambda} \\\nonumber 
    & \psi^\dagger_{\kk+\qq,A,\ovl{\eta},s} \left[ \sum_{j=1}^{3} e^{i\kk\cdot\dd_j} e^{i\eta\frac{2\pi}{3}(j-1)} \left( \uu_{\qq+\eta\KK}^{(B)} e^{i \qq \cdot\dd_j} e^{i\eta\frac{2\pi}{3}(j-1)} - \uu_{\qq+\eta\KK}^{(A)} \right)\cdot\hat{\dd}_j \right] \psi_{\kk,B,\eta,s} + \mrm{H.c.}
\end{align}
We still start with the expression inside the square bracket. 
Even simpler than the case of $\Gamma$-phonons, here it suffices to directly approximate both $e^{i\qq\cdot\dd_j}\approx1$ and $e^{i\kk\cdot\dd_j}\approx1$, because all the in-plane $K$-phonons have finite frequencies ($\gtrsim 100$meV), thus any EPC elements linear in $|\qq|$ can be ignored. 
Thereof, we obtain
\begin{align}
    ~& \left[ \sum_{j=1}^{3} e^{i\kk\cdot\dd_j} e^{i\eta\frac{2\pi}{3}(j-1)} \left( \uu_{\qq+\eta\KK}^{(B)} e^{i \qq \cdot \dd_j} e^{i\eta\frac{2\pi}{3}(j-1)} - \uu_{\qq+\eta\KK}^{(A)} \right)\cdot\hat{\dd}_j \right] \\\nonumber 
    \approx& \sum_{j=1}^{3} \left( \uu_{\qq+\eta\KK}^{(B)} e^{-i\eta\frac{2\pi}{3}(j-1)} - \uu_{\qq+\eta\KK}^{(A)} e^{i\eta\frac{2\pi}{3}(j-1)} \right)\cdot\hat{\dd}_j \\\nonumber 
    =& \left( \frac{3}{2} \right) \left[ (-i\eta) \left(\uu_{\qq+\eta\KK,x}^{(A)} + \uu_{\qq+\eta\KK,x}^{(B)} \right) - \left(\uu_{\qq+\eta\KK,y}^{(A)} - \uu_{\qq+\eta\KK,y}^{(B)}\right) \right] 
\end{align}
To compare this expression with the definition of $K$-phonon modes, Eq. (\ref{eq:uK_q}), we first recover the layer index, and rotate it back to the MATBG coordinates, which generates negligible $O(\theta)$ corrections. Then we exploit $\GGamma' = \KK^{(l)} - \KK_l$, and write $\qq + \eta\KK^{(l)} = \qq + \eta\GGamma' + \eta\KK_l$. Therefore, 
\begin{align}
\label{eq:simplify_K}
    \left( \frac{3}{2} \right) \left[ (-i\eta) \left(\uu_{\qq+\eta\KK^{(l)},x}^{(A)} + \uu_{\qq+\eta\KK^{(l)},x}^{(B)} \right) - \left(\uu_{\qq+\eta\KK^{(l)},y}^{(A)} - \uu_{\qq+\eta\KK^{(l)},y}^{(B)}\right) \right] 
    =& \left( \frac{3}{\sqrt{2}} \right) \left[ u_{\qq+\eta\KK_l, l, \Aone} - i \eta u_{\qq+\eta\KK_l, l, \Bone} \right]
\end{align}
By inserting Eq. (\ref{eq:simplify_K}) into Eq. (\ref{eq:Hepc_SLG_K_1}), and then adding up the two layers, 
\begin{align}
    \hat{H}_{\mrm{epc,SLG},K} 
    =& \frac{1}{\sqrt{N_g}} \left( -\frac{V_\pi}{r_0} \right) \left( \frac{3}{\sqrt{2}} \right) \sum_{\eta,s,l} \sum_{|\kk|, |\qq| \lesssim \Lambda} \psi^\dagger_{\kk+\qq,l,A,\ovl{\eta},s}  \left[ u_{\qq+\eta\KK_l , l, \Aone} - i\eta u_{\qq+\eta\KK_l,l,\Bone} \right] \psi_{\kk,l,B,\eta,s} \\\nonumber 
    +& \frac{1}{\sqrt{N_g}} \left( - \frac{V_\pi}{r_0} \right) \left( \frac{3}{\sqrt{2}}\right) \sum_{\eta,s,l} \sum_{|\kk|, |\qq| \lesssim \Lambda}  \psi^\dagger_{\kk,l,B,{\eta},s}  \left[ u_{-\qq-\eta\KK_l , l, \Aone} + i\eta u_{-\qq-\eta\KK_l,l,\Bone} \right] \psi_{\kk+\qq,l,A,\ovl{\eta},s}
\end{align}
where we again exploited $u_{\qq,l,\chi}^\dagger = u_{-\qq,l,\chi}$ for $\chi@\KK\in\{\Aone, \Bone\}$. we have also written out the Hermitian conjugate terms explicitly. By a relabeling of the summed variables, $\ovl{\eta}\to\eta$, $\kk+\qq\to\kk$, and then $-\qq\to\qq$, 
\begin{align}
    \hat{H}_{\mrm{epc,SLG},K} 
    =& \frac{1}{\sqrt{N_g}} \left( -\frac{V_\pi}{r_0} \right) \left( \frac{3}{\sqrt{2}} \right) \sum_{\eta,l,s} \sum_{|\kk|, |\qq| \lesssim \Lambda} \psi^\dagger_{\kk+\qq,l,A,\ovl{\eta},s}  \left[ u_{\qq+\eta\KK_l , l, \Aone} - i\eta u_{\qq+\eta\KK_l,l,\Bone} \right] \psi_{\kk,l,B,\eta,s} \\\nonumber 
    +& \frac{1}{\sqrt{N_g}} \left( - \frac{V_\pi}{r_0} \right) \left( \frac{3}{\sqrt{2}}\right) \sum_{\eta,l,s} \sum_{|\kk|, |\qq| \lesssim \Lambda} \psi^\dagger_{\kk+\qq,l,B,\ovl{\eta},s}  \left[ u_{\qq+\eta\KK_l , l, \Aone} - i\eta u_{\qq+\eta\KK_l,l,\Bone} \right] \psi_{\kk,l,A,\eta,s}
\end{align}
Using the Pauli's $\sigma_\mu$ and $\tau_\mu$, we obtain, 
\begin{align}
\label{eqn:Hepc_SLG_K_2layer}
    \hat{H}_{\mrm{epc,SLG},K} 
    = \frac{1}{\sqrt{N_g}} \left( -\frac{V_\pi}{r_0} \right) \left( \frac{3}{\sqrt{2}} \right) \sum_{\eta,l,s} \sum_{|\kk|, |\qq| \lesssim \Lambda} \psi^\dagger_{\kk+\qq,l,\alpha',\eta',s}  \left[ u_{\qq+\eta\KK_l , l, \Aone} \left(\sigma_x\tau_x\right) - u_{\qq+\eta\KK_l,l,\Bone} \left(\sigma_x\tau_y\right) \right]_{\alpha'\eta',\alpha\eta} \psi_{\kk,l,\alpha,\eta,s} 
\end{align}
Fouriered to the real-space continuum with Eq. (\ref{eq:psi_r}) and (\ref{eq:u_r}), and replacing $v_F = -\frac{\sqrt{3}}{2} a_0 V_\pi$, 
\begin{align}
\label{eqn:Hepc_SLG_K_r}
    \hat{H}_{\mrm{epc,SLG},K} = \left( \sqrt{6} \frac{ v_F}{a_0r_0} \right) \sum_{\eta,s,l} \int\mrm{d}^2\rr~  \psi^\dagger_{l,\alpha',\eta',s}(\rr) \left[ u_{l,\Aone}(\rr) \left( \sigma_x \tau_x \right) - u_{l,\Bone}(\rr) \left(\sigma_x \tau_y \right) \right]_{\alpha'\eta',\alpha\eta} \psi_{l,\alpha,\eta,s}(\rr) e^{-i\eta\KK_l \cdot\rr} 
\end{align}
Under the gauge choice \cref{eq:psi_TR,eq:uq_TR}, the above expressions are translationally invariant. By re-choosing the gauge for electron fields in accordance with the phonon fields, \cref{eqn:Hepc_SLG_K_r} will take an identical form to that used in Ref. \cite{Basko_2008, Wu_2018_SCop}. 

Using the SLG coordinates, lattice vibration corresponding to the phonon $u_{0,\Aone}$ takes the real-space pattern, 
\begin{align}
    \uu(\RR^{(A)}) = u_{0,\Aone} ( - \sin \KK\cdot\RR^{(A)} , - \cos \KK \cdot \RR^{(A)} ), \quad 
    \uu(\RR^{(B)}) = u_{0,\Aone} ( - \sin \KK\cdot\RR^{(B)} ,  \cos \KK \cdot \RR^{(B)} )
\end{align}
while lattice vibration corresponding to the phonon $u_{0,\Bone}$ takes the real-space pattern, 
\begin{align}
    \uu(\RR^{(A)}) = u_{0,\Bone} ( \cos \KK\cdot\RR^{(A)} , - \sin \KK \cdot \RR^{(A)} ), \quad 
    \uu(\RR^{(B)}) = u_{0,\Bone} ( \cos \KK\cdot\RR^{(B)} ,  \sin \KK \cdot \RR^{(B)} )
\end{align}
It can thus be verified that, both modes change the nearest-neighboring bond lengths, thus induce EPC.

\subsection{Inter-layer moir\'e couplings} \label{app:cont-Moire}

The inter-layer tunnelings are given by, 
\begin{align}
    \hat{H}_{\rm Moire} = \sum_{s} \sum_{\RR^{(1, \alpha)}}  \sum_{\RR^{(2, \alpha')}} \psi_s^\dagger(\RR^{(1, \alpha)})  t(\RR^{(1, \alpha)} + \uu(\RR^{(1, \alpha)}) - \RR^{(2, \alpha')} - \uu(\RR^{(2, \alpha')}))  \psi_s(\RR^{(2, \alpha')}) + \mrm{H.c.}
\end{align}
By projecting to the low-energy Dirac electrons with Eq. (\ref{eq:psi_k}), there is
\begin{align}
    \hat{H}_{\rm Moire} = \sum_{s} \psi^\dagger_{\kk, 1, \alpha, \eta, s} \langle 1,\alpha,\eta,\kk | \hat{H} | 2,\alpha',\eta',\kk'\rangle_\uu \psi_{\kk', 2, \alpha', \eta', s} + \mrm{H.c.}
\end{align}
with 
{\small \begin{align} \label{eq:mat_ele_Moire}
    \langle 1,\alpha,\eta,\kk | \hat{H} | 2,\alpha',\eta',\kk'\rangle_\uu = \frac{1}{N_g} \sum_{\RR^{(1,\alpha)}} \sum_{\RR^{(2,\alpha')}} e^{-i(\kk+\eta\KK^{(1)})\cdot\RR^{(1,\alpha)}} t(\RR^{(1,\alpha)} + \uu(\RR^{(1,\alpha)}) - \RR^{(2,\alpha')} - \uu(\RR^{(2,\alpha')}))    e^{i(\kk'+\eta'\KK^{(2)})\cdot\RR^{(2,\alpha')}} 
\end{align} }
We can separate $\uu = \uu_\parallel + \uu_z$ into the in-plane components and the out-of-plane component, hence
\begin{align}
    t(\RR^{(1,\alpha)} + \uu(\RR^{(1,\alpha)}) - \RR^{(2,\alpha')} - \uu(\RR^{(2,\alpha')})) &= t(\RR^{(1,\alpha)} + \uu_\parallel(\RR^{(1,\alpha)}) - \RR^{(2,\alpha')} - \uu_\parallel(\RR^{(2,\alpha')})) \\\nonumber
    &+ \partial_z t(\RR^{(1,\alpha)} - \RR^{(2,\alpha')})  \left[ \uu_z(\RR^{(1, \alpha)}) - \uu_z(\RR^{(2, \alpha')}) \right]
\end{align}
To proceed, we divide the 3D parameter space of $\rr$ into multiple 2D $\rr_x$-$\rr_y$ planes labeled by $\rr\cdot\hat{\mbf{z}}$. Within each plane, $t(\rr)$ can be Fouriered as (with $\mrm{d}^2\rr = \mrm{d}\rr_x \mrm{d}\rr_y$), 
\begin{align}
    t_{\pp}^{(\rr\cdot\hat{\mbf{z}})} = \int \mrm{d}^2\rr ~ e^{-i\pp\cdot\rr} ~ t(\rr) , \qquad \qquad
    t(\rr) = \int \frac{\mrm{d}^2\pp}{(2\pi)^2} ~ e^{i\pp\cdot\rr} ~ t_{\pp}^{(\rr\cdot\hat{\mbf{z}})}
\end{align}
Note that $\pp$ can run over the entire 2D momentum space. Correspondingly, 
\begin{align}
    \partial_z t_\pp^{(\rr \cdot \hat{\mbf{z}})}  = \int \mrm{d}^2\rr ~ e^{-i\pp \cdot \rr} ~ \partial_z t(\rr), 
    \qquad \qquad \partial_z t(\rr) = \int \frac{\mrm{d}^2 \pp}{(2\pi)^2} ~ e^{i\pp \cdot \rr} ~ \partial_z t_\pp^{(\rr \cdot \hat{\mbf{z}})}
\end{align}
Consequently, 
\begin{align}   \label{eq:tRRuu}
    & t(\RR^{(1,\alpha)} + \uu(\RR^{(1,\alpha)}) - \RR^{(2,\alpha')} - \uu(\RR^{(2,\alpha')})) \\\nonumber
    &= \int \frac{\mrm{d}^2\pp}{(2\pi)^2} ~ t^{(c_0)}_\pp e^{i\pp \cdot (\RR^{(1,\alpha)} - \RR^{2,\alpha'})} e^{i\pp\cdot(\uu_\parallel(\RR^{(1,\alpha)}) -\uu_\parallel(\RR^{(2,\alpha')}))} + \int \frac{\mrm{d}^2\pp}{(2\pi)^2} ~ \partial_z t^{(c_0)}_{\pp} e^{i\pp \cdot (\RR^{(1,\alpha)} - \RR^{2,\alpha'})}  \left[ \uu_z(\RR^{(1, \alpha)}) - \uu_z(\RR^{(2, \alpha')}) \right] \\\nonumber
    &\approx \int \frac{\mrm{d}^2\pp}{(2\pi)^2} ~ \left[ t^{(c_0)}_\pp + t^{(c_0)}_\pp \left[ i\pp \cdot (\uu_{\parallel}(\RR^{(1,\alpha)}) - \uu_\parallel(\RR^{2,\alpha'})) \right]  + \partial_z t^{(c_0)}_\pp \left[ \uu_z(\RR^{(1, \alpha)}) - \uu_z(\RR^{(2, \alpha')}) \right]  \right] e^{i\pp \cdot (\RR^{(1,\alpha)} - \RR^{2,\alpha'})} 
\end{align}
where we have used the fact that $(\RR^{(1,\alpha)} - \RR^{(2,\alpha')}) \cdot \hat{{\mbf{z}}} = c_0$ is the layer spacing for arbitrary $\alpha, \alpha'$, and then approximated the expression to the first order of $\uu_\parallel$ and $\uu_z$. 
The three terms in the line expression of \cref{eq:tRRuu} describe the moir\'e coupling with no vibration, with in-plane vibration, and with out-of-plane vibration, respectively. Inserting them back to \cref{eq:mat_ele_Moire}, they separate the inter-layer matrix element into three terms, 
\begin{align}
    \langle 1,\alpha,\eta,\kk | \hat{H} | 2,\alpha',\eta',\kk'\rangle_\uu = \langle 1,\alpha,\eta,\kk | \hat{H} | 2,\alpha',\eta',\kk'\rangle_{0} + \langle 1,\alpha,\eta,\kk | \hat{H} | 2,\alpha',\eta',\kk'\rangle_{\parallel} + \langle 1,\alpha,\eta,\kk | \hat{H} | 2,\alpha',\eta',\kk'\rangle_{z}
\end{align}
which give rise to the moir\'e coupling in the continuum model $\hat{H}_{\rm 0, Moire}$, and the in-plane and out-of-plane EPC vertices, $\hat{H}_{\rm epc, Moire, \parallel} + \hat{H}_{\rm epc, Moire, z}$. We now examine each term separately.

\subsubsection{\texorpdfstring{$\hat{H}_{0,\mrm{Moire}}$}{H0, moir\'e}} \label{app:cont-Moire-H0}

Inserting the non-vibrating term (the first term) of \cref{eq:tRRuu} into \cref{eq:mat_ele_Moire}, there is
\begin{align}
     \langle 1,\alpha,\eta,\kk | \hat{H} | 2,\alpha',\eta',\kk'\rangle_{0} =  \frac{1}{N_g} \int\frac{\mrm{d}^2\pp}{(2\pi)^2} ~ t_{\pp}^{(c_0)} \left[ \sum_{\RR^{(1,\alpha)}} e^{-i(\kk+\eta\KK^{(1)}-\pp) \cdot \RR^{(1,\alpha)}} \right] \left[ \sum_{\RR^{(2,\alpha')}}  e^{i(\kk'+\eta'\KK^{(2)}-\pp) \cdot \RR^{(2,\alpha')}} \right] 
\end{align}
so that $\sum_{\RR^{(1,\alpha)}}$ and $\sum_{\RR^{(2,\alpha')}}$ are decoupled and thus can be evaluated independently. The following identity will be of great use, 
\begin{align}
\label{eqn:sum_R}
    \frac{1}{N_g} \sum_{\RR^{(l,\alpha)}} e^{i(\kk+\eta\KK^{(l)}-\pp) \cdot \RR^{(l,\alpha)}} = \sum_{\GG^{(l)}} \delta_{\pp,\kk+\eta\KK^{(l)}+\GG^{(l)}} e^{-i\GG^{(l)}\cdot\ttau^{(l,\alpha)}}
\end{align} 
where $\GG^{(l)}$ is the graphene reciprocal lattice vector of layer $l$, and $\ttau^{(l,\alpha)}$ denotes position of the $\alpha$-sublattice modulo graphene lattice vectors $\RR^{(l)}$. Recall that in our scenario, $|\kk|\ll |\KK^{(l)}|$, and $\pp$ can run over the entire 2D momentum space. With these knowledge, by further discretizing the integral $\int\frac{\mrm{d}^2\pp}{(2\pi)^2} \to \frac{1}{N_g\Omega_g}\sum_{\pp}$ (where $\Omega_g$ denotes the area of a graphene unit cell), we obtain, 
\begin{align}
    \langle 1,\alpha,\eta,\kk | \hat{H} | 2,\alpha',\eta',\kk'\rangle_0 = \sum_{\pp} \sum_{\GG^{(1)}} \sum_{\GG^{(2)}} \frac{t_{\pp}^{(c_0)} }{\Omega_g} e^{i\GG^{(1)}\cdot\ttau^{(1,\alpha)}} e^{-i\GG^{(2)}\cdot\ttau^{(2,\alpha')}} \delta_{\pp,\kk+\eta\KK^{(1)}+\GG^{(1)}} \delta_{\pp,\kk'+\eta'\KK^{(2)}+\GG^{(2)}} 
\end{align}
Since $t_\pp^{(c_0)}$ will decay fast with $|\pp|$, it suffices to keep the shortest $\pp$'s. If $\eta=\eta'$, as $|\kk|,|\kk'|\ll|\KK^{(l)}|$, dictated by the $\delta$-factors, the shortest possible $\pp$ is reached by three terms, namely those satisfying $\pp\approx \eta\KK^{(l)} + \GG^{(l)} = \eta C_{3z}^{j-1}\KK^{(l)}$ for $j=1,2,3$. 
For such terms, since $\left|\frac{1}{t_{\pp}^{(c_0)}} \frac{\partial t^{(c_0)}_{\pp}}{\partial \pp}\right| \sim \left|\frac{1}{t(\rr)}\frac{\partial t(\rr)}{\partial \rr}\right|^{-1}  \sim r_0$, and the typical deviation of $\pp$ from $\eta C_{3z}^{j-1} \KK^{(l)}$ can be estimated as $|\pp - \eta C_{3z}^{j-1} \KK^{(l)}| \sim|\kk|\sim|\kk'| \sim k_\theta \ll |\KK^{(l)}|$, and $k_\theta r_0 \ll 1$, therefore, we can approximate $\frac{t^{(c_0)}_{\pp}}{\Omega_g} \approx \frac{t^{(c_0)}_{\eta C_{3z}^{j-1}\KK^{(l)}}}{\Omega_g} = w_1 \approx 112$meV. 
Such an approximation is equivalent to neglecting non-local inter-layer tunnelings. 
These three terms will correspond to the scattering processes that satisfy $\kk - \kk' = \eta C_{3z}^{j-1}\left( \KK^{(2)} - \KK^{(1)} \right)= \eta\qq_j$ (see Fig.\ref{fig:MBZ}), respectively. Moreover, we define and calculate
\begin{align}
    \left[T_{j,\eta}\right]_{\alpha,\alpha'} = e^{i\GG^{(1)}\cdot\ttau^{(1,\alpha)}} e^{-i\GG^{(2)}\cdot\ttau^{(2,\alpha')}} = \begin{pmatrix}
        1 & e^{-i\eta\frac{2\pi}{3}(j-1)} \\
        e^{i\eta\frac{2\pi}{3}(j-1)} &  1 \\
    \end{pmatrix}_{\alpha,\alpha'}, ~~~~\mrm{if}~\GG^{(l)} + \eta\KK^{(l)} = \eta C_{3z}^{j-1} \KK^{(l)} 
\end{align}
\par 

On the other hand, if $\ovl{\eta} = \eta'$, to match the $\delta$-signs, the length of the shortest $\GG^{(l)}$, and therefore, the shortest $\pp$, can be estimated as $|\pp| \sim |\GG^{(l)}| \sim \frac{|\KK^{(l)}|}{\theta}$, rendering the corresponding $t^{(c_0)}_\pp$ vanishingly small. Thereby, it suffices to keep the intra-layer elements,   
\begin{align}
    \langle 1,\alpha,\eta,\kk | \hat{H} | 2,\alpha',\eta',\kk'\rangle_0 \approx \delta_{\eta,\eta'} \sum_{j} w_1 \left[ T_{j,\eta} \right]_{\alpha,\alpha'} \delta_{\kk,\kk'+\eta\qq_j} 
\end{align}
Combined with its Hermitian conjugate, we write down the second-quantized form of $\hat{H}_{0,\mrm{Moire}}$, 
\begin{align}
\label{eqn:H0_Moire_2}
    \hat{H}_{0,\mrm{Moire}} = w_1 \sum_{\eta} \sum_{\kk} \sum_{j=1}^{3} \psi^\dagger_{\kk+\eta\qq_j,1,\alpha,\eta} \left[ T_{j,\eta} \right]_{\alpha,\alpha'} \psi_{\kk,2,\alpha',\eta} + \mrm{H.c.} 
\end{align}
which when inverse-Fouriered to the continuum real-space with \cref{eq:psi_r} reads, 
\begin{align}
\label{eqn:H0_Moire}
    \hat{H}_{0,\mrm{Moire}} = w_1 \sum_{\eta} \int \mrm{d}^2\rr ~ \psi^\dagger_{l,\alpha,\eta}(\rr) \left[T(\rr) \right]_{l\alpha\eta,l'\alpha'\eta'} \psi_{l',\alpha',\eta'}(\rr) 
\end{align}
where $T(\rr)$ is a layer-off-diagonal Hertimian matrix defined as $\left[T(\rr)\right]_{1\alpha\eta,2\alpha'\eta'} = \delta_{\eta,\eta'} [T^{(\eta)}(\rr)]_{\alpha,\alpha'}$, and $T^{(\eta)}(\rr) = \sum_{j=1}^{3} e^{i\eta\qq_j\cdot\rr} [T_{j,\eta}]_{\alpha, \alpha'}$ (namely, \cref{eq:Tr}),  

It can be verified that, at the AA-stacked regions ($e^{i(\qq_{j+1}-\qq_j)\cdot\rr_{AA}} = 1$), only diagonal $[T(\rr_{AA})]_{A\eta,A\eta}$ and $[T_\eta(\rr_{AA})]_{B\eta,B\eta}$ survive, while at the AB-stacked ($e^{i(\qq_{j+1} - \qq_j)\cdot\rr_{AB}} = e^{i\frac{2\pi}{3}}$) or the BA-stacked ($e^{i(\qq_{j+1}-\qq_j)\cdot\rr_{BA}} = e^{-i\frac{2\pi}{3}}$) regions, only off-diagonal $[T_\eta(\rr_{AB})]_{A\eta,B\eta}$ or $[T_\eta(\rr_{BA})]_{B\eta,A\eta}$ survive. Phenomenologically, to account for the lattice corrugation that diminishes the area of the AA-stacked regions and increases the AB/BA-stacked regions, we will introduce a numeric factor $u_0$ to the diagonal elements, as
\begin{align}
    \left[ T_{j,\eta} \right]_{\alpha,\alpha'} = \begin{pmatrix}
        u_0 & e^{-i\eta\frac{2\pi}{3}(j-1)} \\
        e^{i\eta\frac{2\pi}{3}(j-1)} & u_0  \\
    \end{pmatrix}_{\alpha,\alpha'}
\end{align}
We take $u_0\approx 0.8$ for $\theta\approx1.05^\circ$.

\subsubsection{\texorpdfstring{$\hat{H}_{\mrm{epc, Moire,\parallel}}$}{Hepc, moir\'e, in}} \label{app:cont-Moire-in}

By inserting the in-plane vibrating term of \cref{eq:tRRuu} into \cref{eq:mat_ele_Moire}, and Fouriering the phonon fields using \cref{eq:def_uq}, we get
\begin{align}   \label{eq:Hepc_para_pre}
     &~ \langle 1,\alpha,\eta,\kk | \hat{H} | 2,\alpha',\eta',\kk'\rangle_\parallel \\\nonumber 
     &= \frac{1}{N_g^{3/2}} \sum_{\RR^{(1,\alpha)}} \sum_{\RR^{(2,\alpha')}} \int\frac{\mrm{d}^2\pp}{(2\pi)^2} ~ e^{-i (\kk+\eta\KK^{(1)}) \cdot \RR^{(1,\alpha)}} \times \left( t_{\pp}^{(c_0)} \right)  e^{i\pp\cdot(\RR^{(1,\alpha)} - \RR^{(2,\alpha')})} \times \\\nonumber 
    &~ ~~~~~~~~~~~~~~~~~~~~~~~~~~~~~~ i\pp\cdot \left(\sum_{\qq\in\gBZ^{(1)}}  \uu_{\qq}^{(1,\alpha)} e^{i\qq\cdot\RR^{(1,\alpha)}} - \sum_{\qq\in\gBZ^{(2)}} \uu_{\qq}^{(2,\alpha')} e^{i\qq\cdot\RR^{(2,\alpha')}}  \right) \times e^{i(\kk'+\eta'\KK^{(2)}) \cdot \RR^{(2,\alpha')}}  \\ \nonumber 
    &= \frac{1}{N_g^{3/2}} \sum_{\qq\in\gBZ^{(1)}} \int\frac{\mrm{d}^2\pp}{(2\pi)^2} ~ i\pp\cdot\uu_{\qq}^{(1,\alpha)} \left[\sum_{\RR^{(1,\alpha)}} e^{-i(\kk - \qq+ \eta\KK^{(1)} - \pp) \cdot \RR^{(1,\alpha)}} \right] \left[\sum_{\RR^{(2,\alpha')}} e^{i(\kk' + \eta'\KK^{(2)} - \pp) \cdot \RR^{(2,\alpha')}} \right] \\\nonumber 
    &~ ~~ - \frac{1}{N_g^{3/2}} \sum_{\qq\in\gBZ^{(2)}} \int\frac{\mrm{d}^2\pp}{(2\pi)^2} ~ i\pp\cdot\uu_{\qq}^{(2,\alpha')} \left[\sum_{\RR^{(1,\alpha)}} e^{-i(\kk + \eta\KK^{(1)} - \pp) \cdot \RR^{(1,\alpha)}} \right] \left[\sum_{\RR^{(2,\alpha')}} e^{i(\kk' +\qq + \eta'\KK^{(2)} - \pp) \cdot \RR^{(2,\alpha')}} \right] 
\end{align}
Still, $\sum_{\RR^{(1,\alpha)}}$ and $\sum_{\RR^{(2,\alpha')}}$ are decoupled, and Eq. (\ref{eqn:sum_R}) can be applied to each summation. We obtain, 
\begin{align}
    &~ \langle 1,\alpha,\eta,\kk | \hat{H} | 2,\alpha',\eta',\kk'\rangle_\parallel \\\nonumber
    &= \frac{1}{\sqrt{N_g}} \sum_{\pp} \sum_{\GG^{(1)}} \sum_{\GG^{(2)}} (i\pp\cdot\uu_{\qq}^{(1,\alpha)}) \frac{t_{\pp}^{(c_0)} }{\Omega_g} e^{i\GG^{(1)}\cdot\ttau^{(1,\alpha)}} e^{-i\GG^{(2)}\cdot\ttau^{(2,\alpha')}}  \delta_{\pp,\kk-\qq+\eta\KK^{(1)}+\GG^{(1)}} \delta_{\pp,\kk'+\eta'\KK^{(2)}+\GG^{(2)}} \\\nonumber 
    &- \frac{1}{\sqrt{N_g}} \sum_{\pp} \sum_{\GG^{(1)}} \sum_{\GG^{(2)}} (i\pp\cdot\uu_{\qq}^{(2,\alpha')}) \frac{t_{\pp}^{(c_0)} }{\Omega_g} e^{i\GG^{(1)}\cdot\ttau^{(1,\alpha)}} e^{-i\GG^{(2)}\cdot\ttau^{(2,\alpha')}} \delta_{\pp,\kk+\eta\KK^{(1)}+\GG^{(1)}} \delta_{\pp,\kk'+\qq+\eta'\KK^{(2)}+\GG^{(2)}}
\end{align}

For $\GGamma$-phonons, namely those $\uu_\qq$ with $|\qq|\ll|\KK^{(l)}|$, the analysis parallels that of $\hat{H}_{0,\mrm{Moire}}$, namely only three terms that satisfy $\GG^{(l)} + \eta \KK^{(l)} = \eta C^{j-1}_{3z}\KK^{(l)} $ contribute, and they correspond to scatterings that satisfy $\kk = \kk'+\qq+\eta\qq_j$. For these terms, still, we treat $\frac{t_{\pp}^{(c_0)}}{\Omega_g} \approx \frac{t^{(c_0)}_{\eta C_{3z}^{j-1}\KK^{(l)}}}{\Omega_g} = w_1$; and moreover, by observing that $\pp \approx \eta C_{3z}^{j-1}\KK^{(l)} \approx \frac{\eta}{\theta}\hat{\mbf{z}} \times \qq_j $ up to $O(k_\theta)$ momenta, we get
\begin{equation}
\label{eqn:hop_Moire_1}
    \langle 1,\alpha,\eta,\kk | \hat{H} | 2,\alpha',\eta,\kk'\rangle_\parallel = \frac{1}{\sqrt{N_g}} \left(w_1\right) \sum_{j=1}^3 
     i\eta \left(\frac{\hat{\mbf{z}} \times \qq_j}{\theta} \right) \cdot \left(\uu^{(1,\alpha)}_{\qq} - \uu^{(2,\alpha')}_{\qq} \right) \delta_{\kk,\kk'+\qq+\eta\qq_j} [T_{j,\eta}]_{\alpha,\alpha'} , ~~\mrm{if}~|\qq|\ll|\KK^{(l)}|
\end{equation} 
By comparing to \cref{eq:uG_q,eq:uG_q_cr}, such an EPC vertex will involve the $(r,\XA/\YA)$ modes, as well as both $(c,\XO/\YO)$ and $(r,\XO/\YO)$ modes, but does not involve the $(c,\XA/\YA)$ modes. It can be quickly checked that, the EPC strengths to the $(l,\XO/\YO)$ modes are $\frac{w_1|\KK^{(l)}|}{v_F/a_0 r_0} \sim \frac{1}{30}$ times weaker than that in $\hat{H}_{\mrm{epc,SLG},\GGamma}$ (Eq. (\ref{eqn:Hepc_SLG_G_r})), hence negligible.

For $\KK$-phonons, namely those $\uu_{\qq}$ with $\qq\approx \eta\KK^{(l)}$, necessarily $\ovl{\eta} = \eta'$. If we check the $\langle 1,\alpha,\ovl{\eta},\kk|\hat{H}|2,\alpha',\eta,\kk' \rangle_\parallel$ element, then for the $\uu_{\qq}^{(1,\alpha)}$ terms, the shortest $\pp$ is reached by the following three terms, $\pp\approx \eta\KK^{(2)} + \GG^{(2)} = \eta C_{3z}^{j-1}\KK^{(2)}$, or equivalently, $\pp\approx \GG^{(1)} - 2\eta\KK^{(1)} = \eta C_{3z}^{j-1}\KK^{(1)}$; while for the $\uu_{\qq}^{(2,\alpha')}$ terms, the shortest $\pp$ is reached by the following three terms, $\pp\approx -\eta\KK^{(1)} + \GG^{(1)} = -\eta C_{3z}^{j-1}\KK^{(1)}$, or equivalently, $\pp\approx \GG^{(2)} + 2\eta\KK^{(2)} = -\eta C_{3z}^{j-1}\KK^{(2)}$. For all these cases, $\frac{t^{(c_0)}_{\pp}}{\Omega_g}\approx w_1$ as well. Consequently, the EPC strengths to the $\KK$-phonons will also be $\frac{w_1|\KK^{(l)}|}{v_F/a_0 r_0} \sim \frac{1}{30}$ times weaker than that contained in $\hat{H}_{\mrm{epc,SLG},\KK}$ (Eq. (\ref{eqn:Hepc_SLG_K_r})). We also remark that $\hat{H}_{\mrm{epc,Moire}}$ will couple to more $\KK$-phonon modes other than $(l,\Aone)$ and $(l,\Bone)$. Nevertheless, due to their weak coupling strengths, we will neglect them altogether.

Exploiting the definition of $u_{\qq,r,\XA}$ and $u_{\qq,r,\YA}$ as given in \cref{eq:uG_q,eq:uG_q_cr}, we get 
\begin{align}
    \hat{H}_{\mrm{epc,Moire,\parallel}} = \frac{w_1}{\sqrt{N_g}} \sum_{\qq,\kk} \sum_{\eta} \sum_{j=1}^{3} \psi^\dagger_{\kk+\qq+\eta\qq_j,1,\alpha,\eta} i\eta  \left[ \left( \frac{\hat{\mbf{z}}\times\qq_j}{\theta}\right)_{x} u_{\qq,r,\XA} +  \left( \frac{\hat{\mbf{z}}\times\qq_j}{\theta}\right)_{y} u_{\qq,r,\YA}  \right]  \left[ T_{j,\eta} \right]_{\alpha,\alpha'} \psi_{\kk,2,\alpha',\eta} + \mrm{H.c.}
\end{align}
Next we write it in the continuum real-space with \cref{eq:u_r,eq:psi_r}, as
\begin{align}
    \hat{H}_{\mrm{epc,Moire,\parallel}}  =& \left(\frac{w_1}{\theta}\right) \sum_{\eta} \int\mrm{d}^2\rr~ \psi^\dagger_{1,\alpha,\eta}(\rr) \left[ \left( \hat{\mbf{z}} \times \sum_{j=1}^{3} (i\eta\qq_j)  e^{i\eta\qq_j\cdot\rr} \left[ T_{j,\eta} \right]_{\alpha,\alpha'}  \right)_x u_{r,\XA}(\rr) \right. \\\nonumber 
    ~& ~~~~~~ \left. + \left( \hat{\mbf{z}} \times \sum_{j=1}^{3} (i\eta\qq_j)  e^{i\eta\qq_j\cdot\rr} \left[ T_{j,\eta} \right]_{\alpha,\alpha'}  \right)_y u_{r,\YA}(\rr)\right] \psi_{2,\alpha',\eta}(\rr) + \mrm{H.c.}
\end{align}

By noting that $\left[\partial_{\rr} T^{(\eta)}(\rr)\right]_{\alpha,\alpha'} = \sum_{j=1}^3 (i\eta \qq_{j}) e^{i\eta\qq_j\cdot\rr} \left[ T_{j,\eta} \right]_{\alpha,\alpha'}$, and defining $u_r(\rr) = (u_{r,\XA}(\rr), u_{r,\YA}(\rr))$, we arrive at
\begin{align}
\label{eqn:Hepc_Moire_r}
    \hat{H}_{\mrm{epc,Moire,\parallel}} = \left(\frac{w_1}{\theta}\right) \int\mrm{d}^2\rr~  \psi^\dagger_{l,\alpha,\eta}(\rr)  \left[u_r(\rr) \cdot \left(\hat{\mbf{z}} \times \partial_\rr T(\rr)\right) \right]_{l\alpha\eta,l'\alpha'\eta'} \psi_{l',\alpha',\eta'}(\rr) 
\end{align}
Equivalently, by writing the first-quantized form of $\hat{H}_0$ as $\left[ H(\rr) \right]_{l\alpha\eta, l'\alpha'\eta'}$, $\hat{H}_\mrm{epc,Moire}$ can be written as 
\begin{equation}
\label{eqn:Hepc_Moire_r2}
\hat{H}_{\mrm{epc,Moire,\parallel}} = \left( \frac{1}{\theta} \right)
    \int\mrm{d}^2\rr\ \psi^\dagger_{l,\alpha,\eta}(\rr) \left[ u_r(\rr) \cdot \left(\hat{\mbf{z}} \times [\partial_\rr, H(\rr)] \right) \right]_{l\alpha\eta ,l'\alpha'\eta'} \psi_{l',\alpha',\eta'}(\rr)
\end{equation}
where $H(\rr)$ is the first quantized form contained in $\hat{H}_0 = \hat{H}_{0,\mrm{SLG}} + \hat{H}_{0,\mrm{Moire}}$.

\subsubsection{\texorpdfstring{$\hat{H}_{\rm epc, Moire, z}$}{Hepc, moir\'e, z}} \label{app:cont-Moire-z}

Inserting the out-of-plane vibrating term of \cref{eq:tRRuu} into \cref{eq:mat_ele_Moire}, and exploiting the Fourier transformation \cref{eq:def_uq}, 
\begin{align}   \label{eq:Hepc_z_pre}
     & \langle 1,\alpha,\eta,\kk | \hat{H} | 2,\alpha',\eta',\kk'\rangle_{z} \\\nonumber
     &= \frac{1}{N_g^{3/2}} \sum_{\RR^{(1,\alpha)}} \sum_{\RR^{(2,\alpha')}} \int\frac{\mrm{d}^2\pp}{(2\pi)^2}  
     e^{-i (\kk + \eta\KK^{(1)} ) \cdot \RR^{(1,\alpha)}} \times  \left(\partial_{z} t^{(c_0)}_{\pp} \right) e^{i\pp\cdot(\RR^{(1,\alpha)} - \RR^{(2,\alpha')})} \times \\\nonumber
     &~ \qquad\qquad \left[ \sum_{\qq \in\gBZ^{(1)}} \uu^{(1,\alpha)}_{\qq,z} e^{i\qq\cdot\RR^{(1,\alpha)}} -  \sum_{\qq \in\gBZ^{(2)}} \uu^{(2,\alpha')}_{\qq,z} e^{i\qq\cdot \RR^{(2,\alpha')}} \right] \times e^{i(\kk'+\eta'\KK^{(2)} - \pp) \cdot \RR^{(2,\alpha')}} 
\end{align}
By noting that \cref{eq:Hepc_z_pre} takes the identical form as (the first line of) \cref{eq:Hepc_para_pre}, a similar derivation will follow. We arrive at
\begin{align}
    \langle 1,\alpha,\eta,\kk | \hat{H} | 2,\alpha',\eta',\kk'\rangle_{z} 
    &= \frac{1}{\sqrt{N_g}} \sum_{\pp} \sum_{\GG^{(1)}} \sum_{\GG^{(2)}} \uu_{\qq,z}^{(1,\alpha)} \frac{\partial_z t^{(c_0)}_\pp}{\Omega_g} e^{i\GG^{(1)} \cdot \ttau^{(1,\alpha)}} e^{-i\GG^{(2)} \cdot \ttau^{(2,\alpha')}} \delta_{\pp, \kk-\qq+\eta\KK^{(1)}+\GG^{(1)}} \delta_{\pp, \kk'+\eta'\KK^{(2)}+\GG^{(2)}} \\\nonumber
    &- \frac{1}{\sqrt{N_g}} \sum_{\pp} \sum_{\GG^{(1)}} \sum_{\GG^{(2)}} \uu_{\qq,z}^{(2,\alpha')} \frac{\partial_z t^{(c_0)}_\pp}{\Omega_g}  e^{i\GG^{(1)} \cdot \ttau^{(1,\alpha)}} e^{-i\GG^{(2)} \cdot \ttau^{(2,\alpha')}}  \delta_{\pp, \kk+\eta\KK^{(1)}+\GG^{(1)}} \delta_{\pp, \kk'+\qq+\eta'\KK^{(2)}+\GG^{(2)}} 
\end{align}
We check the intra-valley processes ($\eta=\eta'$, related to $\Gamma$-phonons) first. 
Still, it suffices to keep the shortest $\pp$, which satisfies $|\pp| \approx |\KK^{(l)}|$. Correspondingly, we approximate $\frac{\partial_z t^{(c_0)}_{\pp}}{\Omega_g} \approx \frac{\partial_z t^{(c_0)}_{\KK^{(l)}}}{\Omega_g} = -1.11 \times \frac{w_1}{r_0}$. By a similar derivation to \cref{app:cont-Moire-in}, 
\begin{align}   \label{eq:Hepc_z_2}
    \langle 1,\alpha,\eta,\kk | \hat{H} | 2,\alpha',\eta,\kk'\rangle_{z} = \frac{1}{\sqrt{N_g}} \left( \frac{-1.11 w_1}{r_0} \right) \sum_{j=1}^{3} \left( \uu_{\qq,z}^{(1,\alpha)} - \uu_{\qq,z}^{(2,\alpha')} \right) \delta_{\kk,\kk'+\qq+\eta\qq_j} [T_{j,\eta}]_{\alpha,\alpha'} 
\end{align}
Comparing to \cref{eq:uG_q,eq:uG_q_cr}, \cref{eq:Hepc_z_2} can decompose to $(r, \ZA)$ phonons as well as $(c/r, \ZO)$ phonons, with equal overall EPC strengths. However, as detailed in \cref{sec:epc-phn}, $\omega_{\qq,r,\ZA} = 11$meV while $\omega_{\qq,c/r,\ZO} = 100$meV. Correspondingly, the effective electron interactions mediated by $(c/r, \ZO)$ phonons will be $\left( \frac{\omega_{\qq,r,\ZA}}{\omega_{\qq,c/r,\ZO)}}\right)^2 \sim \frac{1}{81}$ times weaker than $(r, \ZA)$. As a result, it suffices to keep the $(r, \ZA)$ component in \cref{eq:Hepc_z_2}. For inter-valley processes ($\eta=\ovl{\eta'}$, related to $K$-phonons), the shortest $\pp$ satisfies $|\pp|\approx|\KK^{(l)}|$ as well, resulting in an overall EPC strength $\frac{w_1}{r_0}$, the same as the intra-valley processes. Since all the $K$-phonon frequencies are much higher than $\omega_{\qq,r,\ZA}$, the corresponding matrix elements can be dropped as well. Summing up, we obtain the EPC vertex induced by out-of-plane phonons as
\begin{align}
    \hat{H}_{\rm epc, Moire,z} =  \frac{-1.11 w_1 / r_0}{\sqrt{N_g}} \sum_{\qq,\kk} \sum_{\eta} \sum_{j=1}^3 
    \psi^\dagger_{\kk+\qq+\eta\qq_j, 1, \alpha, \eta}  u_{\qq, r, \ZA}  [T_{j,\eta}]_{\alpha, \alpha'}  \psi_{\kk, 2, \alpha', \eta}  + \mrm{H.c.}
\end{align}
Note that the EPC vertex is irrelevant of $\qq$, and takes the same form as the inter-valley coupling matrix in $\hat{H}_{\rm 0, Moire}$. Written in terms of the continuum fields, 
\begin{align} \label{eqn:Hepc_Moire_z}
    \hat{H}_{\rm epc, Moire,z} = -1.11 \frac{w_1}{r_0} \int\mrm{d}^2\rr ~ 
    \psi^\dagger_{ l, \alpha, \eta}(\rr)  u_{r, \ZA}(\rr)  [{T}(\rr)]_{l\alpha\eta, l'\alpha'\eta'}  \psi_{l', \alpha', \eta'}(\rr) 
\end{align}

\subsection{Summary}

The EPC vertex $\hat{H}_{\rm epc}$ in the continuum model contains \cref{eqn:Hepc_SLG_G_r,eqn:Hepc_SLG_K_r,eqn:Hepc_Moire_r,eqn:Hepc_Moire_z}, and for a quick index based on phonon modes, please refer to \cref{tab:appF_sum}. 

To convert them to the form displayed in the main text, one first needs to Fourier transform the continuum phonon fields $u_{\chi}(\rr)$ to $u_{\qq,\chi}$ based on \cref{eq:u_r}. Then, we note that \cref{eqn:Hepc_SLG_G_r} has been written in the layer basis $(l, \chi_g@\Gamma)$ with $l=1,2$, hence one needs to convert the layer basis to the co-moving/relative basis $(c/r, \chi_g@\Gamma)$ using \cref{eq:uG_q_cr}. Obtaining the $C$-matrices presented in the main text (\cref{eq:C_c_XA} to \cref{eq:C_l_B1}) will be direct. 
\begin{table}[h]
    \centering
    \begin{tabular}{c|c|c}
    \hline\hline
         & Intra-layer (SLG) & Inter-layer (Moir\'e) \\
        \hline 
        $(c,\XA/\YA)$ & \cref{eqn:Hepc_SLG_G_r} & - \\
        $(c,\XO/\YO)$ & \cref{eqn:Hepc_SLG_G_r} & - \\
        $(r,\XA/\YA)$ & \cref{eqn:Hepc_SLG_G_r} & \cref{eqn:Hepc_Moire_r} \\
        $(r,\XO/\YO)$ & \cref{eqn:Hepc_SLG_G_r} & - \\
        $(r,\ZA)$ & - & \cref{eqn:Hepc_Moire_z} \\
        $(l,\Aone/\Bone)$ & \cref{eqn:Hepc_SLG_K_r} & - \\
    \hline\hline
    \end{tabular}
    \caption{\label{tab:appF_sum} Summary of the tight-binding EPC vertices. }
\end{table}

\section{EPC vertices from deformation potential}  \label{app:deform}

In this section, we also discuss EPC vertices arising from the deformation potential \cite{Suzuura_2002, Hwang_acoustic_2008, Mariani_2010_temperature-dependent, Efetov_2010_controlling}. 
By inducing a local area modulation proportional to $\partial_x u_{l,\XA}(\rr) + \partial_y u_{l,\YA}(\rr)$, the (longitudinal) deformation of a single-layer graphene creates a variation in the background positive charge density, hence a potential to the electrons in the same layer, 
\begin{align}  \label{eq:Hepc_deform}
    \hat{H}_{\rm epc,deform} = g_1 \sum_{l,\alpha,\eta,s} \int\mrm{d}^2\rr ~ \psi^\dagger_{l,\alpha,\eta,s} (\rr) \left[ \partial_x u_{l,\XA}(\rr) + \partial_y u_{l,\YA}(\rr) \right] \psi_{l,\alpha,\eta,s}(\rr) 
\end{align}
Note that \cref{eq:Hepc_deform} is a mechanism different from the tight-binding hopping variation analyzed in \cref{app:cont}. 
\cref{eq:Hepc_deform} induces intra-layer scatterings that are \textit{diagonal} in the sublattice index $\alpha$. 
The strength of $g_1$ has been estimated to be in the range of $g_1=16\sim30$eV \cite{Suzuura_2002, Hwang_acoustic_2008, Mariani_2010_temperature-dependent, Efetov_2010_controlling}. 
We now show that, for an $f$-site, \cref{eq:G3} simply contributes a negative on-site Hubbard that can be absorbed to the much larger Coulomb $U$, and does not affect the multiplet splitting, nor the selection of symmetry-breaking correlated states.

Using \cref{eq:u_r,eq:uG_q_cr}, we can write \cref{eq:Hepc_deform} in the form of \cref{eq:Hepc_cont}, so as to obtain the corresponding $C$-matrices. We find (where the superscript $d$ is introduced to emphasize the deformation potential origin)
\begin{align}
    C^{(c,\XA/\YA),d}(\qq;\rr) = g_1 \frac{(i\qq_{x/y})}{\sqrt{2}} \rho_0 , \qquad \quad
    C^{(r,\XA/\YA),d}(\qq;\rr) = g_1 \frac{(i\qq_{x/y})}{\sqrt{2}} \rho_z  
\end{align}
Crucially, for the relative modes, $C^{(r,\XA/\YA),d}(\qq;\rr)$ anti-commute with the emergent symmetry $\td{S} = \rho_y \sigma_z$ defined in \cref{sec:vertex-G-1}. 
Therefore, similar to the tight-binding EPC vertices of $(c,\XA/\YA/\XO/\YO)$ (see \cref{sec:vertex-G-1}), the relative in-plane acoustic modes will have vanishing coupling to the $f$-orbitals through the deformation potential mechanism. It thus suffices to examine the form and strength induced by the co-moving modes. 

Since $C^{(c,\XA/\YA),d}(\qq;\rr)$ are proportional to identity in the $l,\alpha,\tau,s$ indices, the deformation potential simply couples the co-moving phonon modes to the total electron density $\sum_{l,\alpha,\eta,s} \psi^\dagger_{l,\alpha,\eta,s}(\rr) \psi_{l,\alpha,\eta,s}(\rr)$, which highly resembles the Coulomb interaction \cref{eq:HC}. 
The projected EPC vertex thus reads
\begin{align}  \label{eq:G3}
    M^{(c,\XA/\YA),d}_{\beta\eta,\beta'\eta'}(\qq) &= g_1 \cdot \delta_{\eta,\eta'} \cdot \frac{(i\qq_{x/y})}{\sqrt{2}} \cdot 
    \int\mrm{d}^2\rr ~ e^{i\qq\cdot\rr} \sum_{l,\alpha} \left[  w^{(\eta)*}_{l\alpha,\beta}(\rr) ~ w^{(\eta)}_{l\alpha,\beta'}(\rr) \right] \\\nonumber
    &= g_1 \cdot \delta_{\eta,\eta'} \delta_{\beta,\beta'} \cdot \frac{(i\qq_{x/y})}{\sqrt{2}} \cdot 
    \int\mrm{d}^2\rr ~ e^{i\qq\cdot\rr} \sum_{l,\alpha} \left[  w^{(\eta)*}_{l\alpha,\beta}(\rr) ~ w^{(\eta)}_{l\alpha,\beta}(\rr) \right] \\\nonumber
    &= g_1 \cdot \delta_{\eta,\eta'} \delta_{\beta,\beta'} \cdot \frac{(i\qq_{x/y})}{\sqrt{2}} \cdot \left[ \alpha_1^2 e^{-\frac{|\qq|^2 \lambda_1^2}{4}} + \alpha_2^2 e^{-\frac{|\qq|^2\lambda_2^2}{4}} \left( 1 - \frac{|\qq|^2}{4} \right) \right] \\\nonumber
    &\approx g_1 \cdot \delta_{\eta,\eta'} \delta_{\beta,\beta'} \cdot \frac{(i\qq_{x/y})}{\sqrt{2}} \cdot e^{-\frac{|\qq|^2{\lambda'}^2}{4}}
\end{align}
where the integral can be understood as a projected density form factor. Below we first explain the results in \cref{eq:G3}. 

The 2nd line of \cref{eq:G3} derives from the following fact: as analyzed in Ref. \cite{Song_2022}, due to the $C_{2z}T$ and $P$ symmetries, the density form factors, hence \cref{eq:G3}, must be proportional to $\delta_{\beta,\beta'}$. 
To be concrete, $C_{2z}T$ (\cref{eq:w_Gauss_2}) and $P$ (\cref{eq:DP_eig}) dictate that
\begin{align}
    w^{(+)*}_{l\alpha,\beta}(\rr) = w^{(+)}_{l\ovl{\alpha},\ovl{\beta}}(-\rr) = (-i) (-1)^{\beta} (-1)^l w^{(+)}_{\ovl{l}\ovl{\alpha},\ovl{\beta}}(\rr)
\end{align}
Therefore, for the square bracket in \cref{eq:G3}, if $\beta'=\ovl{\beta}$, there will be
\begin{align}
    \sum_{l,\alpha} \left[  w^{(+)*}_{l\alpha,\beta}(\rr) ~ w^{(+)}_{l\alpha,\ovl{\beta}}(\rr) \right] &= (-i)(-1)^{\beta} \sum_{l,\alpha} \left[ (-1)^l w^{(+)}_{\ovl{l}\ovl{\alpha},\ovl{\beta}}(\rr)  w^{(+)}_{l \alpha,\ovl{\beta}}(\rr) \right] \\\nonumber
    &= (-i)(-1)^{\beta} \sum_{l} \left[ (-1)^l w^{(+)}_{\ovl{l}\ovl{\alpha},\ovl{\beta}}(\rr)  w^{(+)}_{l \alpha,\ovl{\beta}}(\rr) + (-1)^l w^{(+)}_{\ovl{l} {\alpha},\ovl{\beta}}(\rr)  w^{(+)}_{l \ovl{\alpha},\ovl{\beta}}(\rr) \right] \\\nonumber
    &= (-i)(-1)^{\beta} \sum_{l} \left[ (-1)^l w^{(+)}_{\ovl{l}\ovl{\alpha},\ovl{\beta}}(\rr)  w^{(+)}_{l \alpha,\ovl{\beta}}(\rr) - (-1)^l w^{(+)}_{\ovl{l} \ovl{\alpha},\ovl{\beta}}(\rr) w^{(+)}_{l {\alpha},\ovl{\beta}}(\rr)  \right] = 0
\end{align}
where we first explicitly wrote out the summation over $\alpha$, and then swapped the positions $w^{(+)}_{\ovl{l} \alpha,\ovl{\beta}}(\rr)$ and $w^{(+)}_{l \ovl{\alpha},\ovl{\beta}}(\rr)$, and finally changed the dummy variable $l\to\ovl{l}$.

To obtain the 3rd line in \cref{eq:G3}, we explicitly calculated the density form factor using the Gaussian fits \cref{eq:w_Gauss_A,eq:w_Gauss_B}, and to obtain the 4th line in \cref{eq:G3}, we approximated the density form factor using one Gaussian packet that approaches $1$ as $\qq \to 0$, and spans a width $\lambda'$ as an average of $\lambda_{1,2}$. 

Using \cref{eq:G3}, we can calculate the zero-frequency interaction, which thus takes a density-density form as the on-site Hubbard $U$. 
The frequency dependence can be argued as un-important due to the slow kinetic energy of the $f$-electrons, similar to the tight-binding hopping variation mechanism. In terms of definition \cref{eq:V}, the effective interaction induced by \cref{eq:G3} reads
\begin{align}   \label{eq:G6}
    V^{\eta' \eta}_{\beta'\alpha', \beta\alpha} &= - \frac{\delta_{\beta',\beta} \delta_{\alpha'\alpha}}{N_g m_0} \sum_{\chi = (c,\XA/\YA)} \sum_{|\qq| \lesssim \Lambda} \frac{M^{(\chi), d}_{\beta\eta',\beta\eta'}(-\qq)  M^{(\chi), d}_{\alpha\eta,\alpha\eta}(\qq)}{v^2_s |\qq|^2} \\\nonumber
    &= - \delta_{\beta',\beta} \delta_{\alpha'\alpha} \cdot \frac{(\frac{g_1}{v_s})^2}{2 m_0} \cdot \theta^2\Omega_M \int \frac{\mrm{d}^2\qq}{(2\pi)^2} ~ e^{-\frac{|\qq|^2 {\lambda'}^2}{2}} = - \delta_{\beta',\beta} \delta_{\alpha'\alpha} \cdot \frac{(\frac{g_1}{v_s})^2}{2 m_0} \cdot 1.35 \times 10^{-3}
\end{align}
In the last line, we have replaced $\frac{1}{N_g} \sum_{\qq} \to \Omega_g \int\frac{\mrm{d}^2\qq}{(2\pi)^2}$. 
If assuming $g_1=19$eV \cite{Hwang_acoustic_2008}, $V^{\eta'\eta}_{\beta\alpha,\beta\alpha} = - 6.75$meV. 
In terms of the symmetry-parametrized form \cref{eq:Vee,eq:Ve}, one can verify that \cref{eq:G6} does \textit{not} contribute to any multiplet-splitting parameters $J_{\rm a,b,d,e}$, but only to a Hubbard $J_{\rm c}$ that can be absorbed to the repulsive $U$.

\end{document}